# Slater–Condon Rules and Spin–Orbit Couplings: 2-(2-(2,5-dimethoxybenzylidene)hydrazineyl)-4-(trifluoromethyl)thiazole a test case

Hasnain Mehmood,[1] Tashfeen Akhtar,[1] Jesús Espinosa-Romero,[2] Mauricio Maldonado-Domínguez,*[2] Jakub Višňák[3,4] and Mirza Wasif Baig*[4]

[1] Department of Chemistry, Mirpur University of Science and Technology (MUST), 10250-Mirpur (AJK) Pakistan.

[2] Facultad de Química, Departamento de Química Orgánica, Universidad Nacional Autónoma de México, 04510 Ciudad de México, México.

[3] Faculty of Mathematics and Physics, Charles University, Ke Karlovu 3, 12116 Prague, Czech Republic

[4] J. Heyrovský Institute of Physical Chemistry of the Czech Academy of Sciences, Dolejškova 2155/3, 18223 Prague 8, Czech Republic.

**Abstract**

Light-atom chromophores can display properties often associated with heavy-atom compounds, such as intersystem crossing leading to phosphorescence and singlet oxygen generation, yet their use remains comparatively underexplored. Here, we report the synthesis of HM610, a derivative of the benzylidenehydrazinylthiazole light-atom chromphore backbone. Spin–orbit couplings (SOCs), computed with the sf-X2C-S-TDDFT method, follow Slater–Condon rules and predict moderate values. Trajectory surface hopping simulations further illustrate the role of dynamical effects in promoting ISC, yet these results together establish that HM610 has only limited potential as a triplet sensitizer without further structural modification, such as heavy atom substitution. Based on the benchmarked (TD)DFT protocol, a computational set studying six systematic analogues allowed us to study the influence of electron-donating (–OMe) and electron-withdrawing (–$CF_3$) substituents on the common backbone, revealing the impact of substitution on the geometry and photophysics of light atom analogues of HM610 and pave the way for future studies where the introduction of heavy atoms and their impact on triplet sensitization by this family of chromophores can be probed.

## Introduction

Thiazole derivatives represent a versatile class of heterocycles with broad relevance in medicinal chemistry and functional materials.[1-5] Their electronic properties, dictated by the N,S-containing polarizable framework, make them attractive scaffolds for exploring photo-induced processes, where intersystem crossing (ISC) and spin–orbit coupling (SOC) may strongly influence reactivity and application.[6,7] Despite this importance, the excited-state dynamics of thiazole derivatives remains underexplored compared to other heteroaromatics, and their potential to serve as photon-to-chemistry energy transducers is mostly unknown.

Theoretical approaches, such as density functional theory (DFT), have been successful in computing the energetics of a broad range of molecular systems from small metal clusters[10] to medium-sized molecules[11,12] ultimately leading to complex condensed-phase molecular assemblies.[13] Its time-dependent extension (TDDFT)[14] provides practical tools for mapping electronic excitations,[15] while higher-level correlated methods like coupled cluster singles and doubles (CC2) and the algebraic diagrammatic construction scheme through second order ADC(2) are increasingly used as benchmarks for validating functional performance in the description of excited states for single-reference molecules.[16–20] Recent work shows that SOC can be appreciable even in molecules composed solely of light atoms,[21] raising questions about their role in ISC of organic chromophores. Spin–vibronic mechanism can also be crucial for inducing ISC in molecules with small SOCs.[22] Trajectory surface hopping (TSH)[23] which is a mixed quantum-classical method that is extensively used to simulate chemical reactions[24a,b] and excited-state dynamics[24c,d] of photo-active molecules can implicitly include first order spin–vibronic effects.[22a,24d]

In this study, we report the synthesis of 2-(2-(2,5-dimethoxybenzylidene)hydrazinyl)-4-(trifluoromethyl)thiazole (**HM610**), a derivative of the benzylidenehydrazinylthiazole organic backbone, and investigate its excited states through a combined experimental–computational strategy. Benchmarking multiple functionals against CC2 and ADC(2) allows us to assess the reliability of TDDFT descriptions. Relativistic two-component sf-X2C-S-TDDFT method[25] has been employed to explore SOCs between states of different multiplicities. TSH simulations have been employed to reveal influence of dynamical effects on SOCs in this light-atom based chromophore. TSH simulations for **HM610** evaluates dependence of SOCs on nuclear coordinates.

To place this molecule within a broader context, we constructed a systematic series of six analogues derived from the same thiazole–hydrazone–aryl backbone. The set spans different combinations of electron-donating (–OMe) and electron-withdrawing (–$CF_3$) substituents at three defined positions, ranging from the unsubstituted parent to fully substituted variants, including HM610. Taken together, these results establish HM610 and its analogues as reference systems for probing ISC in light-atom heterocycles, illustrating how donor–acceptor substitution patterns modulate their geometrical and photophysical landscape, and allowing future comparisons with other light-atom and with heavy-atom containing chromophores.

**Experimental Methods and Materials**

High purity solvents and reagents were used in synthesis and were procured from well reputed chemical suppliers (Sigma-Aldrich, Merck and Fischer Scientific). Precoated aluminium sheets (Kiesel gel 60, F254, E. Merck, Germany) were used to monitor reaction progress via TLC. UV-vis absorption spectrum was recorded on Shimadzu Ultraviolet 1800 spectrophotometer using acetone as solvent. Bruker OPUS ATR was used to record IR spectrum. NMR experiments were performed on Bruker DPX-400 MHz spectrometer. Bruker Micro TOF-ESI mass spectrometer was used to record mass spectrum.

**Synthesis of 2-(2-(2,5-dimethoxybenzylidene)hydrazineyl)-4-(trifluoromethyl)thiazole (HM610)**

An equimolar mixture of 2-(2,5-dimethoxybenzylidene)hydrazine-1-carbothioamide (0.001 mol) and 3-bromo-1,1,1-trifluoroacetone (0.001 mol) was refluxed in ethanol for 4h. The progress of reaction and product purity was checked by TLC. Upon completion, workup was done in ice cold water which resulted in precipitation. The precipitates were washed with copious water, filtered and dried. UV spectrum was recorded in acetone and NMR spectrum was recorded in $CDCl_3$.

**2-(2-(2,5-dimethoxybenzylidene)hydrazineyl)-4-(trifluoromethyl)thiazole (HM610)**

Light yellow solid; Yield: 74%; $R_f$: 0.90 (acetone/*n*-hexane, 1:1); $\lambda_{max}$= 329 nm; FTIR (ATR) $cm^{-1}$: 3178 (N-H stretching), 2958 (C-H aliphatic stretching), 1614 (thiazole skeletal vibrations), 1566 (C=N stretching), 1521, 1373, 1342, 1226 (C=C aromatic ring stretching), 1421 (C-H aliphatic bending), 1110 (C-O stretching), 1008-704 (characteristic thiazole vibrations); $^1$H NMR (400 MHz, $CDCl_3$): δ (ppm) 10.26 (s, 1H, H-N-), 8.27 (s, 1H, H-C=N-), 7.49 (d, 1H, Ar-H, $J$ = 3.0 Hz), 7.14 (s 1H, thiazole-H), 6.94 (dd, 1H, Ar-H, $J$ = 9.0, 3.0 Hz), 6.87 (d, 1H, Ar-H, $J$ = 9.0 Hz) 3.86 (s, 3H, $-OCH_3$), 3.85 (s, 3H, $-OCH_3$); $^{19}F$ NMR (375 MHz) δ -64.65 ($-CF_3$); $^{13}C$ NMR (100 MHz) δ 170.7 (thiazole C2), 153.7 (Ar-C-OMe), 152.5 (Ar-C-OMe), 140.7 (q, $^{2}J_{C\text{-}CF3}$ = 36.7 Hz, thiazole C4), 139.4 (-C=N-), 122.4 (q, $^{1}J_{C\text{-}F}$ = 271 Hz, $-CF_3$), 119.2, 117.2 (Ar-C), 112.2 (q, $^{3}J_{C\text{-}CF3}$ = 4.0 Hz, thiazole C5), 110.3 (Ar-C), 56.2 ($-OCH_3$), 55.8 ($-OCH_3$); HRMS: m/z calculated for $C_{13}H_{12}F_3N_3O_2S$, $[M+H]^+$: 332.0681, found: 332.0674, $[M+Na]^+$: Calculated: 354.0500, found: 354.0491, $[2M+Na]^+$: Calculated: 685.1102, found: 685.1114.

## Computational Methodology

All quantum chemical calculations were performed using TURBOMOLE 7.2.1,[26] BDF,[27] and Gaussian 09.[28] Excited-state dynamics were simulated with Newton-X.[29]

**Geometry optimization.** Ground-state geometries of all studied thiazole derivatives were optimized with the meta-hybrid functional M062X,[30] and the def2-TZVP basis set.[31] Optimizations were followed by vibrational frequency analyses to confirm that the structures corresponded to true minima on the potential energy surface, verified by the absence of imaginary frequencies. Calculations employed an ultrafine integration grid (int=ultrafine), were carried out in the gas phase without explicit solvent effects, and used reduced symmetry conditions (nosymm) to avoid artificial restrictions. Grimme's D3 dispersion correction was included in all DFT optimizations.[32]

**Excited-state calculations.** Vertical excitation energies were obtained with correlated wavefunction methods [CC2 and ADC(2)] using the RI approximation in TURBOMOLE.[33] In parallel, TDDFT calculations were performed with six functionals (M062X, BHLYP,[34] CAM-B3LYP,[35] ωB97, ωB97X, and ωB97X-D[36]) and the def2-TZVP basis set. For each compound, twenty singlet excited states were computed on both singlet and triplet optimized geometries, enabling characterization of excitation energies, oscillator strengths, natural transition orbitals, and singlet–triplet gaps. Using a consistent methodological scheme ensured direct comparison between ground- and excited-state properties. Solvent effects on excitation energies were also explored with the polarizable continuum model (PCM).[37]

**Spin–orbit coupling.** Spin–orbit couplings (SOCs) were evaluated with the sf-X2C-S-TDDFT approach as implemented in BDF.[38] This method incorporates a Douglas–Kroll–Hess (DKH1)-like spin–orbit Hamiltonian, with one-electron terms included explicitly and two-electron contributions treated via the molecular mean-field approximation. The CIS-like wave function is constructed from TDDFT excitation vectors (X+Y) with normalization, providing a reliable treatment of relativistic effects in molecules containing light to medium atoms.[39]

**DMRG calculations.** DMRG calculations have been done using the program MOLMPS[40] employing the def-TZVP basis set to evaluate multireference character[41] and orbital entropies[42] in HM610. DMRG calculations have been used in recent years to evaluate multi-reference character of different molecular systems.[43] DMRG calculations have been performed in the active space of 20 electrons in 20 orbitals around the HOMO and that the initial DFT orbitals were split-localised.

**Surface Hopping Molecular Dynamics.** Excited-state dynamics were simulated with Newton-X. The nuclear ensemble approach (NEA) was applied to simulate absorption spectra,[44] while mixed quantum–classical trajectory surface hopping (Tully's FSSH) was employed for nonadiabatic dynamics including intersystem crossing.[45,46] Both time-derivative couplings and SOCs were included using a three-step integrator scheme. The Tamm–Dancoff approximation (TDA) was used within TDDFT to minimize triplet instabilities[47,48] and improve the description of conical intersections.[49] Dynamics trajectories were propagated at the BHLYP-D3/dhf-TZVP level of theory as employed in our previous study.[6b]

**NMR chemical shifts.** $^{1}H$ and $^{13}C$ chemical shifts were calculated with the gauge-including atomic orbital (GIAO) method, using reference TMS shielding constants at the same theoretical level.[50] Calculations were carried out at B3LYP-D3/cc-pVQZ in both the gas phase and in $CDCl_3$ (PCM). Additional calculations with M062X/cc-pVQZ and M062X/def2-QZVPP (PCM, Gaussian) were performed for comparison.[51] The use of B3LYP for NMR predictions is supported by previous reports of its accuracy.[52]

## Results and Discussion

### Synthesis of 2-(2-(2,5-dimethoxybenzylidene)hydrazineyl)-4-(trifluoromethyl)thiazole (3)

The synthesis of 2-(2-(2,5-dimethoxybenzylidene)hydrazineyl)-4-(trifluoromethyl)thiazole was achieved by literature reported method[53] as shown in scheme 1. An equimolar mixture of 2-(2,5-

dimethoxybenzylidene)hydrazine-1-carbothioamide and 3-bromo-1,1,1-trifluoroacetone was refluxed in ethanol for 4h.

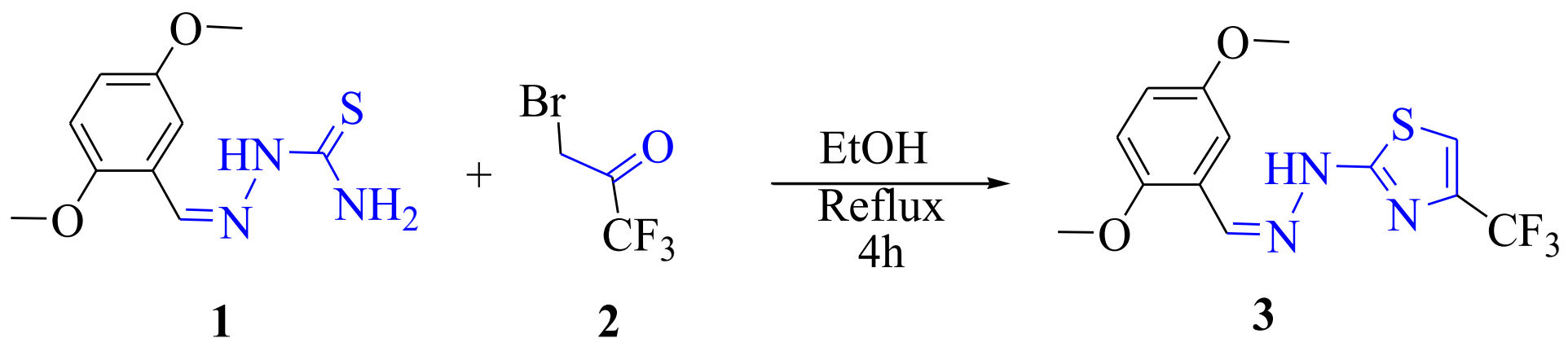

**Scheme 1**: synthesis of 2-(2-(2,5-dimethoxybenzylidene)hydrazineyl)-4-(trifluoromethyl)-thiazole

The synthesized compound was initially characterised by UV-Vis and IR spectroscopy. The structure of synthesized compound was established through NMR spectroscopy. The hydrazone functionality is indicated by characteristic amine 3178 $cm^{-1}$ stretching.[54] This connectivity was further supported by characteristic (azomethine) one proton singlet at 8.27 ppm in $^{1}$H-NMR spectrum. $^{13}$C-NMR spectroscopy added confirmation to the presence of azomethine linkage by a signal at 139.4 ppm. Similarly, thiazole moiety was indicated by characteristic thiazole vibrations[54] in the range 1008-704 $cm^{-1}$ and further confirmed by $^{13}$C-NMR where thiazole C2 signal was observed at 170.7 ppm, thiazole C4 at 140.7 ppm and thiazole C5 at 112.2 ppm. The aromatic protons and carbons appeared within the usually reported ranges.

**Benchmarking of density functionals**

Coupled-cluster-based wavefunction methods, such as CC2 and ADC(2), have been shown to yield comparable band gap energies across a wide range of organic dyes[19] and provide reliable estimates for excited-state decay rate constants.[55] In this study, these methods are employed to benchmark the TDDFT functionals under investigation. For the HM610 molecule, all vertical excitation energies computed using wavefunction methods (CC2 and ADC(2)) and TDDFT functionals (M062X; ωB97; ωB97X; ωB97X-D; BHLYP and CAM-B3LYP) are summarized in Table 1 and 2. At first glance, the data suggest that M062X yields excitation energies more closely aligned with those from CC2 and ADC(2) than other functionals, in agreement with previous findings.[55] Vertical excitation energies of most important states are tabulated in tables S1–S3. A more detailed comparison is provided in Tables S4–S6 which shows the differences in excitation energies across the various methods. The maximum absolute energy difference between ADC(2) and CC2 for the first seven singlet excited states is 0.05 eV, while the energies for the first ten triplet states are identical, an agreement consistent with trends reported for other organic molecules.[19]

From the comparison of six functionals (M062X, ωB97, ωB97X, ωB97X-D, CAM-B3LYP, and BHLYP) against CC2/ADC(2) benchmarks, clear trends emerge. For the first excited state ($S_1$), M062X and CAM-B3LYP give the closest agreement, with CAM-B3LYP slightly outperforming all others. Across the entire set of singlet and triplet states, however, ωB97X/ωB97X-D and M062X show the most balanced accuracy, while BHLYP and ωB97 tend to overestimate deviations. CAM-B3LYP performs very well for $S_1$ but displays larger systematic errors for triplets. Overall, CAM-B3LYP is best if $S_1$ is the priority, while M062X and ωB97X/ωB97X-D are more reliable for a balanced description of the full excited-state manifold. Nonetheless, the electronic state characters of the first two singlets, the first five triplets, and the two highest triplet states remain consistent across all methods.

Notably, some differences in excited-state character are observed. For singlet states, the character of $S_3$ and $S_4$ is swapped between $(\pi;\sigma^*)^1$ and $(\pi;\pi^*)^1$ configurations for M062X relative to CC2 and ADC(2). Similarly, for $S_5$ and $S_6$, the character is reversed between $(\sigma;\pi^*)^1$ and $(\pi;\pi^*)^1$ configurations in both M062X and ωB97 compared to CC2 and ADC(2). Among the triplets, $T_7$ and $T_8$ exhibit a character swap between $(\pi;\sigma^*)^3$ and $(\pi;\pi^*)^3$ configurations in ADC(2) and M062X relative to CC2. A similar swap occurs for $T_6$ and $T_7$ between $(\sigma;\pi^*)^3$ and $(\pi;\pi^*)^3$configurations in ωB97 compared to ADC(2). For the singlet states, ωB97X predicts the same ordering as CC2 and ADC(2) for $S_3$ and $S_4$, both retaining the $(\pi;\pi^*)^1$and $(\pi;\sigma^*)^1$ characters respectively. The $S_5$ and $S_6$ states are also consistent with CC2/ADC(2), where $S_5$ is $(\sigma;\pi^*)^1$and $S_6$ is $(\pi;\pi^*)^1$, so no character swaps occur among the singlets relative to the reference methods. In contrast, ωB97X-D, CAM-B3LYP, and BHLYP retain the CC2/ADC(2) ordering for $S_3$ and $S_4$, but they exhibit a reversal for $S_5$ and $S_6$: these functionals assign $S_5$ as $(\pi;\pi^*)^1$ and $S_6$ as $(\sigma;\pi^*)^1$, opposite to the character distribution found with CC2/ADC(2). Among the triplets, ωB97X differs from CC2/ADC(2) by assigning $T_6$ as $(\pi;\pi^*)^3$ and $T_7$ as $(\sigma;\pi^*)^3$, thereby reversing the $(\sigma;\pi^*)^3$/ $(\pi;\pi^*)^3$ordering seen with CC2/ADC(2). The same $T_6$/$T_7$ reversal is also observed with ωB97X-D, CAM-B3LYP, and BHLYP, which assign $T_6$ as $(\pi;\pi^*)^3$ and $T_7$ as $(\sigma;\pi^*)^3$instead of the CC2/ADC(2) pattern. Furthermore, these three functionals predict a distinct rearrangement of $T_7$ and $T_8$ compared to both CC2 and ADC(2); rather than alternating $(\pi;\sigma^*)^3$ and $(\pi;\pi^*)^3$ as in CC2, or $(\pi;\pi^*)^3$ and $(\pi;\sigma^*)^3$ as in ADC(2)/M062X, they place $T_7$ as $(\sigma;\pi^*)^3$and $T_8$ as $(\pi;\sigma^*)^3$. As mentioned above, M062X and CAM-B3LYP show the closest agreement with high-level wavefunction methods.

Based on the presented results, the effect of solvent on vertical excitation energies were evaluated using only the M062X and CAM-B3LYP functionals for the HM610 molecule, employing the PCM solvation model with the dielectric constant of dimethylsulfoxide (DMSO). The vertical excitation energies of HM610 presented in Table S7 show minimal differences between the gas phase and DMSO, with most shifts within ±0.05 eV. As shown in Table S8, both M062X and CAM-B3LYP functionals maintain consistent excitation character across environments, with only a few state-specific changes (e.g., $S_3$, $S_4$). These results indicate that solvent effects on the excited-state properties of HM610 are relatively minor.

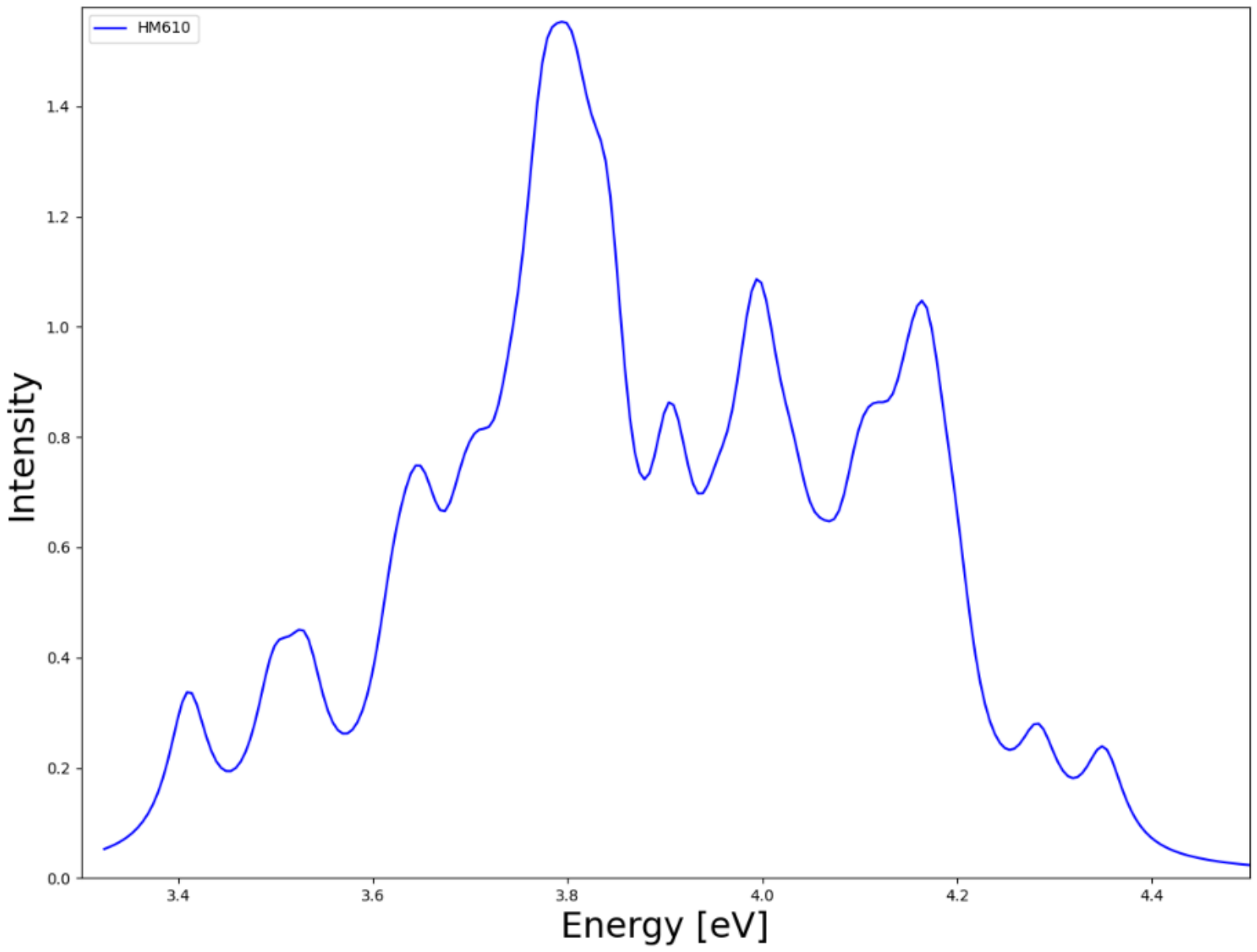

**Figure 1.** Simulated absorption spectrum of HM610 computed at BHLYP/def2-TZVP level of theory.

The simulated absorption spectra of HM610 in the gas phase were calculated at the TDDFT/BHLYP-D3/dhf-TZVP level of theory using the nuclear ensemble approach (NEA) presented in Figure 1. In NEA, photoabsorption cross-sections are averaged over configuration-space points, either sampled from phase-space distributions or collected during a ground-state MD simulation. The cross-section for each geometry depends on the sum of oscillator strengths of transitions, combined with Gaussian line shapes centered at their respective vertical excitation energies. The calculations used nuclear configurations sampled from the Wigner distribution of a canonical ensemble of quantum harmonic oscillators at 298 K. Seven of the lowest singlet excited states of HM610 were included in the simulation.

**Spin-Orbit Coupling Calculations for HM610**

Spin-orbit couplings (SOCs) between first eight singlets (including ground state $S_0$) and ten triplets are computed by relativistic two-component time-dependent sf-X2C-S-TD-DFT method employing M062X; ωB97; ωB97X; ωB97X-D; BHLYP and CAM-B3LYP functionals. The magnitudes of spin–orbit couplings (SOCs) for all six functionals are presented in Tables 3–8, while the corresponding data for the M062X and CAM-B3LYP functionals in DMSO solvent are provided in Tables S9 and S10. There are only few numbers of coupled states (states whose magnitude of SOCs is greater than or equal to 30 $cm^{-1}$) in HM610 that have been tabulated in Table S11–S18 computed by all six functionals considered in this work. Appreciable magnitudes of SOCs in HM610 will be explained by rules devised in our earlier work.[56] According to them, when both the singlet state ($\varphi_i^S \rightarrow \varphi_f^S$) and the triplet state ($\varphi_i^T \rightarrow \varphi_f^T$) differ by a single pair of molecular orbitals or by both pairs, it is expected that these states of different multiplicity will exhibit strong spin–orbit coupling (SOC). First considering the case when singlet state ($\varphi_i^S \rightarrow \varphi_f^S$) and triplet state ($\varphi_i^T \rightarrow \varphi_f^T$) have either one of the orbitals different i.e.

$$\varphi_i^S \neq \varphi_i^T \tag{1}$$

Or;

$$\varphi_f^S \neq \varphi_f^T \tag{2}$$

Then, there will be appreciable SOC between these respective singlet and triplet states. Above mentioned rules can be simply explained with Slater-Condon rules[57] i.e.

$$|S\rangle = |\ldots\varphi_i\varphi_n\ldots\rangle \tag{3}$$

$$|T\rangle = |\ldots\varphi_j\varphi_n\ldots\rangle \tag{4}$$

$$\langle S|h_{SOC}|T\rangle = \langle\varphi_i|h_{SOC}|\varphi_j\rangle \tag{5}$$

If the singlet and triplet states differ by only one orbital, then the magnitude of the spin-orbit coupling (SOC) between them is given by the spin-orbit Hamiltonian $h_{SOC}$, evaluated between the differing orbitals of the respective singlet and triplet states as shown in eq 5. So, the first rule that says states of different multiplicities differ by a single pair of molecular orbitals will have appreciable SOC is well justified by Slater-Condon rule. In the conventional Slater–Condon formulation, spin orbitals are used to construct Slater determinants. In the present discussion, however, we employ molecular orbitals (i.e., the spatial parts of spin orbitals) "$\varphi_z|_{z=i\ldots n}$" in eq 3-5 as the spin dependence is explicitly treated through the spin–orbit Hamiltonian. Now, considering the case where the singlet state ($\varphi_i^S \rightarrow \varphi_f^S$) and the triplet state ($\varphi_i^T \rightarrow \varphi_f^T$) simultaneously differs by both pairs of orbitals, i.e.,

$$\varphi_i^S \neq \varphi_i^T \tag{6}$$

And;

$$\varphi_f^S \neq \varphi_f^T \tag{7}$$

In that case as well, there will be again significant spin–orbit coupling (SOC) between the singlet and triplet states. This rule can also be explained with Slater–Condon rules[57] i.e.

$$|S\rangle = |\ldots\varphi_i\varphi_j\ldots\rangle \tag{8}$$

$$|T\rangle = |\ldots\varphi_k\varphi_l\ldots\rangle \tag{9}$$

$$\langle S|h_{SOC}|T\rangle = \langle\varphi_i|h_{SOC}|\varphi_k\rangle\langle\varphi_j|\varphi_l\rangle + \langle\varphi_i|h_{SOC}|\varphi_l\rangle\langle\varphi_j|\varphi_k\rangle + \langle\varphi_j|h_{SOC}|\varphi_k\rangle\langle\varphi_i|\varphi_l\rangle + \langle\varphi_j|h_{SOC}|\varphi_l\rangle\langle\varphi_i|\varphi_k\rangle \tag{10}$$

$$\langle\varphi_j|\varphi_l\rangle = S_{jl} \tag{11}$$

$$\langle\varphi_j|\varphi_k\rangle = S_{jk} \tag{12}$$

$$\langle\varphi_i|\varphi_l\rangle = S_{il} \tag{13}$$

$$\langle \varphi_i | \varphi_k \rangle = S_{ik} \tag{14}$$

All these *Overlap matrix* in eq 10 would be zero if $\varphi_z|_{z=i\ldots n}$ in eq 11-14 would be spin–orbitals. Considering only one pair let assume $\langle \varphi_j | \varphi_l \rangle = S_{jl}$ *overlap matrix* "$S_{jl}$" will be non-zero i.e. $S_{jl} \neq 0$ if "$\varphi_j$" and "$\varphi_l$" are drawn from two different ("state dependent") orbital sets (one for bra determinant *S*, another for ket determinant *T*), thus resulting in a non-zero magnitude of spin-orbit coupling between respective singlet and triplet, i.e.

$$\langle S | h_{SOC} | T \rangle \neq 0 \tag{15}$$

In fact, we are working with a single set of orthonormal orbitals, so within that set $S_{ij} = \delta_{ij}$. The exact reasoning for non-zero SOC matrix elements between *S* and *T* states differing by more than one orbital is that these states are, in fact, CIS-like and could be expanded into multiple Slater determinants (though the expansion is, in most cases, dominated by the reference configuration).

$$|T\rangle = \Sigma_j \, c_{T,j} \, |\Phi_j\rangle \tag{16}$$

$$\langle S| = \Sigma_i \, c_{S,i}^* \, \langle \Phi_i| \tag{17}$$

Inserting such for both S and T, there could be a bilinear expression in

$$\langle S|h_{SOC}|T\rangle = \Sigma_{ij} \, c_{T,j} \, c_{S,i}^* \, \langle \Phi_i|h_{SOC}|\Phi_j\rangle \tag{18}$$

always with (several) terms where individual $\langle \Phi_i|h_{SOC}|\Phi_j\rangle$ are with $\Phi_i$ and $\Phi_j$ differing by a single orbital (consider monoexcitation from one of these "different orbitals in reference" into the same orbital). After that, there is only a single orbital difference, and normal Slater–Condon rules apply, resulting in $\langle \Phi_i|h_{SOC}|\Phi_j\rangle \neq 0$ and thus (likely, except improbable case of mutual cancel-out) $\langle S|h_{SOC}|T\rangle \neq 0$. We can introduce "effective state-specific rotated" MO basis different for *S* and *T* (with some numerical error, not affecting the quality of the argument) and stick to "non-orthogonal overlap" explanation, though. For convenience, coupled states differing by a single pair of molecular orbitals are referred to as "*true-coupled states*" whereas those differing by both pairs of molecular orbitals are denoted as "*pseudo-coupled states*". Now first considering coupled states computed by M062X functional at sf-X2C-S-TD-DFT-SOC/x2c-TZVPPall level of theory presented in Table S11. All coupled states computed by M062X functional can be explained by the first rule explained above i.e. singlet and triplet state differ by a single pair of molecular orbitals. So, they are all true-coupled states. Now considering coupled states computed by ωB97 functional at sf-X2C-S-TD-DFT-SOC/x2c-TZVPPall level of theory presented in Table S12. All coupled states computed by ωB97 functional can be explained by the first rule mentioned above in eq 1 and eq 2. Except coupled state $S_3T_8$ (ωB97) this can be explained by the second rule explained above in light of Slater–Condon rules. Now considering $S_3T_8$ (ωB97) being pseudo-coupled state whose respective singlet and triplet differ by both pair of orbitals simultaneously then one has to expand them into multiple Slater determinants which will result in bilinear expression differing only by a single orbital difference, and normal Slater–Condon rules apply. Now considering coupled states computed by ωB97X and ωB97X-D functionals at sf-X2C-S-TD-DFT-SOC/x2c-TZVPPall level of theory presented in Table S13 and S14. First two coupled states are well explained by the first rule that if a singlet state $(\varphi_i^S \rightarrow \varphi_f^S)$ and triplet state $(\varphi_i^T \rightarrow \varphi_f^T)$ differ by just single pair of orbitals, then they will result in a true-coupled states. From Tables S13 and S14 it can be inferred that $S_3T_8$ and $S_4T_3$ coupled states of ωB97X and ωB97X-D functionals are are pseudo-coupled, which can be confirmed by expanding these states into multiple Slater determinants. Now considering coupled states computed by CAM-B3LYP and BHLYP functionals at sf-X2C-S-TD-DFT-SOC/x2c-TZVPPall level of theory presented in Table S15 and S16. The first three coupled states of the CAM-B3LYP functional being true-coupled states can be simply explained by the first rule.[52] For pseudo-coupled state $S_4T_3$ (CAM-B3LYP) one has to follow the procedure mentioned in eq 16-18. Similarly, the first two and the fourth coupled states of the BHLYP functional can again be simply explained by the first rule.[56] For the pseudo-coupled state $S_3T_8$ (BHLYP) again same methodology mentioned above in eq 16-18 has to be considered. Finally considering coupled states computed by M062X and CAM-B3LYP functionals at sf-X2C-S-TD-DFT-SOC/x2c-TZVPP all level of theory in DMSO solvent employing the PCM solvation model presented in Tables S17 and S18. All coupled states can be explained in light of the first rule that singlet and triplet states differ by a single orbital pair.[56] For the pseudo-coupled state $S_4T_3$ (M062X and CAM-B3LYP) again bilinear expansion has to be considered. Thus, all coupled states computed by all six functionals (even in DMSO) summarized in Tables S11–S18 are well explained by rules devised earlier[56] which can be traced back to Slater-Condon rules as shown in this work. The significant magnitude of spin–orbit coupling (SOC) observed when only a single pair of spin orbitals differs between singlet and triplet states has been previously reported.[6b,58] However, a comprehensive justification for the occurrence of similarly large SOC values when both orbital pairs differ in singlet and triplet configurations is presented here for the first time where CIS-like states should be expanded into multiple Slater determinants.[59] When the orbitals involved in transitions between states of different multiplicities differ by their type (e.g., $n \rightarrow \pi^*$ vs. $\pi \rightarrow \pi^*$), the resulting large spin–orbit coupling (SOC) arises from an orientational change[60] during the transition. In contrast, when the orbitals differ only by subscripts or energy levels (e.g., $\pi_1 \rightarrow \pi^*$ vs. $\pi_2 \rightarrow \pi^*$), the large SOC primarily results from a spatial change in the involved orbitals.[56a] El-Sayed rules,[61] normally employed to explain SOCs in organic molecules is just a special case of the first rule explained above. Shapes of other molecular orbitals of HM610 are also presented in Table S19. Thus, to comprehensively explain SOCs in a molecule, a threshold value for SOC (30 $cm^{-1}$ in the present work) should be defined. States of different spin multiplicities exhibiting SOC values equal to or greater than this threshold can then be designated as *coupled states*. For such coupled states, the electronic character and the dominant orbitals involved in the corresponding transitions should be explicitly analyzed.

**DMRG results**

DMRG calculations were performed to confirm that the system does not exhibit multireference character and that the chosen methodology is appropriate. Two indicators were used for this purpose: the occupation numbers of the DMRG natural orbitals and the single-orbital entropies. The latter quantify the importance of individual orbitals in the wave function expansion and have been previously employed in CAS-based analyses. The DMRG natural orbital occupations 1.94 for the HOMO and 0.06 for the LUMO indicate a predominantly single-reference character, which is consistent with the single-orbital entropies reported in the Supporting Information in Table S20.

**Simulation of Dynamical Effects on Spin–orbit Coupling in HM610**

To explore the dynamical effects (first order spin–vibronic effects) on SOCs trajectory surface hopping (TSH) simulations of HM610 were run at BHLYP-D3/dhf-TZVP level of theory. An ensemble of FSSH molecular dynamics (MD) simulations, consisting of 11 trajectories, was performed. The initial conditions nuclear coordinates and momenta were sampled from the Wigner distribution corresponding to a canonical ensemble of non-interacting quantum harmonic oscillators at the optimized ground-state geometry at 298 K. Each trajectory was propagated for 100 fs, employing a scaling factor of 5. Scaling approach for TSH simulations was employed since SOCs in HM610 are not large enough with few numbers of coupled states. Results of TSH simulations presented in Figures 2 clearly shows that first order spin–vibronic effects i.e.

dependence of SOCs on nuclear coordinates[22a,d] can't be an effective mechanism for inducing appreciable intersystem crossing in HM610. Even when large scaling factor of 5 was used for SOCs no appreciable filling of triplet manifold is observed during the entire simulation run. To elucidate the contribution of individual singlet states to the increasing total singlet population observed in Figure 2, the average populations of each singlet state and the aggregated triplet population across all trajectories are presented in Figure 3. Figure 3 shows that within the singlet manifold, $S_2$ is the primary contributor, corresponding to the second-brightest state. Apart from direct decay from $S_1$ via fluorescence, there is a possibility that the molecule transfers part of its population to $S_2$ through a radiationless transition.[62] Despite this population transfer, the final decay pathway of $S_2$ remains radiative relaxation to the ground state, typically occurring within 1–10 ns,[63] consistent with known fluorescence timescales. Mixed quantum-classical dynamics methods, such as surface hopping, are computationally prohibitive for simulating these longer timescales associated with fluorescence. Therefore, the chosen TSH simulation timescale is appropriate and confirms the absence of significant triplet population transfer via spin–vibronic coupling.

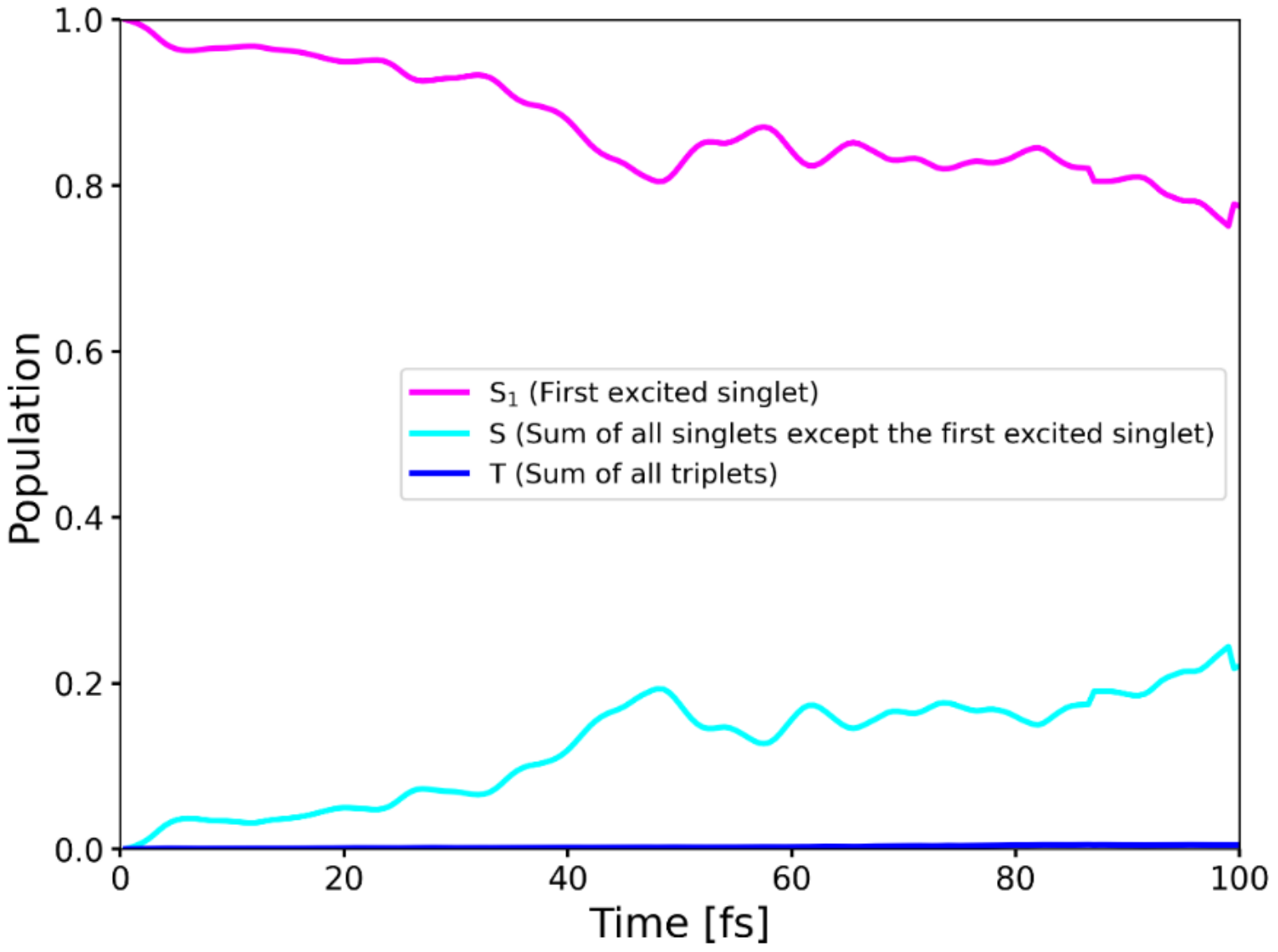

**Figure 2.** Average populations of singlets and triplets (back-transformed from the spin-adiabatic basis) in the dynamics of HM610 started in $S_1$ state, with a scaling factor α=5.

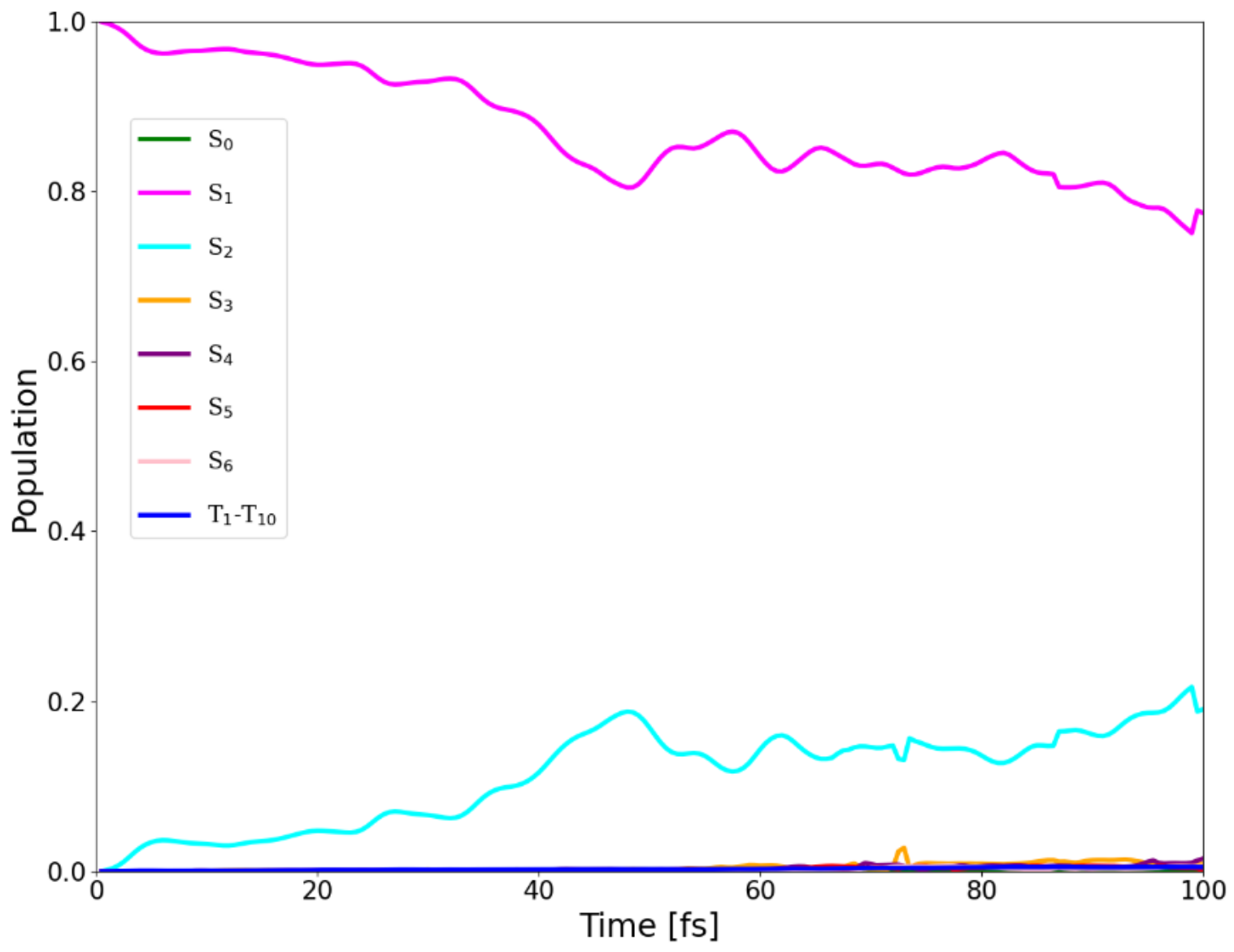

**Figure 3.** Average populations of individual singlet states and the combined triplet population ($T_1$–$T_{10}$) in the dynamics of HM610 started in the $S_1$ state, with a scaling factor $\alpha = 5$.

**Calculation of NMR chemical shifts**

The computed $^1H$ and $^{13}C$ NMR chemical shifts in both the gas phase and in $CDCl_3$ (using PCM) are summarized in Tables S21 and S22. Overall, the experimental spectra exhibit close agreement with theory, validating the structural assignments. For the $^1H$ NMR spectra, the aromatic resonances (H21–H26) are well reproduced by all methods. The N–H proton (H22, 10.26 ppm) remains more deshielded in experiment, with deviations up to ~1.8 ppm, though this error is slightly reduced with larger basis sets. Quantitatively, the RMSD analysis shows that B3LYP/Def2QZVPP ($CDCl_3$) achieves the best agreement with experiment (0.51 ppm), outperforming both B3LYP/cc-pVQZ and M062X methods. The azomethine proton (H25, 8.27 ppm) is predicted within 0.5 ppm of experiment across all levels. Aliphatic protons (H29–H34) in the 3.2–4.2 ppm region are also consistently matched, confirming reliable treatment of methoxy and methylene environments. For the $^{13}C$ NMR spectra, deshielded carbons such as the carbonyl C2 (170.7 ppm) and imine carbons C11 and C14 (~153 ppm) are reproduced with reasonable accuracy, while aromatic carbons (C5, C8–C13, C17) are captured in the correct 110–140 ppm range. Methoxy carbons (C27, C28) are predicted near 56 ppm with excellent consistency across all functionals. The RMSD analysis indicates that B3LYP/cc-pVQZ ($CDCl_3$) provides the best overall agreement (8.43 ppm). These findings align with previous benchmarking, which established B3LYP as a reliable choice for NMR prediction in organic systems. Literature reports also suggest that M062X, particularly with larger basis sets such as Def2QZVPP, can offer improved accuracy for conjugated and hydrogen-bonded systems.[64] However, in the present case, M062X methods show significantly larger deviations (RMSD ≥ 18 ppm for $^{13}C$), indicating that they do not outperform B3LYP for this molecule. Importantly, our results demonstrate that basis set expansion within the B3LYP framework (Def2QZVPP, $CDCl_3$) improves proton predictions, while B3LYP/cc-pVQZ ($CDCl_3$) remains superior for carbons. The strong correlation between theory and experiment therefore validates the computational approach and provides robust support for the proposed molecular structure.

**Computational Exploration of HM610 Analogues**

To assess how donor–acceptor substitution modulates the geometry and photophysics of the benzylidenehydrazinylthiazole backbone, we constructed a systematic set of six derivatives (Figure 4). The series spans the unsubstituted parent (system 1), electron-donating OMe substitution at the aryl ring (systems 2 and 4), electron-withdrawing $CF_3$ substitution at the thiazole ring (system 3), fully substituted analogues (systems 4 and 5), and the experimentally synthesized compound HM610 (system 6). Geometry optimizations were performed for both singlet and triplet states at the M062X/def2TZVP level, followed by TDDFT calculations of vertical excitations. All optimized geometries are provided in SI. Total SCF energies, zero-point vibrational energies, and thermal contributions to Gibbs energy are provided in Table S23. Results from TDDFT calculations (absorption wavelengths, energy gaps, and oscillator strengths) are condensed in Table S24.

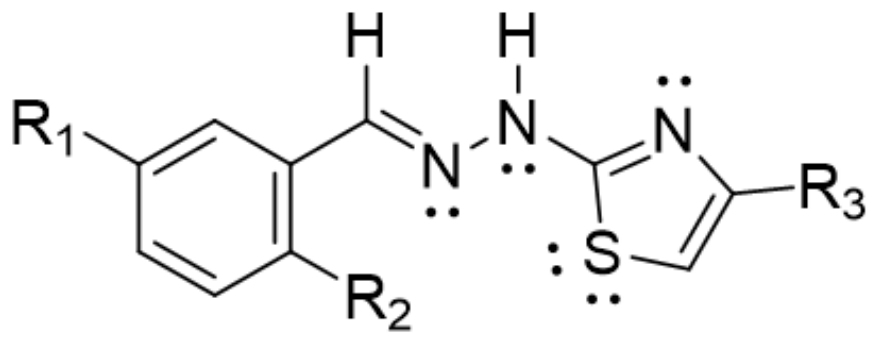

| System | $R_1$ | $R_2$ | $R_3$ |
|---|---|---|---|
| **1** | H | H | H |
| **2** | MeO | MeO | H |
| **3** | H | H | $CF_3$ |
| **4** | MeO | MeO | MeO |
| **5** | $CF_3$ | $CF_3$ | $CF_3$ |
| **HM610** | MeO | MeO | $CF_3$ |

**Figure 4**. General structure of the benzylidenehydrazinylthiazole scaffold with variable substituents (R1–R3). The systematic set of six derivatives spans the unsubstituted parent (1), donor-substituted analogues (2, 4), acceptor-substituted analogues (3, 5), and the experimentally synthesized compound HM610.

The results show that substitution strongly influences singlet–triplet energetics and absorption profiles. OMe groups, as electron donors, generally red-shift the lowest-energy transitions and increase oscillator strengths. In contrast, $CF_3$ groups act as electron-withdrawing substituents, leading to larger blue-shifted excitations. The fully substituted analogues (systems 4 and 5) illustrate the extreme cases of donor- or acceptor-dominated substitution, while HM610 (system 6) lies in between, combining donor and acceptor groups in a push–pull arrangement. Natural transition orbital (NTO) analysis illustrates how donor–acceptor substitution patterns modulate the photophysics of the studied thiazole–hydrazone derivatives. For the lowest singlet excitations (Figure 5), the excitation energies span 3.86–4.19 eV across the series. Systems bearing electron-donating OMe groups (2, 4, and HM610) display a modest red-shift relative to the unsubstituted parent (1), while strong electron-withdrawing substitution with $CF_3$ (system 3) induces a slight blue-shift. The fully substituted analogues represent limiting cases: all-donor substitution (system 4) lowers the excitation energy to 3.86 eV, whereas all-acceptor substitution (system 5) maintains values close to the parent. Importantly, all molecules are essentially planar, allowing for maximal π overlap.

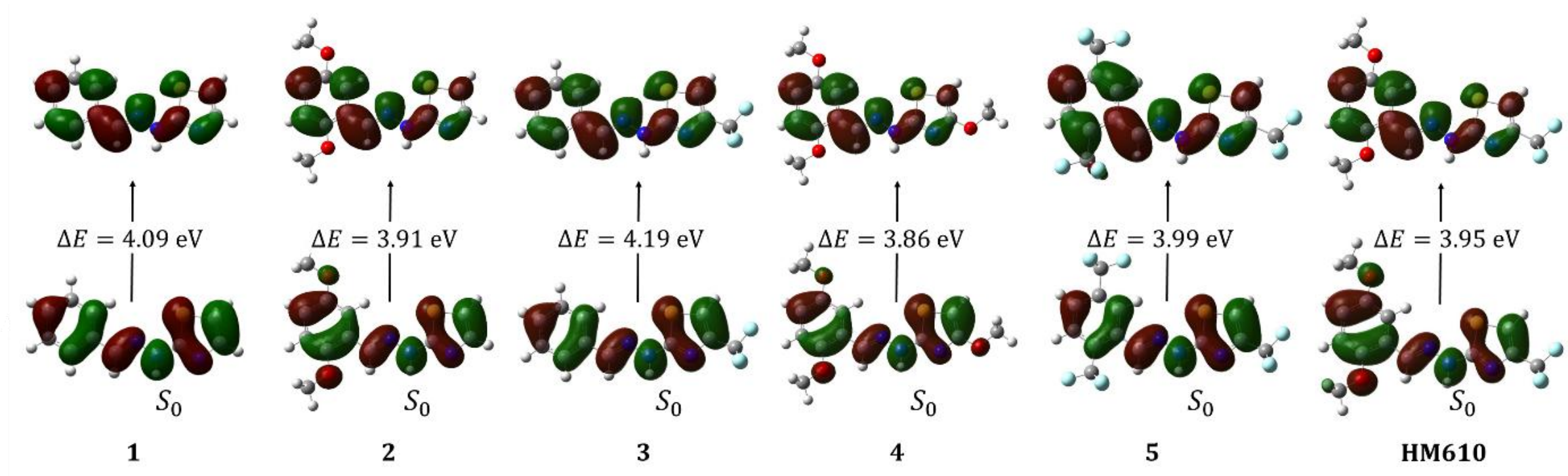

**Figure 5.** Natural transition orbitals representing the hole (bottom) and particle (top) orbitals for systems 1–5 and HM610 in the singlet spin state, with associated excitation energies ($\Delta E$, eV). All values were calculated at the td-M062X/def2TZVP level *in vacuo*, with 20 states in the CIS procedure.

The NTOs reveal that the lowest excitations are dominated by π→π* transitions in every case, delocalized across the aryl–hydrazone–thiazole backbone, with substitution patterns affecting the relative localization of the hole and particle orbitals only slightly. The triplet state for all compounds is nonplanar excitations (Figure 6), and the relevant orbitals lie either on the thiazole subunit (compound 1) or on the phenyl ring (compounds 2-5 and HM610). For the lowest triplet, the series spans 2.67–3.45 eV. As in the singlet manifold, donor substitution tends to stabilize the triplet states, lowering excitation energies (systems 2, 4, and HM610). Conversely, $CF_3$ substitution (system 3) increases energy gaps relative to the parent. In the triplet space, the lowest energy gap is predicted for HM610 (2.67 eV).

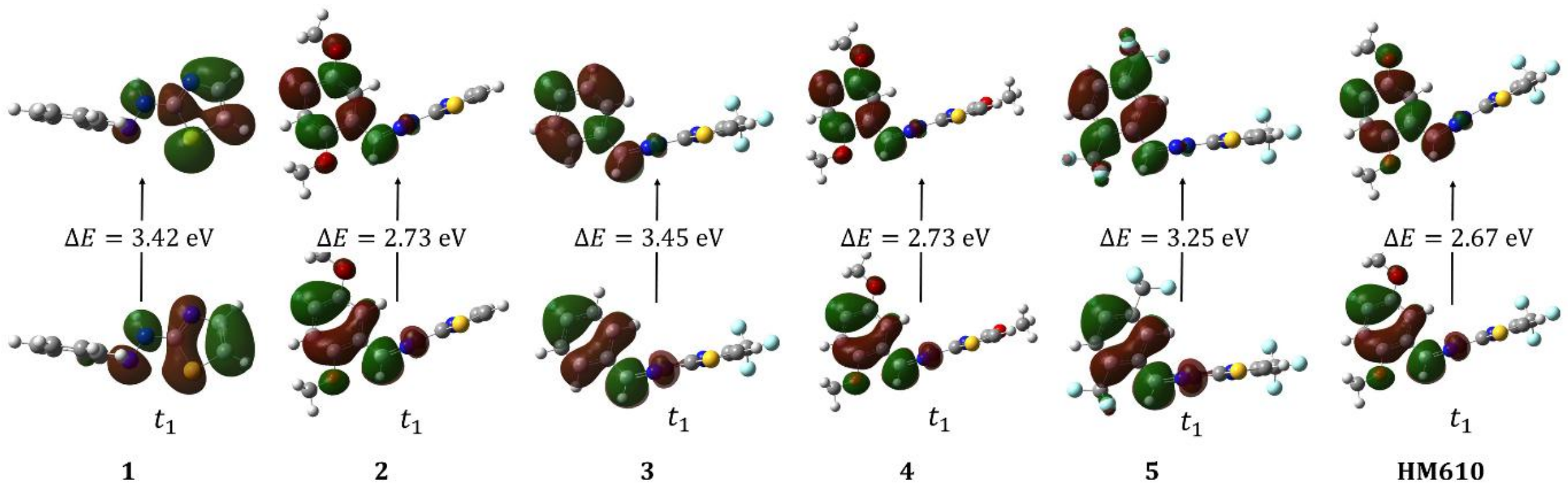

**Figure 6**. Natural transition orbitals representing the hole (bottom) and particle (top) orbitals for systems 1–5 and HM610 in the triplet spin state, with associated excitation energies ($\Delta E$, eV).

All values were calculated at the td-M062X/def2TZVP level *in vacuo*, with 20 states in the CIS procedure. Overall, the systematic comparison of electron-donating and electron-withdrawing substitutions on the thiazole–hydrazone backbone shows that simple donor/acceptor modifications induce only moderate changes in excitation energies, oscillator strengths, and singlet–triplet gaps. Relativistic calculations confirm that spin–orbit couplings remain small, consistent with expectations for light-atom systems. These results indicate that, while substitution can fine-tune the photophysics of this scaffold, heavy-atom incorporation is likely essential to achieve efficient intersystem crossing and practical triplet sensitization. This perspective sets the stage for ongoing efforts aimed at introducing heavy substituents to enhance ISC in this family of chromophores.

## Conclusions

In this work, we established the synthetic feasibility of the thiazole–hydrazone scaffold through the preparation of HM610 and applied a combined computational strategy to investigate its photophysics alongside a systematic set of analogues. Benchmarking TDDFT against correlated wavefunction methods (CC2 and ADC(2)) demonstrated close agreement in vertical excitation energies and state character, with M062X performing reliably relative to ωB97. Spin–orbit couplings (SOCs), evaluated with the sf-X2C-S-TDDFT method, were analyzed in terms of orbital contributions and found to follow the Slater–Condon rules: appreciable couplings occur when singlet and triplet states differ by either one pair or both pair of orbitals. Consistently, the SOC magnitudes predicted for HM610 are modest, suggesting limited intersystem crossing efficiency in the absence of heavy-atom effects.

Our systematic DFT study of donor- and acceptor-substituted derivatives further revealed that electron-donating (–OMe) and electron-withdrawing (–$CF_3$) groups produce only moderate shifts in excitation energies, oscillator strengths, and singlet–triplet gaps. Taken together, these findings indicate that heavy-atom substitution will likely be required to promote efficient triplet sensitization in this family of chromophores, providing clear motivation for ongoing studies directed at enhancing intersystem crossing by targeted functionalization.

**Acknowledgements** M. W. Baig acknowledges the support by the Czech Science Foundation, project No. 23-06364S. M. Maldonado-Domínguez acknowledges the support by PAPIIT (grant IA201024) and LANCAD-UNAM-DGTIC-462. M.W. Baig also thanks to Dr. Libor Veis for providing MOLMPS code and helping in DMRG calculations. This work is dedicated by M. W. Baig to the memory of the late Prof. Sheng Hsien Lin.

**Author contribution** H.M. carried out the synthesis and characterization, interpreted the results, and contributed to the manuscript writing. T.A. conceived the research idea, supervised the project, and was responsible for proofreading and improving the manuscript. J.E.-R. performed the DFT and TDDFT calculations for the molecular series, including the optimization of singlet and triplet states, evaluation of spin-state energetics, and characterization of electronic excitations. M.M.-D. was responsible for the conceptualization, planning, and analysis of the DFT and TDDFT results, as well as for manuscript writing. J.E.-R. and M.M.-D. contributed to the discussion of the findings and approved the final version of the manuscript. J.V. helped with analyzing the DMRG calculations and contributed to the interpretation and writing of spin–orbit coupling computations. M.W.B. performed the majority of the quantum chemical calculations, conducted the TSH simulations, carried out all sf-X2C-S-TD-DFT computations and their analysis, and was also responsible for writing and finalizing the manuscript.

**Table 1:** Vertical excitation energies computed with four different electronic structure methods at $S_0$ optimized geometry.

| States | CC(2) | Character | ADC(2) | Character | M062X | Character | ωB97 | Character |
|---|---|---|---|---|---|---|---|---|
| $S_1$ | 3.79 (0.50) | $(\pi;\pi^*)^1$ | 3.75 (0.46) | $(\pi;\pi^*)^1$ | 3.96 (0.66) | $(\pi;\pi^*)^1$ | 4.17 (0.61) | $(\pi;\pi^*)^1$ |
| $S_2$ | 4.21 (0.33) | $(\pi;\pi^*)^1$ | 4.16 (0.24) | $(\pi;\pi^*)^1$ | 4.40 (0.09) | $(\pi;\pi^*)^1$ | 4.62 (0.16) | $(\pi;\pi^*)^1$ |
| $S_3$ | 5.41 (0.09) | $(\pi;\pi^*)^1$ | 5.38 (0.12) | $(\pi;\pi^*)^1$ | 5.43 (0.00) | $(\pi;\sigma^*)^1$ | 5.60 (0.03) | $(\pi;\pi^*)^1$ |
| $S_4$ | 5.44 (0.02) | $(\pi;\sigma^*)^1$ | 5.47 (0.00) | $(\pi;\sigma^*)^1$ | 5.46 (0.11) | $(\pi;\pi^*)^1$ | 5.85 (0.00) | $(\pi;\sigma^*)^1$ |
| $S_5$ | 5.58 (0.03) | $(\pi;\pi^*)^1$ | 5.53 (0.04) | $(\pi;\pi^*)^1$ | 5.52 (0.00) | $(\sigma;\pi^*)^1$ | 5.87 (0.00) | $(\sigma;\pi^*)^1$ |
| $S_6$ | 5.75 (0.00) | $(\sigma;\pi^*)^1$ | 5.70 (0.00) | $(\sigma;\pi^*)^1$ | 5.60 (0.00) | $(\pi;\pi^*)^1$ | 5.98 (0.36) | $(\pi;\pi^*)^1$ |
| $S_7$ | 5.87 (0.19) | $(\pi;\pi^*)^1$ | 5.86 (0.21) | $(\pi;\pi^*)^1$ | 5.95 (0.04) | $(\pi;\sigma^*)^1$ | 6.39 (0.01) | $(\pi;\pi^*)^1$ |
| $T_1$ | 2.99 (0.00) | $(\pi;\pi^*)^3$ | 2.99 (0.00) | $(\pi;\pi^*)^3$ | 2.80 (0.00) | $(\pi;\pi^*)^3$ | 2.40 (0.00) | $(\pi;\pi^*)^3$ |
| $T_2$ | 3.62 (0.00) | $(\pi;\pi^*)^3$ | 3.62 (0.00) | $(\pi;\pi^*)^3$ | 3.53 (0.00) | $(\pi;\pi^*)^3$ | 3.39 (0.00) | $(\pi;\pi^*)^3$ |
| $T_3$ | 4.00 (0.00) | $(\pi;\pi^*)^3$ | 4.00 (0.00) | $(\pi;\pi^*)^3$ | 3.86 (0.00) | $(\pi;\pi^*)^3$ | 3.46 (0.00) | $(\pi;\pi^*)^3$ |
| $T_4$ | 4.42 (0.00) | $(\pi;\pi^*)^3$ | 4.42 (0.00) | $(\pi;\pi^*)^3$ | 4.31 (0.00) | $(\pi;\pi^*)^3$ | 3.87 (0.00) | $(\pi;\pi^*)^3$ |
| $T_5$ | 4.81(0.00) | $(\pi;\pi^*)^3$ | 4.81(0.00) | $(\pi;\pi^*)^3$ | 4.71 (0.00) | $(\pi;\pi^*)^3$ | 4.61 (0.00) | $(\pi;\pi^*)^3$ |
| $T_6$ | 5.21 (0.00) | $(\sigma;\pi^*)^3$ | 5.21 (0.00) | $(\sigma;\pi^*)^3$ | 4.90 (0.00) | $(\sigma;\pi^*)^3$ | 4.99 (0.00) | $(\pi;\pi^*)^3$ |
| $T_7$ | 5.28 (0.00) | $(\pi;\sigma^*)^3$ | 5.28 (0.00) | $(\pi;\pi^*)^3$ | 5.13 (0.00) | $(\pi;\pi^*)^3$ | 5.10 (0.00) | $(\sigma;\pi^*)^3$ |
| $T_8$ | 5.32 (0.00) | $(\pi;\pi^*)^3$ | 5.32 (0.00) | $(\pi;\sigma^*)^3$ | 5.25 (0.00) | $(\pi;\sigma^*)^3$ | 5.51 (0.00) | $(\pi;\sigma^*)^3$ |
| $T_9$ | 5.72 (0.00) | $(\pi;\pi^*)^3$ | 5.72 (0.00) | $(\pi;\pi^*)^3$ | 5.47 (0.00) | $(\pi;\pi^*)^3$ | 5.68 (0.00) | $(\pi;\pi^*)^3$ |
| $T_{10}$ | 5.89 (0.00) | $(\pi;\pi^*)^3$ | 5.89 (0.00) | $(\pi;\pi^*)^3$ | 5.63 (0.00) | $(\pi;\pi^*)^3$ | 5.81 (0.00) | $(\pi;\pi^*)^3$ |

**Table 2:** Vertical excitation energies computed with four different electronic structure methods at $S_0$ optimized geometry.

| States | ωB97X | Character | ωB97X-D | Character | CAM-B3LYP | Character | BHLYP | Character |
|---|---|---|---|---|---|---|---|---|
| $S_1$ | 4.09 (0.62) | $(\pi;\pi^*)^1$ | 3.93 (0.61) | $(\pi;\pi^*)^1$ | 3.91 (0.61) | $(\pi;\pi^*)^1$ | 4.06 (0.68) | $(\pi;\pi^*)^1$ |
| $S_2$ | 4.54 (0.15) | $(\pi;\pi^*)^1$ | 4.37 (0.16) | $(\pi;\pi^*)^1$ | 4.33 (0.14) | $(\pi;\pi^*)^1$ | 4.49 (0.07) | $(\pi;\pi^*)^1$ |
| $S_3$ | 5.55 (0.05) | $(\pi;\pi^*)^1$ | 5.44 (0.06) | $(\pi;\pi^*)^1$ | 5.42 (0.07) | $(\pi;\pi^*)^1$ | 5.50 (0.08) | $(\pi;\pi^*)^1$ |
| $S_4$ | 5.70 (0.00) | $(\pi;\sigma^*)^1$ | 5.48 (0.00) | $(\pi;\sigma^*)^1$ | 5.45 (0.00) | $(\pi;\sigma^*)^1$ | 5.60 (0.00) | $(\pi;\sigma^*)^1$ |
| $S_5$ | 5.79 (0.00) | $(\sigma;\pi^*)^1$ | 5.65 (0.12) | $(\pi;\pi^*)^1$ | 5.55 (0.08) | $(\pi;\pi^*)^1$ | 5.65 (0.06) | $(\pi;\pi^*)^1$ |
| $S_6$ | 5.89 (0.28) | $(\pi;\pi^*)^1$ | 5.66 (0.00) | $(\sigma;\pi^*)^1$ | 5.67 (0.00) | $(\sigma;\pi^*)^1$ | 5.90 (0.00) | $(\sigma;\pi^*)^1$ |
| $S_7$ | 6.25 (0.13) | $(\pi;\pi^*)^1$ | 5.98 (0.18) | $(\pi;\pi^*)^1$ | 5.90 (0.20) | $(\pi;\pi^*)^1$ | 6.06 (0.21) | $(\pi;\pi^*)^1$ |
| $T_1$ | 2.44 (0.00) | $(\pi;\pi^*)^3$ | 2.50 (0.00) | $(\pi;\pi^*)^3$ | 2.38 (0.00) | $(\pi;\pi^*)^3$ | 2.06 (0.00) | $(\pi;\pi^*)^3$ |
| $T_2$ | 3.39 (0.00) | $(\pi;\pi^*)^3$ | 3.34 (0.00) | $(\pi;\pi^*)^3$ | 3.29 (0.00) | $(\pi;\pi^*)^3$ | 3.16 (0.00) | $(\pi;\pi^*)^3$ |
| $T_3$ | 3.46 (0.00) | $(\pi;\pi^*)^3$ | 3.51 (0.00) | $(\pi;\pi^*)^3$ | 3.40 (0.00) | $(\pi;\pi^*)^3$ | 3.36 (0.00) | $(\pi;\pi^*)^3$ |
| $T_4$ | 3.92 (0.00) | $(\pi;\pi^*)^3$ | 3.99 (0.00) | $(\pi;\pi^*)^3$ | 3.86 (0.00) | $(\pi;\pi^*)^3$ | 3.67 (0.00) | $(\pi;\pi^*)^3$ |
| $T_5$ | 4.57 (0.00) | $(\pi;\pi^*)^3$ | 4.52 (0.00) | $(\pi;\pi^*)^3$ | 4.46 (0.00) | $(\pi;\pi^*)^3$ | 4.49 (0.00) | $(\pi;\pi^*)^3$ |
| $T_6$ | 4.95 (0.00) | $(\pi;\pi^*)^3$ | 4.91 (0.00) | $(\pi;\pi^*)^3$ | 4.83 (0.00) | $(\pi;\pi^*)^3$ | 4.73 (0.00) | $(\pi;\pi^*)^3$ |
| $T_7$ | 5.01 (0.00) | $(\sigma;\pi^*)^3$ | 4.92 (0.00) | $(\sigma;\pi^*)^3$ | 4.87 (0.00) | $(\sigma;\pi^*)^3$ | 4.98 (0.00) | $(\sigma;\pi^*)^3$ |
| $T_8$ | 5.40 (0.00) | $(\pi;\sigma^*)^3$ | 5.22 (0.00) | $(\pi;\sigma^*)^3$ | 5.17 (0.00) | $(\pi;\sigma^*)^3$ | 5.27 (0.00) | $(\pi;\sigma^*)^3$ |
| $T_9$ | 5.60 (0.00) | $(\pi;\pi^*)^3$ | 5.46 (0.00) | $(\pi;\pi^*)^3$ | 5.36 (0.00) | $(\pi;\pi^*)^3$ | 5.42 (0.00) | $(\pi;\pi^*)^3$ |
| $T_{10}$ | 5.72 (0.00) | $(\pi;\pi^*)^3$ | 5.58 (0.00) | $(\pi;\pi^*)^3$ | 5.49 (0.00) | $(\pi;\pi^*)^3$ | 5.62 (0.00) | $(\pi;\pi^*)^3$ |

**Table 3:** Spin-orbit couplings (cm$^{-1}$) between first eight singlets and ten triplets in **HM610** employing M062X (TDDFT) functional at $S_0$ optimized geometry.

| $\hat{H}_{DKH}$ | $\|T_1\rangle$ | $\|T_2\rangle$ | $\|T_3\rangle$ | $\|T_4\rangle$ | $\|T_5\rangle$ | $\|T_6\rangle$ | $\|T_7\rangle$ | $\|T_8\rangle$ | $\|T_9\rangle$ | $\|T_{10}\rangle$ |
|---|---|---|---|---|---|---|---|---|---|---|
| $\langle S_0\|$ | 1 | 1 | 0 | 1 | 3 | 33 | 11 | 98 | 3 | 2 |
| $\langle S_1\|$ | 0 | 0 | 0 | 1 | 1 | 7 | 2 | 15 | 1 | 0 |
| $\langle S_2\|$ | 0 | 0 | 1 | 1 | 1 | 3 | 2 | 15 | 1 | 0 |
| $\langle S_3\|$ | 14 | 13 | 36 | 9 | 7 | 0 | 9 | 8 | 3 | 8 |
| $\langle S_4\|$ | 3 | 2 | 6 | 1 | 1 | 5 | 5 | 38 | 2 | 2 |
| $\langle S_5\|$ | 10 | 2 | 7 | 3 | 2 | 1 | 1 | 2 | 12 | 6 |
| $\langle S_6\|$ | 1 | 1 | 2 | 0 | 1 | 7 | 1 | 1 | 0 | 1 |
| $\langle S_7\|$ | 1 | 1 | 1 | 0 | 1 | 7 | 0 | 1 | 0 | 1 |

**Table 4:** Spin-orbit couplings (cm$^{-1}$) between first eight singlets and ten triplets in **HM610** employing $\omega$B97 (TDDFT) functional at $S_0$ optimized geometry.

| $\hat{H}_{DKH}$ | $\|T_1\rangle$ | $\|T_2\rangle$ | $\|T_3\rangle$ | $\|T_4\rangle$ | $\|T_5\rangle$ | $\|T_6\rangle$ | $\|T_7\rangle$ | $\|T_8\rangle$ | $\|T_9\rangle$ | $\|T_{10}\rangle$ |
|---|---|---|---|---|---|---|---|---|---|---|
| $\langle S_0\|$ | 0 | 1 | 0 | 0 | 1 | 1 | 29 | 110 | 2 | 0 |
| $\langle S_1\|$ | 1 | 0 | 1 | 1 | 0 | 1 | 7 | 14 | 1 | 0 |
| $\langle S_2\|$ | 1 | 1 | 0 | 1 | 0 | 1 | 5 | 20 | 1 | 1 |
| $\langle S_3\|$ | 0 | 0 | 1 | 0 | 0 | 1 | 6 | 37 | 1 | 1 |
| $\langle S_4\|$ | 15 | 34 | 12 | 13 | 1 | 11 | 1 | 1 | 13 | 4 |
| $\langle S_5\|$ | 8 | 14 | 6 | 11 | 2 | 7 | 0 | 1 | 10 | 1 |
| $\langle S_6\|$ | 1 | 2 | 1 | 1 | 0 | 1 | 2 | 21 | 1 | 0 |
| $\langle S_7\|$ | 0 | 1 | 0 | 0 | 0 | 1 | 6 | 17 | 1 | 0 |

**Table 5:** Spin-orbit couplings ($cm^{-1}$) between first eight singlets and ten triplets in **HM610** employing $\omega$B97X (TDDFT) functional at $S_0$ optimized geometry.

| $\hat{H}_{DKH}$ | $\vert T_1\rangle$ | $\vert T_2\rangle$ | $\vert T_3\rangle$ | $\vert T_4\rangle$ | $\vert T_5\rangle$ | $\vert T_6\rangle$ | $\vert T_7\rangle$ | $\vert T_8\rangle$ | $\vert T_9\rangle$ | $\vert T_{10}\rangle$ |
|---|---|---|---|---|---|---|---|---|---|---|
| $\langle S_0\vert$ | 0 | 1 | 0 | 0 | 1 | 1 | 30 | 101 | 1 | 0 |
| $\langle S_1\vert$ | 1 | 0 | 1 | 1 | 0 | 1 | 7 | 15 | 1 | 0 |
| $\langle S_2\vert$ | 1 | 0 | 1 | 1 | 0 | 1 | 4 | 20 | 1 | 1 |
| $\langle S_3\vert$ | 0 | 1 | 1 | 0 | 0 | 1 | 7 | 38 | 1 | 1 |
| $\langle S_4\vert$ | 14 | 27 | 30 | 14 | 1 | 12 | 1 | 1 | 10 | 5 |
| $\langle S_5\vert$ | 11 | 3 | 4 | 7 | 2 | 3 | 0 | 1 | 11 | 1 |
| $\langle S_6\vert$ | 1 | 1 | 1 | 1 | 0 | 1 | 1 | 19 | 1 | 0 |
| $\langle S_7\vert$ | 1 | 1 | 1 | 0 | 0 | 1 | 7 | 11 | 1 | 0 |

**Table 6:** Spin-orbit couplings ($cm^{-1}$) between first eight singlets and ten triplets in **HM610** employing $\omega$B97X-D (TDDFT) functional at $S_0$ optimized geometry.

| $\hat{H}_{DKH}$ | $\vert T_1\rangle$ | $\vert T_2\rangle$ | $\vert T_3\rangle$ | $\vert T_4\rangle$ | $\vert T_5\rangle$ | $\vert T_6\rangle$ | $\vert T_7\rangle$ | $\vert T_8\rangle$ | $\vert T_9\rangle$ | $\vert T_{10}\rangle$ |
|---|---|---|---|---|---|---|---|---|---|---|
| $\langle S_0\vert$ | 0 | 1 | 0 | 0 | 1 | 11 | 27 | 88 | 1 | 1 |
| $\langle S_1\vert$ | 0 | 0 | 1 | 1 | 0 | 3 | 6 | 16 | 1 | 0 |
| $\langle S_2\vert$ | 0 | 0 | 1 | 1 | 0 | 2 | 3 | 20 | 1 | 0 |
| $\langle S_3\vert$ | 2 | 2 | 4 | 1 | 1 | 3 | 6 | 39 | 1 | 1 |
| $\langle S_4\vert$ | 16 | 17 | 38 | 13 | 3 | 12 | 6 | 3 | 1 | 7 |
| $\langle S_5\vert$ | 1 | 1 | 2 | 1 | 0 | 2 | 3 | 17 | 1 | 1 |
| $\langle S_6\vert$ | 11 | 3 | 4 | 4 | 2 | 2 | 1 | 1 | 13 | 4 |
| $\langle S_7\vert$ | 1 | 1 | 1 | 0 | 0 | 4 | 7 | 4 | 1 | 1 |

**Table 7:** Spin-orbit couplings ($cm^{-1}$) between first eight singlets and ten triplets in **HM610** employing CAM-B3LYP (TDDFT) functional at $S_0$ optimized geometry.

| $\hat{H}_{DKH}$ | $\lvert T_1\rangle$ | $\lvert T_2\rangle$ | $\lvert T_3\rangle$ | $\lvert T_4\rangle$ | $\lvert T_5\rangle$ | $\lvert T_6\rangle$ | $\lvert T_7\rangle$ | $\lvert T_8\rangle$ | $\lvert T_9\rangle$ | $\lvert T_{10}\rangle$ |
|---|---|---|---|---|---|---|---|---|---|---|
| $\langle S_0\rvert$ | 0 | 1 | 1 | 0 | 2 | 2 | 30 | 91 | 4 | 2 |
| $\langle S_1\rvert$ | 1 | 0 | 1 | 1 | 0 | 1 | 7 | 16 | 1 | 0 |
| $\langle S_2\rvert$ | 0 | 0 | 1 | 1 | 0 | 1 | 3 | 18 | 1 | 0 |
| $\langle S_3\rvert$ | 3 | 4 | 6 | 2 | 1 | 2 | 7 | 40 | 3 | 2 |
| $\langle S_4\rvert$ | 14 | 19 | 34 | 13 | 2 | 13 | 3 | 5 | 2 | 9 |
| $\langle S_5\rvert$ | 1 | 2 | 4 | 1 | 0 | 2 | 4 | 16 | 1 | 1 |
| $\langle S_6\rvert$ | 12 | 3 | 4 | 5 | 2 | 3 | 0 | 1 | 13 | 6 |
| $\langle S_7\rvert$ | 1 | 1 | 1 | 0 | 0 | 1 | 8 | 4 | 0 | 1 |

**Table 8:** Spin-orbit couplings ($cm^{-1}$) between first eight singlets and ten triplets in **HM610** employing BHLYP (TDDFT) functional at $S_0$ optimized geometry.

| $\hat{H}_{DKH}$ | $\lvert T_1\rangle$ | $\lvert T_2\rangle$ | $\lvert T_3\rangle$ | $\lvert T_4\rangle$ | $\lvert T_5\rangle$ | $\lvert T_6\rangle$ | $\lvert T_7\rangle$ | $\lvert T_8\rangle$ | $\lvert T_9\rangle$ | $\lvert T_{10}\rangle$ |
|---|---|---|---|---|---|---|---|---|---|---|
| $\langle S_0\rvert$ | 0 | 1 | 0 | 0 | 1 | 1 | 31 | 96 | 9 | 2 |
| $\langle S_1\rvert$ | 1 | 0 | 0 | 0 | 0 | 1 | 8 | 15 | 2 | 0 |
| $\langle S_2\rvert$ | 1 | 1 | 0 | 1 | 0 | 1 | 3 | 14 | 2 | 0 |
| $\langle S_3\rvert$ | 1 | 2 | 1 | 1 | 0 | 1 | 7 | 36 | 4 | 1 |
| $\langle S_4\rvert$ | 10 | 35 | 2 | 11 | 1 | 11 | 2 | 3 | 5 | 11 |
| $\langle S_5\rvert$ | 3 | 11 | 1 | 3 | 0 | 4 | 4 | 16 | 0 | 3 |
| $\langle S_6\rvert$ | 13 | 3 | 3 | 7 | 2 | 3 | 0 | 1 | 11 | 8 |
| $\langle S_7\rvert$ | 1 | 1 | 1 | 0 | 0 | 0 | 6 | 4 | 1 | 0 |

## References

**1**. (a) Gupta, V.; Kant, V., A review on biological activity of imidazole and thiazole moieties and their derivatives. *Sci. Int.* **2013**, 1 (7), 253-260; (b) Abdu-Rahem, L. R.; Ahmad, A. K.; Abachi, F. T., Synthesis and medicinal attributes of thiazole derivatives: A review. *Sys. Rev. Pharm.* **2021**, 12, 290-295.

**2**. (a) Mishra, C. B.; Kumari, S.; Tiwari, M., Thiazole: A promising heterocycle for the development of potent CNS active agents. *Eur. J. Med. Chem.* **2015**, 92, 1-34. (b) Modrić, M.; Božičević, M.; Odak, I.; Talić, S.; Barić, D.; Mlakić, M.; Raspudić, A.; Škorić, I., The structure–activity relationship and computational studies of 1, 3-thiazole derivatives as cholinesterase inhibitors with anti-inflammatory activity. *C. R. Chim.* **2022**, 25 (G1), 267-279; (c) Mabkhot, Y. N.; Alharbi, M. M.; Al-Showiman, S. S.; Ghabbour, H. A.; Kheder, N. A.; Soliman, S. M.; Frey, W., Stereoselective synthesis, X-ray analysis, computational studies and biological evaluation of new thiazole derivatives as potential anticancer agents. *Chem. Cent. J.* **2018**, 12, 1-9; (d) Yadav, C. K.; Nandeshwarappa, B.; Pasha, K. M., Synthesis, computational study, solvatochromism and biological studies of thiazole-owing hydrazone derivatives. *Chim. Tech. Acta* **2023**, 10 (1), 202310110.

**3**. (a) Siddiqui, N.; Arshad, M. F.; Ahsan, W.; Alam, M. S., Thiazoles: a valuable insight into the recent advances and biological activities. *Int. J. Pharm. Sci. Drug Res* **2009**, 1 (3), 136-143; (b) Kashyap, M.; Mazumder, M. U.; Patowary, P.; Talukdar, A.; Sahariah, B. J.; Majumder, M., An Overview of Synthetic Derivatives of Thiazole and Their Role in Therapeutics. Fabad *J. Pharm. Sci.* **2024**, 49 (3), 603-626; (c) Singh, I. P.; Gupta, S.; Kumar, S., Thiazole compounds as antiviral agents: An update. *Med. Chem.* **2020**, 16 (1), 4-23; (d) Yurttaş, L.; Özkay, Y.; Karaca Gençer, H.; Acar, U., Synthesis of some new thiazole derivatives and their biological activity evaluation. *J. Chem.* **2015** (1), 464379; (e) Arora, P.; Narang, R.; Bhatia, S.; Nayak, S. K.; Singh, S. K.; Narasimhan, B., Synthesis, molecular docking and QSAR studies of 2, 4-disubstituted thiazoles as antimicrobial agents. *J. App. Pharm. Sci.* **2015**, 5 (2), 028-042; (f) Fogel, N., Tuberculosis: a disease without boundaries. *Tuberculosis* **2015**, 95 (5), 527-531; (g) Nagireddy, P. K. R.; Kommalapati, V. K.; Siva Krishna, V.; Sriram, D.; Tangutur, A. D.; Kantevari, S., Imidazo [2, 1-b] thiazole-coupled natural noscapine derivatives as anticancer agents. *ACS Omega* **2019**, 4 (21), 19382-19398; (h) Wu, J.; Ma, Y.; Zhou, H.; Zhou, L.; Du, S.; Sun, Y.; Li, W.; Dong, W.; Wang, R.,

Identification of protein tyrosine phosphatase 1B (PTP1B) inhibitors through De Novo Evoluton, synthesis, biological evaluation and molecular dynamics simulation. *Biochem. Biophys. Res. Comm.* **2020**, 526 (1), 273-280; (i) Manju, S., Identification and development of thiazole leads as COX-2/5-LOX inhibitors through in-vitro and in-vivo biological evaluation for anti-inflammatory activity. *Bioorg. Chem.* **2020**, 100, 103882; (j) Afifi, O. S.; Shaaban, O. G.; Abd El Razik, H. A.; El, S. E.-D. A. S.; Ashour, F. A.; El-Tombary, A. A.; Abu-Serie, M. M., Synthesis and biological evaluation of purine-pyrazole hybrids incorporating thiazole, thiazolidinone or rhodanine moiety as 15-LOX inhibitors endowed with anticancer and antioxidant potential. *Bioorg. Chem.* **2019**, 87, 821-837; (k) Farghaly, T. A.; Alfaifi, G. H.; Gomha, S. M., Recent literature on the synthesis of thiazole derivatives and their biological activities. *Mini Rev. Med. Chem.* **2024**, 24 (2), 196-251.

**4**. (a) Chen, L.; Liu, X. Y.; Zou, Y. X., Recent Advances in the Construction of Phosphorus-Substituted Heterocycles, 2009–2019. *Adv. Synt. Cat.* **2020**, 362 (9), 1724-1818; (b) Eltyshev, A. K.; Dzhumaniyazov, T. H.; Suntsova, P. O.; Minin, A. S.; Pozdina, V. A.; Dehaen, W.; Benassi, E.; Belskaya, N. P., 3-Aryl-2-(thiazol-2-yl) acrylonitriles assembled with aryl/hetaryl rings: Design of the optical properties and application prospects. *Dyes Pigm.* **2021**, 184, 108836; (c) Abdulrahman, B. S.; Nadr, R. B.; Omer, R. A.; Azeez, Y. H.; Kareem, R. O.; Safin, D. A., Synthesis, characterization and computational studies of a series of the thiazole-thiazolidinone hybrids. *J. Mol. Struc.* **2024**, 140806; (d) Siddique choudhry, S.; Mehmood, H.; Haroon, M.; Akhtar, T.; Tahir, E.; Ehsan, M.; Musa, M., Structure-Activity Relationship of Hydrazinylthiazole-5-Carbaldehydes as Potential Anti-Diabetic Agents. *Chem. Biodivers.* **2024**, 21(11), e202400305; (e) Mehmood, H.; Haroon, M.; Akhtar, T.; Jamal, S.; Akhtar, M. N.; Khan, M. U.; Alhokbany, N., Exploring the optical properties of novel N-benzylated thiazoles-based chromophores: Spectroscopic insights and computational analysis. *Synth. Met.* **2024**, 307, 117701.

**5**. Roure, B.; Alonso, M.; Lonardi, G.; Berna Yildiz, D.; Buettner, C. S.; dos Santos, T.; Xu, Y.; Bossart, M.; Derdau, V.; Méndez, M.; Llaveria, J.; Ruffoni, A.; Leonori, D. Photochemical permutation of thiazoles, isothiazoles and other azoles, *Nature*, **2025**, 637, 860–867.

**6**. (a) Srnec, M.; Chalupsky, J.; Fojta, M.; Zendlová, L.; Havran, L.; Hocek, M.; Kyvala, M.; Rulisek, L., Effect of Spin− Orbit Coupling on Reduction Potentials of Octahedral Ruthenium (II/III) and Osmium (II/III) Complexes. *J. Am. Chem. Soc.* **2008**, 130 (33), 10947-10954; (b) Wasif Baig, M.; Pederzoli, M.; Kývala, M.; Pittner, J., Quantum Chemical and Trajectory Surface

Hopping Molecular Dynamics Study of Iodine-based BODIPY Photosensitizer. *J. Comp. Chem.* **2025**, 46 (7), 70026. (c) Wasif Baig, M.; Pederzoli, M.; Kývala, M. r.; Cwiklik, L.; Pittner, J., Theoretical investigation of the effect of alkylation and bromination on intersystem crossing in BODIPY-based photosensitizers. *J. Phys. Chem. B* **2021**, 125 (42), 11617-11627. (d) Plasser, F.; Lischka, H.; Shepard, R.; Szalay, P. G.; Pitzer, R. M.; Alves, R. L.; Aquino, A. J.; Autschbach, J.; Barbatti, M.; Carvalho, J. R., COLUMBUS─ An Efficient and General Program Package for Ground and Excited State Computations Including Spin–Orbit Couplings and Dynamics. *J. Phys. Chem. A* **2025**, 129 (28), 6482-6517.

**7**. (a) Penfold, T.; Worth, G., The effect of molecular distortions on spin–orbit coupling in simple hydrocarbons. *Chem. Phys.* **2010**, 375 (1), 58-66; (b) Danilov, D.; Jenkins, A. J.; Bearpark, M. J.; Worth, G. A.; Robb, M. A., Coherent Mixing of Singlet and Triplet States in Acrolein and Ketene: A Computational Strategy for Simulating the Electron–Nuclear Dynamics of Intersystem Crossing. *J. Phys. Chem. Lett.* **2023**, 14 (26), 6127-6134.

**8**. Pal, R.; Chattaraj, P. K. Chemical Reactivity from a Conceptual Density Functional Theory Perspective. J. Indian Chem. Soc. 2021, 98 (1), 100008.

**9**. Baerends, E. J.; Gritsenko, O. V., A quantum chemical view of density functional theory. *J. Phys. Chem. A* **1997**, 101 (30), 5383-5403.

**10**. Malik, A. M.; Rohrer, J.; Albe, K., Theoretical study of thermodynamic and magnetic properties of transition metal carbide and nitride MAX phases. *Phys. Rev. Mater.* **2023**, 7 (4), 044408. (b) Hussain, A.; Baig, M. W.; Mustafa. N., DFT Studies of Indium Nanoclusters, Nanotubes and their Interaction with Molecular Hydrogen. *The Nucleus* **2015**, 52 (4), 185-191. c) Alotaibi, N. O.; Abdulhussein, H. A.; Alamri, S. M.; Hamza, N. A.; Abo Nasria, A. H.; Computational insights into the physico-chemical properties of pure and single-atom copper–indium sub-nanometre clusters: a DFT-genetic algorithm approach. *RSC advances* **2025**, 15 (8), 5856-5875.

**11**. (a) Wasif Baig, M.; Pederzoli, M.; Jurkiewicz, P.; Cwiklik, L.; Pittner, J., Orientation of Laurdan in phospholipid bilayers influences its fluorescence: Quantum mechanics and classical molecular dynamics study. *Molecules* **2018**, 23 (7), 1707; (b) Ahmad, K.; Khan, B. A.; Akram, B.;

Khan, J.; Mahmood, R.; Roy, S. K., Theoretical investigations on copper catalyzed CN cross-coupling reaction between aryl chlorides and amines. *Comput. Theor. Chem.* **2018**, 1134, 1-7.

**12**. (a) Amir, M. K.; Hogarth, G.; Khan, Z.; Imran, M., Platinum (II) dithiocarbamate complexes [Pt (S2CNR2) Cl (PAr3)] as anticancer and DNA-damaging agents. *Inorg. Chim. Acta* **2020**, 512, 119853; (b) Ahmad, K.; Khan, B. A.; Akhtar, T.; Khan, J.; Roy, S. K., Deciphering the mechanism of copper-catalyzed N-arylation between aryl halides and nitriles: a DFT study. *New J. Chem.* **2019**, 43 (48), 19200-19207. (c) Ahmad, K.; Khan, B. A.; Roy, S. K.; Mahmood, R.; Khan, J.; Ashraf, H. Theoretical Insights on the C–C Bond Reductive Elimination from Co(III) Center. *Comput. Theor. Chem.* **2018**, *1130*, 140–147.

**13.** (a) Dinpajooh, M.; Intan, N. N.; Duignan, T. T.; Biasin, E.; Fulton, J. L.; Kathmann, S. M.; Schenter, G. K.; Mundy, C. J., Beyond the Debye–Hückel limit: Toward a general theory for concentrated electrolytes. *J. Chem. Phys.* **2024**, 161 (23). (b) Baig, M. W.; Siddiq, M., Quantum mechanics of in situ synthesis of metal nanoparticles within anionic microgels. *J. Theor. Chem.* **2013**, 2013 (1), 410417. (c) Chew, P. Y.; Reinhardt, A., Phase diagrams—Why they matter and how to predict them. *J. Chem. Phys.* **2023**, 158 (3). (c) Bui, A. T.; Cox, S. J. Learning Classical Density Functionals for Ionic Fluids. *Phys. Rev. Lett.* **2025**, *134* (14), 148001.

**14.** Laurent, A. D.; Jacquemin, D. TD-DFT Benchmarks: A Review. *Int. J. Quantum Chem.* 2013, *113* (17), 2019–2039.

**15**. (a) Kupfer, S.; Wächtler, M.; Guthmuller, J.; Popp, J.; Dietzek, B.; González, L. A Novel Ru(II) Polypyridine Black Dye Investigated by Resonance Raman Spectroscopy and TDDFT Calculations. *J. Phys. Chem. C* 2012, *116*, 19968–19977 (b) Pederzoli, M.; Wasif Baig, M.; Kyvala, M.; Pittner, J.; Cwiklik, L., Photophysics of BODIPY-based photosensitizer for photodynamic therapy: surface hopping and classical molecular dynamics. *J. Chem. Theory Comput.* 2019, 15 (9), 5046-5057.

**16**. Christiansen, O.; Koch, H.; Jørgensen, P., The second-order approximate coupled cluster singles and doubles model CC2. *Chem. Phys. Lett.* **1995**, 243 (5-6), 409-418.

**17**. Schirmer, J., Beyond the random-phase approximation: A new approximation scheme for the polarization propagator. *Phys. Rev. A* **1982**, 26 (5), 2395.

**18**. Loos, P.-F.; Scemama, A.; Jacquemin, D., The quest for highly accurate excitation energies: A computational perspective. *J. Phys. Chem. Lett.* **2020**, 11 (6), 2374-2383.

**19**. (a) Sülzner, N.; Hättig, C., Role of Singles Amplitudes in ADC (2) and CC2 for Low-Lying Electronically Excited States. *J. Chem. Theory Comput.* **2024**, 20 (6), 2462-2474; (b) Jacquemin, D.; Duchemin, I.; Blase, X., 0–0 energies using hybrid schemes: Benchmarks of TD-DFT, CIS (D), ADC (2), CC2, and BSE/GW formalisms for 80 real-life compounds. *J. Chem. Theory Comput.* **2015**, 11 (11), 5340-5359.

**20**. (a) Szabla, R.; Góra, R. W.; Šponer, J., Ultrafast excited-state dynamics of isocytosine. *Phys. Chem. Chem. Phys.* **2016**, 18 (30), 20208-20218; (b) Plasser, F.; Crespo-Otero, R.; Pederzoli, M.; Pittner, J.; Lischka, H.; Barbatti, M., Surface hopping dynamics with correlated single-reference methods: 9H-adenine as a case study. *J. Chem. Theory Comput.* **2014**, 10 (4), 1395-1405.

**21.** Minns, R.; Parker, D.; Penfold, T.; Worth, G.; Fielding, H., Competing ultrafast intersystem crossing and internal conversion in the "channel 3" region of benzene. *Physical Chemistry Chemical Physics* **2010,** *12* (48), 15607-15615.

**22**. (a) Penfold, T. J.; Gindensperger, E.; Daniel, C.; Marian, C. M., Spin-vibronic mechanism for intersystem crossing. *Chem. Rev.* **2018**, 118 (15), 6975-7025. (b) Liao, C.; Rosenbaum, C.; Glaudin, A. M.; Taub, M.; Banerjee Ghosh, R.; Pristash, S.; Schlenker, C. W.; Li, X., Spin–Vibronic Coupling Enhanced Intersystem Crossing beyond El-Sayed Restrictions. *J. Am. Chem. Soc.* **2025**. (c) Tatchen, J.; Gilka, N.; Marian, C. M., Intersystem crossing driven by vibronic spin–orbit coupling: a case study on psoralen. *Phys. Chem. Chem. Phys.* **2007**, 9 (38), 5209-5221. (d) Albrecht, A. C. Vibronic—Spin-Orbit Perturbations and the Assignment of the Lowest Triplet State of Benzene. *J. Chem. Phys.* **1963**, *38* (2), 354–365.

**23**. (a) Richter, M.; Marquetand, P.; González-Vázquez, J.; Sola, I.; González, L., SHARC: ab initio molecular dynamics with surface hopping in the adiabatic representation including arbitrary couplings. *J. Chem. Theory Comput.* **2011**, 7 (5), 1253-1258. (b) Mai, S.; Marquetand, P.; González, L., A general method to describe intersystem crossing dynamics in trajectory surface hopping. *Int. J. Quantum Chem.* **2015**, 115 (18), 1215-1231.

**24**. (a) Sit, M. K.; Das, S.; Samanta, K., Semiclassical dynamics on machine-learned coupled multireference potential energy surfaces: application to the photodissociation of the simplest criegee intermediate. *J. Phys. Chem. A* **2023**, 127 (10), 2376-2387. (b) Sit, M. K.; Das, S.; Samanta, K. Machine Learning-Assisted Mixed Quantum-Classical Dynamics without Explicit Nonadiabatic Coupling: Application to the Photodissociation of Peroxynitric Acid. *J. Phys. Chem. A* **2024**, *128* (38), 8244–8253. (c) Heindl, M.; Hongyan, J.; Hua, S.-A.; Oelschlegel, M.; Meyer, F.; Schwarzer, D.; González, L., Excited-state dynamics of [Ru (S–Sbpy)(bpy)2] 2+ to form long-lived localized triplet states. *Inorg. Chem.* **2021**, 60 (3), 1672-1682. (d) Atkins, A. J.; González, L., Trajectory surface-hopping dynamics including intersystem crossing in [Ru (bpy) 3] 2+. *J. Phys. Chem. Lett.* **2017**, 8 (16), 3840-3845.

**25.** Liu, W.; Xiao, Y. Relativistic Time-Dependent Density Functional Theories. *Chem. Soc. Rev.* **2018**, 47 (12), 4481–4509.

**26**. Turbomole, V., 7.3, a development of University of Karlsruhe and Forschungszentrum Karlsruhe GmbH, 1989–2007. TURBOMOLE GmbH, since 2007, 2010.

**27**. (a) Liu, W.; Hong, G.; Dai, D.; Li, L.; Dolg, M., The Beijing four-component density functional program package (BDF) and its application to EuO, EuS, YbO and YbS. *Theor. Chem. Acc.* **1997**, 96, 75-83; (b) Liu, W.; Wang, F.; Li, L., The Beijing density functional (BDF) program package: methodologies and applications. *J. Theor. Comp. Chem.* **2003**, 2 (02), 257-272; (c) Liu, W.; Wang, F.; Li, L., Relativistic density functional theory: The BDF program package. Recent advances in relativistic molecular theory **2004**, 257-282; (d) Zhang, Y.; Suo, B.; Wang, Z.; Zhang, N.; Li, Z.; Lei, Y.; Zou, W.; Gao, J.; Peng, D.; Pu, Z., BDF: A relativistic electronic structure program package. *J. Chem. Phys.* **2020**, 152 (6), 064113.

**28**. Gaussian 09, Revision D.01, M. J. Frisch, G. W. Trucks, H. B. Schlegel, G. E. Scuseria, M. A. Robb, J. R. Cheeseman, G. Scalmani, V. Barone, G. A. Petersson, H. Nakatsuji, X. Li, M. Caricato, A. Marenich, J. Bloino, B. G. Janesko, R. Gomperts, B. Mennucci, H. P. Hratchian, J. V. Ortiz, A. F. Izmaylov, J. L. Sonnenberg, D. Williams-Young, F. Ding, F. Lipparini, F. Egidi, J. Goings, B. Peng, A. Petrone, T. Henderson, D. Ranasinghe, V. G. Zakrzewski, J. Gao, N. Rega, G. Zheng, W. Liang, M. Hada, M. Ehara, K. Toyota, R. Fukuda, J. Hasegawa, M. Ishida, T. Nakajima, Y. Honda, O. Kitao, H. Nakai, T. Vreven, K. Throssell, J. A. Montgomery, Jr., J. E. Peralta, F. Ogliaro, M. Bearpark, J. J. Heyd, E. Brothers, K. N. Kudin, V. N. Staroverov, T. Keith, R. Kobayashi, J. Normand, K. Raghavachari, A. Rendell, J. C. Burant, S. S. Iyengar, J. Tomasi, M. Cossi, J. M.

Millam, M. Klene, C. Adamo, R. Cammi, J. W. Ochterski, R. L. Martin, K. Morokuma, O. Farkas, J. B. Foresman, and D. J. Fox, Gaussian, Inc., Wallingford CT, 2016.

**29**. (a) Barbatti, M.; Ruckenbauer, M.; Plasser, F.; Pittner, J.; Granucci, G.; Persico, M.; Lischka, H., Newton-X: a surface-hopping program for nonadiabatic molecular dynamics. *Wiley Interdiscip. Rev. Comput. Mol. Sci.* **2014**, 4 (1), 26-33. (b) Barbatti, M.; Bondanza, M.; Crespo-Otero, R.; Demoulin, B.; Dral, P. O.; Granucci, G.; Kossoski, F.; Lischka, H.; Mennucci, B.; Mukherjee, S., Newton-X platform: New software developments for surface hopping and nuclear ensembles. *J. Chem. Theory Comput.* **2022**, 18 (11), 6851-6865.

**30**. (a) Zhao, Y.; Truhlar, D. G., The M06 suite of density functionals for main group thermochemistry, thermochemical kinetics, noncovalent interactions, excited states, and transition elements: two new functionals and systematic testing of four M06-class functionals and 12 other functionals. *Theor. Chem. Acc.* **2008**, 120, 215-241; (b) Seeger, Z. L.; Izgorodina, E. I., A systematic study of DFT performance for geometry optimizations of ionic liquid clusters. *J. Chem. Theory Comput.* **2020**, 16 (10), 6735-6753; (c) Walker, M.; Harvey, A. J. A.; Sen, A.; Dessent, C. E. H. Performance of M06, M062X, and M06-HF Density Functionals for Conformationally Flexible Anionic Clusters: M06 Functionals Perform Better than B3LYP for a Model System with Dispersion and Ionic Hydrogen-Bonding Interactions. *J. Phys. Chem. A* **2013**, 117 (47), 12590–12600.

**31**. (a) Weigend, F., Accurate Coulomb-fitting basis sets for H to Rn. *Phys. Chem. Chem. Phys.* **2006**, 8 (9), 1057-1065; (b) Zheng, J.; Xu, X.; Truhlar, D. G., Minimally augmented Karlsruhe basis sets. *Theo. Chem. Acc.* **2011**, 128, 295-305.

**32**. Grimme, S.; Antony, J.; Ehrlich, S.; Krieg, H. A Consistent and Accurate Ab Initio Parametrization of Density Functional Dispersion Correction (DFT-D) for the 94 Elements H–Pu. *J. Chem. Phys.* **2010**, 132 (15), 154104.

**33**. Krause, K.; Klopper, W., Implementation of the Bethe− Salpeter equation in the TURBOMOLE program. Wiley Online Library: **2017**.

**34**. (a) Becke, A. D., A new mixing of Hartree-Fock and local density-functional theories. *J. Chem. Phys.* **1993**, 98 (2), 1372-1377. (b) Weigend, F.; Baldes, A., Segmented contracted basis sets for one-and two-component Dirac–Fock effective core potentials. *J. Chem. Phys.* **2010**, 133 (17).

**35**. Yanai, T.; Tew, D. P.; Handy, N. C., A new hybrid exchange–correlation functional using the Coulomb-attenuating method (CAM-B3LYP). *Chem. Phys. Lett.* **2004**, 393 (1-3), 51-57.

**36**. (a) Chai, J.-D.; Head-Gordon, M., Systematic optimization of long-range corrected hybrid density functionals. *J. Chem. Phys.* **2008**, 128 (8):084106. (b) Jacquemin, D.; Perpete, E. A.; Ciofini, I.; Adamo, C., Assessment of the ωB97 family for excited-state calculations. *Theor. Chem. Acc.* **2011**, 128 (1), 127-136.

**37**. Cossi, M.; Rega, N.; Scalmani, G.; Barone, V., Energies, structures, and electronic properties of molecules in solution with the C-PCM solvation model. *J. Comp. Chem.* **2003**, 24 (6), 669-681.

**38**. (a) Li, Z.; Suo, B.; Zhang, Y.; Xiao, Y.; Liu, W., Combining spin-adapted open-shell TD-DFT with spin–orbit coupling. *Mol. Phys.* **2013**, 111 (24), 3741-3755; (b) Zhou, Q.; Suo, B., New implementation of spin-orbit coupling calculation on multi-configuration electron correlation theory. *Int. J. Quantum Chem*. **2021**, 121 (20), e26772.

**39**. Luzanov, A.; Zhikol, O., Electron invariants and excited state structural analysis for electronic transitions within CIS, RPA, and TDDFT models. *Int. J. Quantum Chem.* **2010**, 110 (4), 902-924.

**40.** Brabec, J.; Brandejs, J.; Kowalski, K.; Xantheas, S.; Legeza, Ö.; Veis, L. Massively Parallel Quantum Chemical Density Matrix Renormalization Group Method. *J. Comput. Chem.* **2021**, *42* (8), 534–544.

**41.** Kurashige, Y. Multireference Electron Correlation Methods with Density Matrix Renormalisation Group Reference Functions. *Mol. Phys.* 2014, *112* (11), 1485–1494.

**42.** Pittner, J. Spin-Free Orbital Entropy, Mutual Information and Correlation Analysis. *Mol. Phys.* **2025**, *e2500632*.

**43.** (a) Macarios, C. M.; Pittner, J.; Prasad, V. K.; Fekl, U. Heteroatom-Vacancy Centres in Molecular Nanodiamonds: A Computational Study of Organic Molecules Possessing Triplet Ground States through σ-Overlap. *Phys. Chem. Chem. Phys.* **2024**, *26* (39), 25412–25417. (b) Vrška, D.; Urban, M.; Neogrády, P.; Pittner, J.; Blaško, M.; Pitoňák, M. DFT Modeling of Polyethylene Chains Cross-Linked by Selected ns1 and ns2 Metal Atoms. *J. Phys. Chem. A* **2024**, *128* (36), 7634–7647. (c) Višňák, J., Brandejs, J., Máté, M., Visscher, L., Legeza, Ö., & Pittner, J. (2024). DMRG-tailored coupled cluster method in the 4c-relativistic domain: General implementation and application to the NUHFI and $NUF_3$ molecules. *Journal of Chemical Theory and Computation, 20*(20), 8862–8875. (d) Višňák, J.; Brandejs, J.; Máté, M.; Visscher, L.; Legeza, Ö.; Pittner, J. DMRG-Tailored Coupled Cluster Method in the 4c-Relativistic Domain:

General Implementation and Application to the NUHFI and $NUF_3$ Molecules. *J. Chem. Theory Comput.* **2024**, *20* (20), 8862–8875.

**44**. Crespo-Otero, R.; Barbatti, M., Spectrum simulation and decomposition with nuclear ensemble: formal derivation and application to benzene, furan and 2-phenylfuran. In Marco Antonio Chaer Nascimento: A Festschrift from Theoretical Chemistry Accounts, Springer: 2012; pp 89-102.

**45**. (a) Barbatti, M., Nonadiabatic dynamics with trajectory surface hopping method. *Wiley Interdiscip. Rev. Comput. Mol. Sci.* **2011**, 1 (4), 620-633. (b) Pittner, J.; Lischka, H.; Barbatti, M., Optimization of mixed quantum-classical dynamics: Time-derivative coupling terms and selected couplings. *Chem. Phys.* **2009**, 356 (1-3), 147-152.

**46**. Pederzoli, M.; Pittner, J., A new approach to molecular dynamics with non-adiabatic and spin-orbit effects with applications to QM/MM simulations of thiophene and selenophene. *J. Chem. Phys.* **2017**, 146 (11).

**47**. Hirata, S.; Head-Gordon, M., Time-dependent density functional theory within the Tamm–Dancoff approximation. *Chem. Phys. Lett.* **1999**, 314 (3-4), 291-299.

**48**. Peach, M. J.; Williamson, M. J.; Tozer, D. J., Influence of triplet instabilities in TDDFT. *J. Chem. Theory Comput.* **2011**, 7 (11), 3578-3585.

**49**. (a) Yang, Y.; Shen, L.; Zhang, D.; Yang, W., Conical intersections from particle–particle random phase and Tamm–Dancoff approximations. *J. Phys. Chem. Lett.* **2016**, 7 (13), 2407-2411. (b) Park, W.; Komarov, K.; Lee, S.; Choi, C. H. Mixed-Reference Spin-Flip Time-Dependent Density Functional Theory: Multireference Advantages with the Practicality of Linear Response Theory. *J. Phys. Chem. Lett.* **2023**, *14* (39), 8896–8908.

**50**. (a) Cakmak, I., GIAO calculations of chemical shifts in enantiometrically pure 1-trifluoromethyl tetrahydroisoquinoline alkaloids. *J. Mol. Struc. THEOCHEM* **2005**, 716 (1-3), 143-148; (b) Tellgren, E. I.; Reine, S. S.; Helgaker, T., Analytical GIAO and hybrid-basis integral derivatives: application to geometry optimization of molecules in strong magnetic fields. *Phys. Chem. Chem. Phys.* **2012**, 14 (26), 9492-9499; (c) Wolinski, K.; Hinton, J. F.; Pulay, P., Efficient implementation of the gauge-independent atomic orbital method for NMR chemical shift calculations. *J. Am. Chem. Soc.* **1990**, 112 (23), 8251-8260; (d) Vieille, L.; Berlu, L.; Comborieu,

B.; Hoggan, P., A Quantum Chemistry GIAO molecular site approach of NMR chemical shifts generalized to the whole periodic table. *J. Theor. Comput. Chem.* **2002**, 1 (02), 295-308.

**51**. (a) Hättig, C., Optimization of auxiliary basis sets for RI-MP2 and RI-CC2 calculations: Core–valence and quintuple-ζ basis sets for H to Ar and QZVPP basis sets for Li to Kr. *Phys. Chem. Chem. Phys.* **2005**, 7 (1), 59-66. (b) Weigend, F.; Ahlrichs, R. Balanced Basis Sets of Split Valence, Triple Zeta Valence and Quadruple Zeta Valence Quality for H to Rn: Design and Assessment of Accuracy. *Phys. Chem. Chem. Phys.* **2005**, 7 (18), 3297–3305.

**52**. Iron, M. A., Evaluation of the Factors Impacting the Accuracy of 13C NMR Chemical Shift Predictions using Density Functional Theory The Advantage of Long-Range Corrected Functionals. *J. Chem. Theory Comput.* **2017**, 13 (11), 5798-5819.

**53.** (a) Abed, A. A.; Abid, M. A.; Turki, A.; Haroon, M.; Mehmood, H.; Akhtar, T.; Woodward, S., Synthesis of Substituted Arylidene Hydrazinyl Trifluoromethyl Thiazole Derivatives and their Antibacterial Studies using different Genes Expression. *ChemistrySelect* **2024**, 9 (17), e202401206; (b) Haroon, M.; Akhtar, T.; Mehmood, H.; da Silva Santos, A. C.; da Conceição, J. M.; Brondani, G. L.; Silva Tibúrcio, R. d.; Galindo Bedor, D. C.; Viturino da Silva, J. W.; Sales Junior, P. A., Synthesis of hydrazinyl–thiazole ester derivatives, in vitro trypanocidal and leishmanicidal activities. *Future Med. Chem.* **2024**, 16 (3), 221-238; (c) Mehmood, H.; Akhtar, T.; Haroon, M.; Shah, M.; Rashid, U.; Woodward, S., Synthesis of hydrazinylthiazole carboxylates: a mechanistic approach for treatment of diabetes and its complications. *Future Med. Chem.* **2023**, 15 (13), 1149-1165.

**54**. Taurins, A.; Fenyes, J.; Jones, R. N., Thiazoles: iii. Infrared spectra of methylthiazoles. *Can. J. Chem.* **1957**, 35 (5), 423-427.

**55**. Veys, K.; Bousquet, M. H.; Jacquemin, D.; Escudero, D., Modeling the fluorescence quantum yields of aromatic compounds: benchmarking the machinery to compute intersystem crossing rates. *J. Chem. Theory Comput.* **2023**, 19 (24), 9344-9357.

**56**. (a) Baig, M. W.; Mehmood, H.; Akhtar, T., Relativistic two-component density functional study of ethyl 2-(2-Iodobenzylidenehydrazinyl) thiazole-4-carboxylate. *Comp. Theor. Chem.* **2024**, 114670. (b) Mirza, W. B. *Ab Initio Molecular Dynamics of Photo-Active Organic Molecules with Non-Adiabatic and Spin-Orbit Effects*; Ph.D. Dissertation, Charles University, 2025. (c) Haroon,

M.; Baig, M. W.; Akhtar, T.; Tahir, M. N.; Ashfaq, M., Relativistic two-component time dependent density functional studies and Hirshfeld surface analysis of halogenated arylidenehydrazinylthiazole derivatives. *J. Mol. Struc.* **2023**, 1287, 135692.

**57.** (a) Verbeek, J.; Van Lenthe, J. H., The generalized Slater–Condon rules. International journal of quantum chemistry 1991, 40 (2), 201-210. (b) Burton, H. G., Generalized nonorthogonal matrix elements: Unifying Wick's theorem and the Slater–Condon rules. *J. Chem. Phys.* **2021**, 154 (14). (c) Višňák, J. *A Mathematical Introduction and Identity Used for the Study of the Functional of the Mean Quadratic Fluctuation of Energy in an Initio of Quantum Mechanical Calculations*; Bachelor's Thesis, Charles University, 2010.

**58.** Talbot, J. J.; Cheshire, T. P.; Cotton, S. J.; Houle, F. A.; Head-Gordon, M., The Role of Spin–Orbit Coupling in the Linear Absorption Spectrum and Intersystem Crossing Rate Coefficients of Ruthenium Polypyridyl Dyes. *J. Phys. Chem. A* **2024**, 128 (37), 7830-7842.

**59.** (a) McClure, D. S. Spin-Orbit Interaction in Aromatic Molecules. J. Chem. Phys. 1952, 20 (4), 682–686. (b) Mishima, K.; Kinoshita, T.; Hayashi, M.; Jono, R.; Segawa, H.; Yamashita, K., Theoretical investigation of [Ru (tpy) 2] 2+,[Ru (tpy)(bpy)(H2O)] 2+ and [Ru (tpy)(bpy)(Cl)]+ complexes in acetone revisited: Inclusion of strong spin–orbit couplings to quantum chemistry calculations. *J. Theor. Comput. Chem.* **2016**, 15 (01), 1650001. (c) Lin, S. H. Isotope Effect, Energy Gap Law and Temperature Effect in Resonance Energy Transfer. *Mol. Phys.* **1971**, *21* (4), 853–863.

**60.** (a) Komarov, K.; Park, W.; Lee, S.; Zeng, T.; Choi, C. H. Accurate Spin–Orbit Coupling by Relativistic Mixed-Reference Spin-Flip-TDDFT. *J. Chem. Theory Comput.* **2023**, *19* (3), 953–964. (b) Krylov, A. I. From Orbitals to Observables and Back. *J. Chem. Phys.* **2020**, *153* (8), 080901.

**61.** (a) Pokhilko, P.; Krylov, A. I. Quantitative El Sayed Rules for Many Body Wave Functions from Spinless Transition Density Matrices. *J. Phys. Chem. Lett.* **2019**, 10 (17), 4857–4862. (b) Lower, S. K.; El Sayed, M. A. The Triplet State and Molecular Electronic Processes in Organic Molecules. *Chem. Rev.* **1966**, 66 (2), 199–241. (c) Shimakura, N.; Fujimura, Y.; Nakajima, T. Theory of Intersystem Crossing in Aromatic Compounds: Extension of the El Sayed Rule. *Chem. Phys.* **1977**, 19 (2), 155–163.

**62.** (a) Lin, S. H. Rate of Interconversion of Electronic and Vibrational Energy. *J. Chem. Phys.* **1966**, *44* (10), 3759–3767. (b) Heller, E. J.; Brown, R. C. Radiationless Transitions in a New Light. *J. Chem. Phys.* **1983**, *79* (7), 3336–3351. (c) Manian, A.; Chen, Z.; Sullivan, H. T.; Russo, S. P. The Ups and Downs of Internal Conversion. *Rev. Mod. Phys.* 2025, *97* (3), 035003.

**63.** (a) Zhao, Z.; Cao, S.; Li, H.; Li, D.; He, Y.; Wang, X.; Knutson, J. R. Ultrafast Excited-State Dynamics of Thiazole Orange. *Chem. Phys.* **2022**, *553*, 111392. (b) Berezin, M. Y.; Achilefu, S. Fluorescence Lifetime Measurements and Biological Imaging. *Chem. Rev.* **2010**, *110* (5), 2641–2684.

**64.** (a) Cheeseman, J. R.; Trucks, G. W.; Keith, T. A.; Frisch, M. J. A Comparison of Models for Calculating Nuclear Magnetic Resonance Shielding Tensors. *J. Chem. Phys.* **1996**, 104 (14), 5497–5509. (b) Barone, V. Structure, Magnetic Properties, and Reactivities of Open-Shell Species from Density Functional and Self-Consistent Hybrid Methods. *J. Chem. Phys.* **1995**, 102 (1), 364–377. (c) Zhao, Y.; Truhlar, D. G. The M06 Suite of Density Functionals for Main Group Thermochemistry, Thermochemical Kinetics, Noncovalent Interactions, Excited States, and Transition Elements: Two New Functionals and Systematic Testing of Four M06-Class Functionals and 12 Other Functionals. *Phys. Chem. Chem. Phys.* **2008**, 10 (15), 10757–10816.

# Slater–Condon Rules and Spin–Orbit Couplings: 2-(2-(2,5-dimethoxybenzylidene)hydrazineyl)-4-(trifluoromethyl)thiazole a test case

Hasnain Mehmood,[1] Tashfeen Akhtar,[1] Jesús Espinosa-Romero,[2] Mauricio Maldonado-Domínguez,*[2] Jakub Višňák[3,4] and Mirza Wasif Baig*[4]

[1] Department of Chemistry, Mirpur University of Science and Technology (MUST), 10250-Mirpur (AJK) Pakistan.

[2] Facultad de Química, Departamento de Química Orgánica, Universidad Nacional Autónoma de México, 04510 Ciudad de México, México.

[3] Faculty of Mathematics and Physics, Charles University, Ke Karlovu 3, 12116 Prague, Czech Republic

[4] J. Heyrovský Institute of Physical Chemistry of the Czech Academy of Sciences, Dolejškova 2155/3, 18223 Prague 8, Czech Republic.

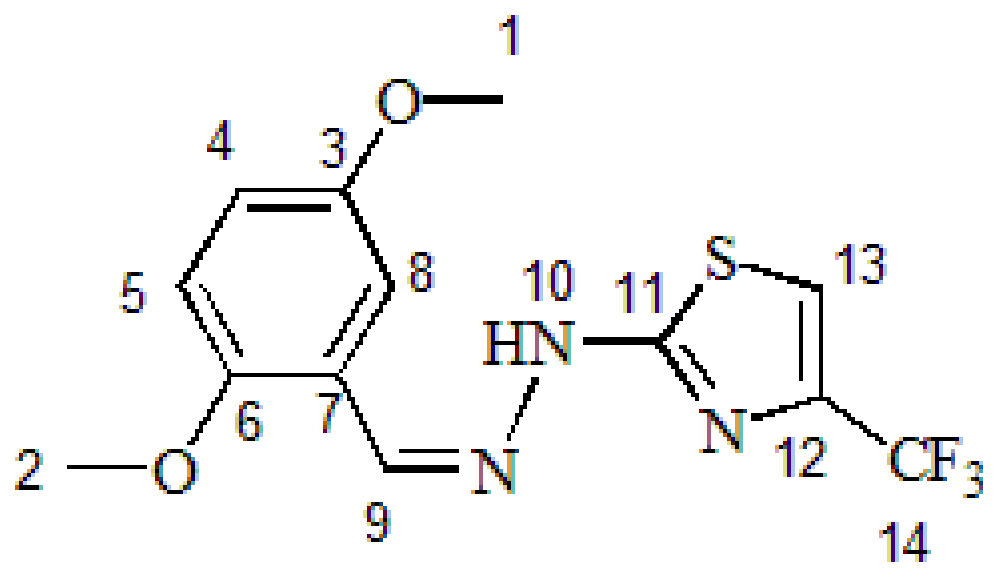

**Figure S1: Molecular Structure of 2-(2-(2,5-dimethoxybenzylidene)hydrazineyl)-4-(trifluoromethyl)thiazole (HM610)**

**UV spectrum of HM610**

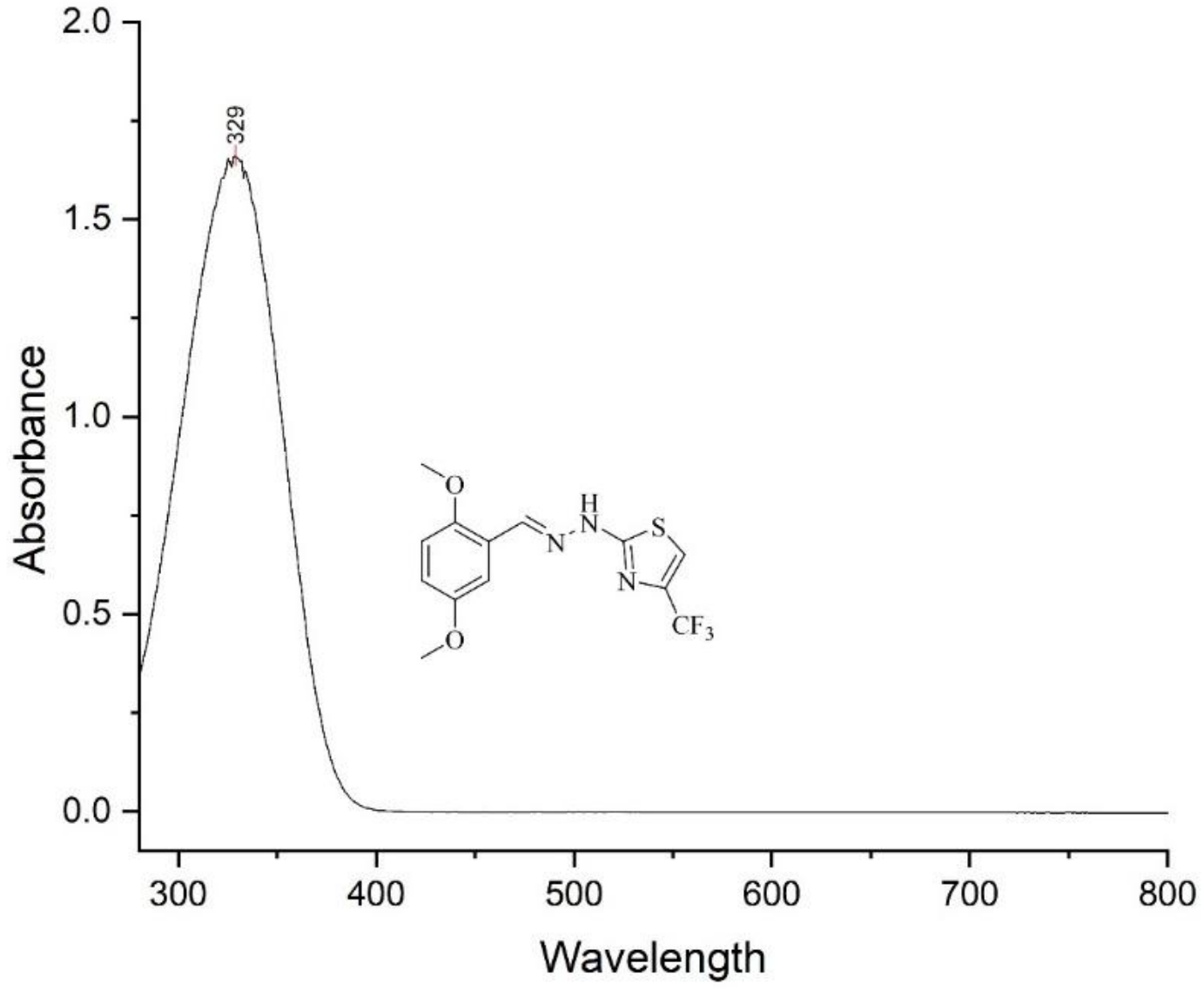

**Simulated spectrum of HM610 in gas phase**

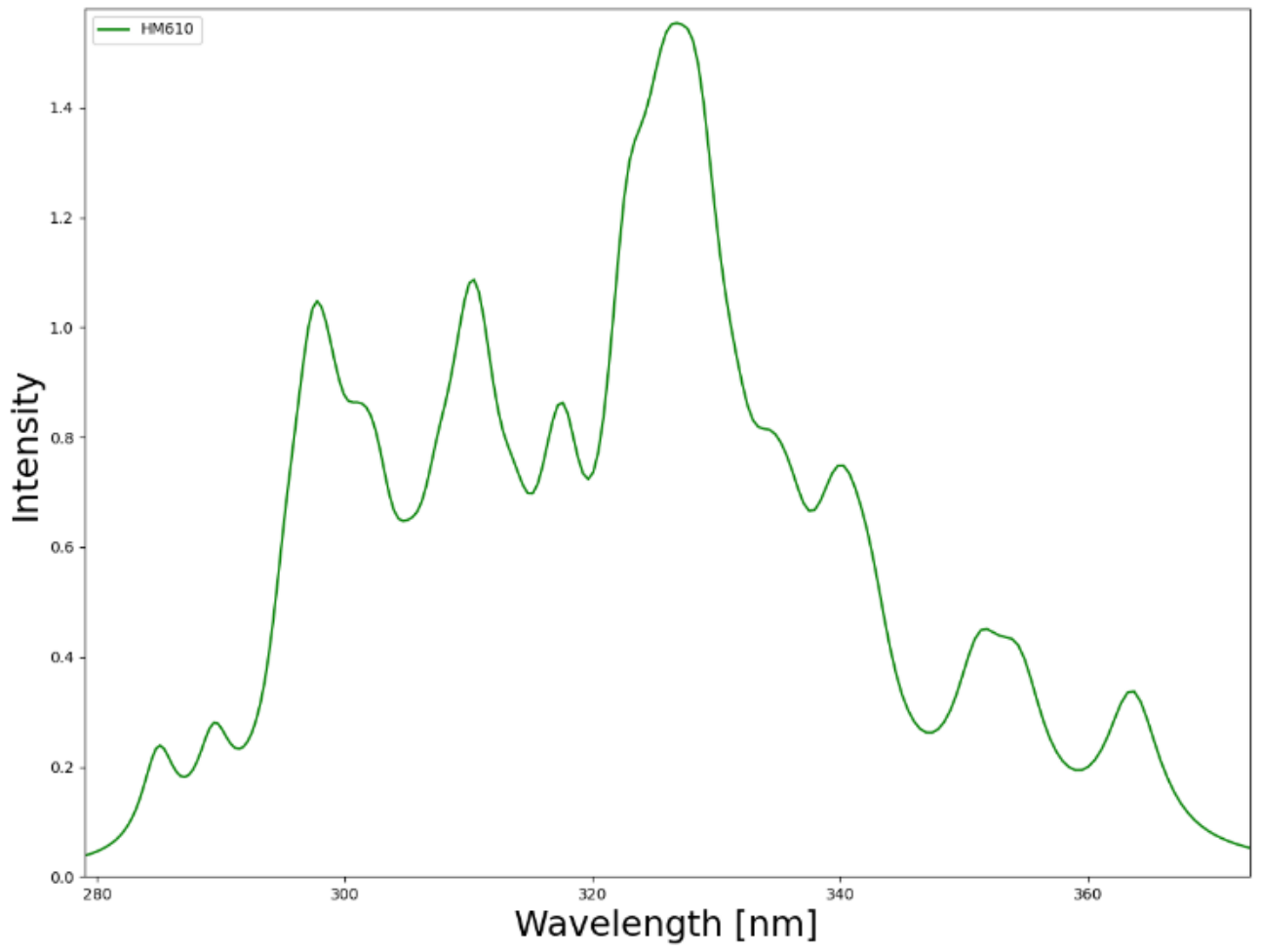

*** $\lambda_{max}$ of HM610 for computed and simulated absorption almost matches i.e 329 nm**

**IR spectrum of HM610**

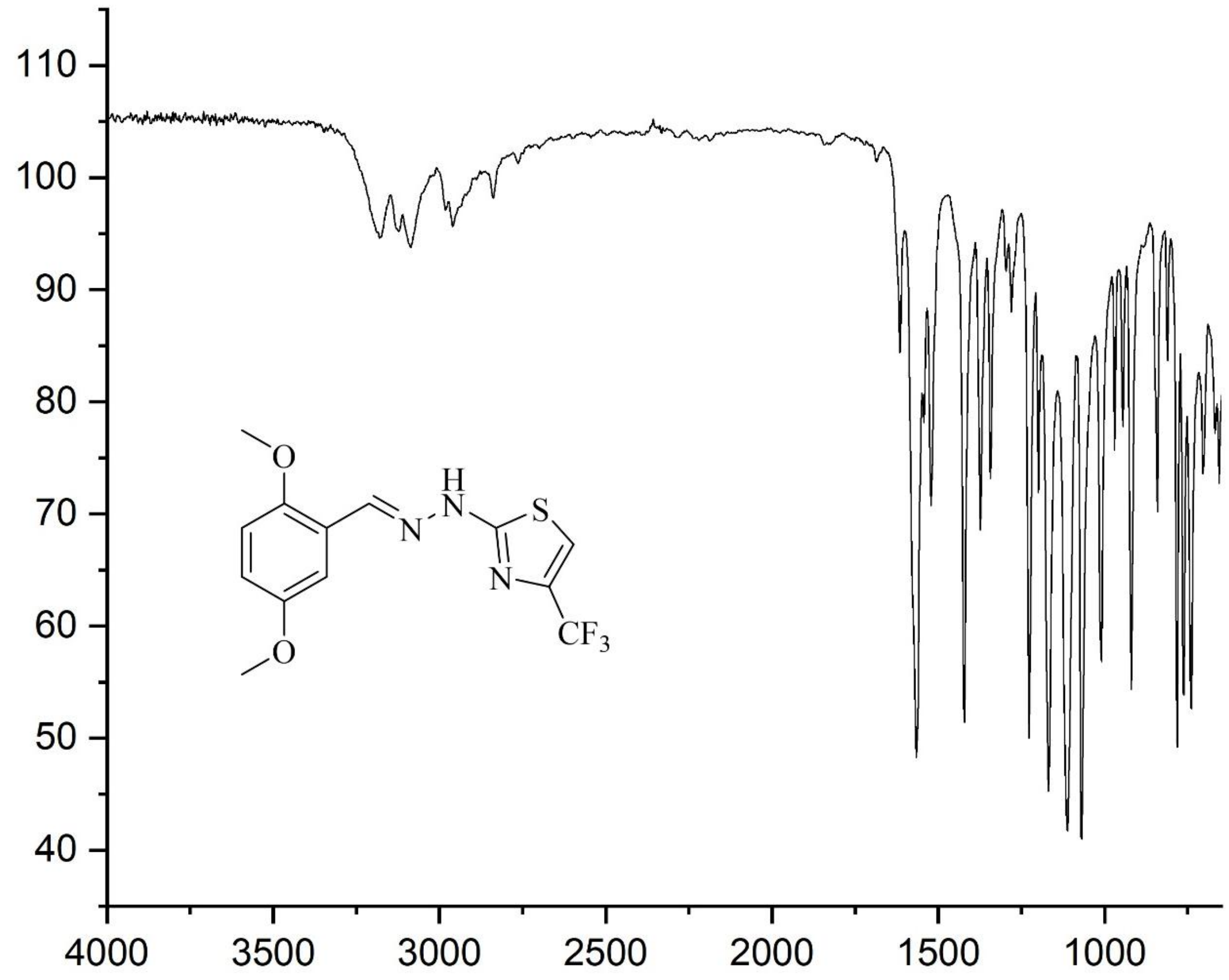

## $^{1}$H-NMR spectrum of HM610

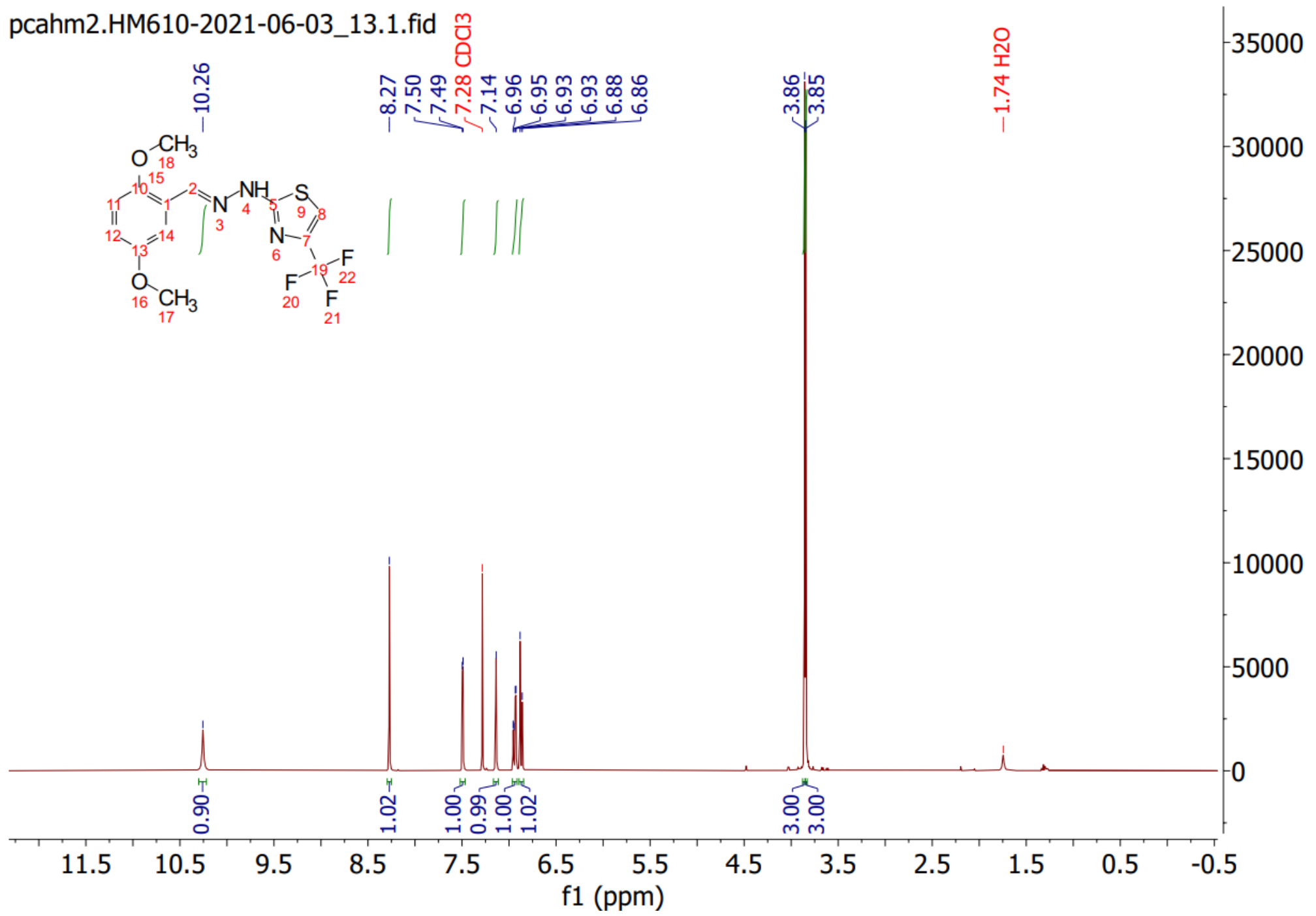

## $^{19}$F-NMR spectrum of HM610

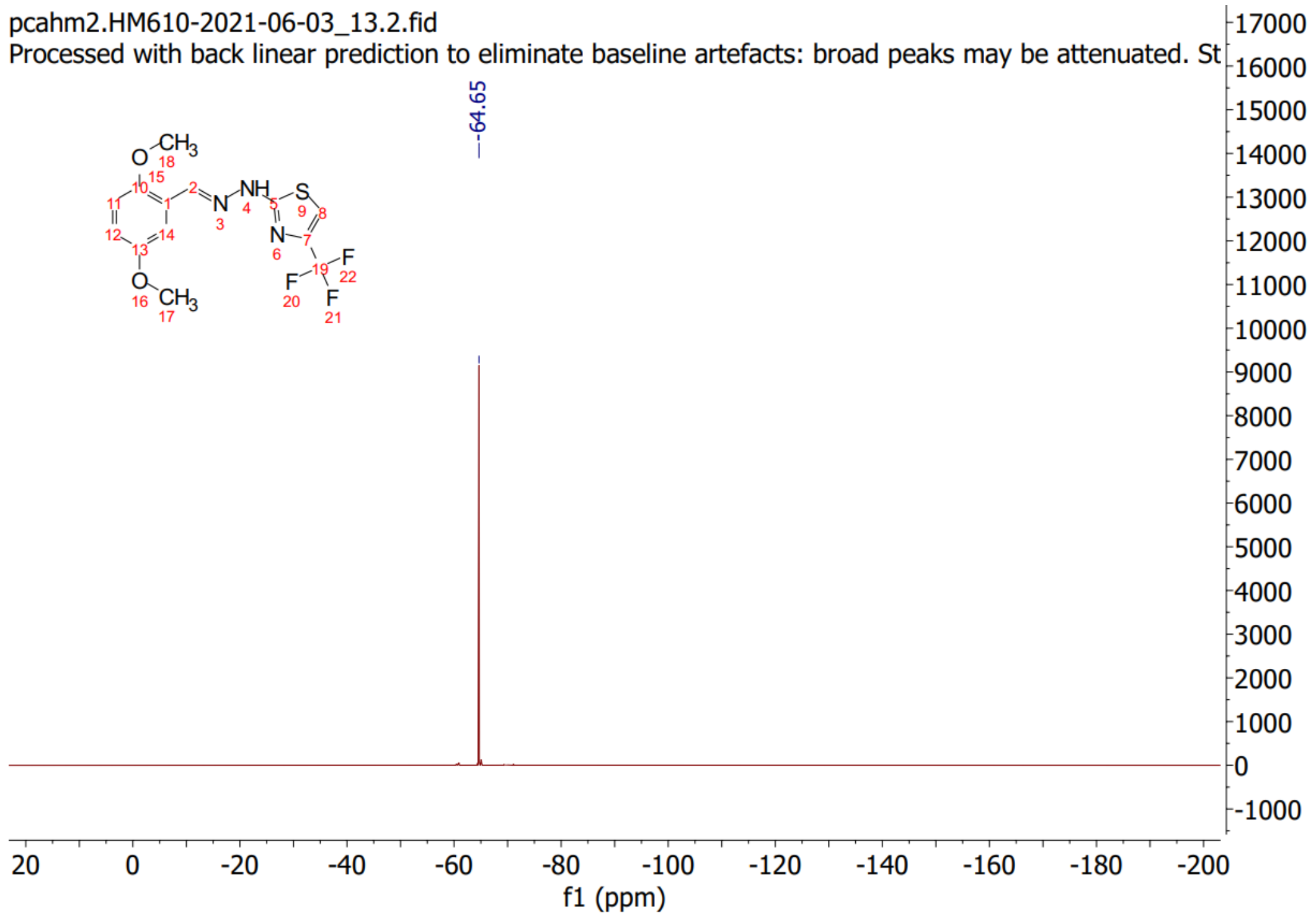

## $^{13}$C-NMR spectrum of HM610

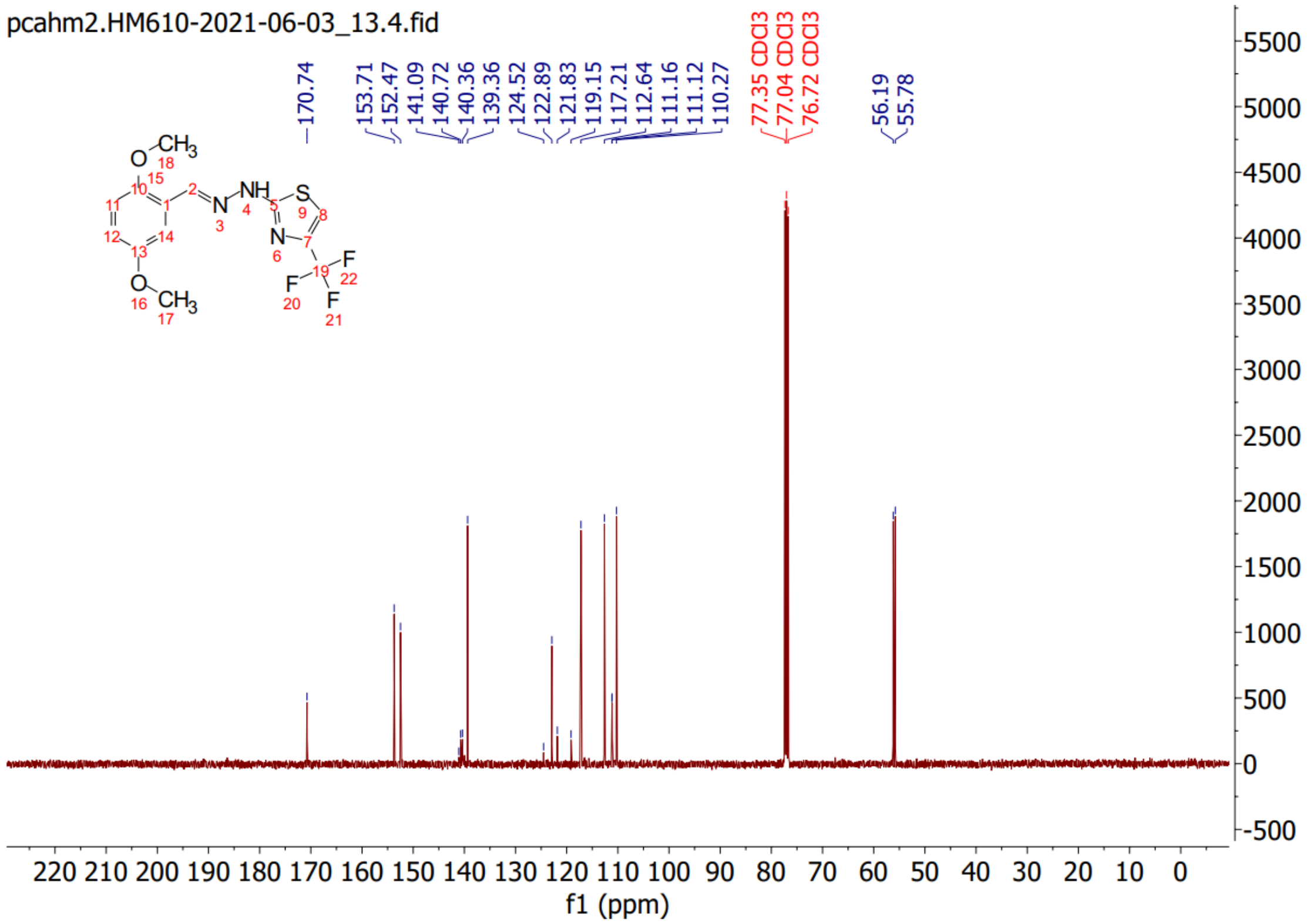

## Mass spectrum of HM610

**Mass Spectrometry**
**Analytical Services**
**School of Chemistry**

**The University of Nottingham**

| | | | |
|---|---|---|---|
| Sample-ID | h_meh_HM610 | Lab | C13 |
| Submitter | Hasnain Mehmood (pcahm2) | Supervisor | Simon Woodward |
| Analysis Name | h_meh_HM610_616969_56_01_32724.d | Acquisition Date | 6/9/2021 9:00:15 AM |
| Ionisation Mode | ESI Positive | Instrument | Bruker MicroTOF |

+MS, 0.7-0.9min #41-52

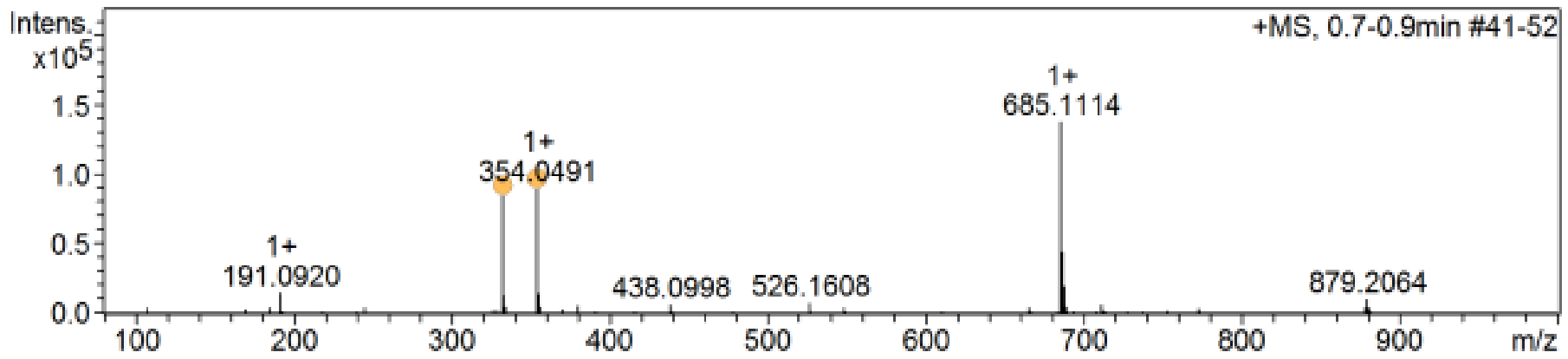

| # | m/z | I % |
|---|---|---|
| 1 | 107.0364 | 3.3 |
| 2 | 191.0920 | 11.1 |
| 3 | 245.1358 | 3.3 |
| 4 | 332.0674 | 61.5 |
| 5 | 333.0700 | 9.4 |
| 6 | 334.0654 | 3.6 |
| 7 | 354.0491 | 65.4 |
| 8 | 355.0522 | 10.7 |
| 9 | 356.0477 | 3.8 |
| 10 | 380.0859 | 4.0 |
| 11 | 438.0998 | 4.4 |
| 12 | 526.1608 | 5.7 |
| 13 | 685.1114 | 100.0 |
| 14 | 686.1142 | 32.6 |
| 15 | 687.1116 | 14.5 |
| 16 | 688.1119 | 3.5 |
| 17 | 711.1485 | 4.7 |
| 18 | 878.1110 | 3.2 |
| 19 | 879.2064 | 7.2 |
| 20 | 880.2092 | 3.2 |

**Generate Molecular Formula Parameters**

| Charge | Tolerance | sigma limit | H/C Ratio | Electron Conf. | Nitrogen Rule | Chrom.BackGround | Calibration |
|---|---|---|---|---|---|---|---|
| +1 | 6 ppm | 0.08 | 3 - 0 | both | false | false | TRUE |

**Expected Formula** C13 H12 F3 N3 O2 S1 **Adduct(s):** H, Na, NH4, C3H5N2, radical

| # | meas. m/z | theo. m/z | \|Err\|[ppm] | Sigma | Formula | Adduct | Adduct Mass |
|---|---|---|---|---|---|---|---|
| 1 | 332.0674 | 332.0675 | 0.40 | 0.0044 | C13H13F3N3O2S | M+H | 1.0078 |
| 1 | 354.0491 | 354.0495 | 0.90 | 0.0018 | C13H12F3N3NaO2S | M+Na | 22.9898 |

Note: Sigma fits < 0.05 indicates high probability of correct MF

**Table S1:** Vertical excitation energies of most important excited states computed with four different electronic structure methods at $S_0$ optimized geometry.

| States | CC(2) | ADC(2) | M062X | $\omega$B97 |
|---|---|---|---|---|
| $(\pi;\pi^*)^1$ | 3.79 (0.50) | 3.75 (0.46) | 3.96 (0.66) | 4.17 (0.61) |
| $(\pi;\sigma^*)^1$ | 5.44 (0.02) | 5.47 (0.00) | 5.43 (0.00) | 5.85 (0.00) |
| $(\sigma;\pi^*)^1$ | 5.75 (0.00) | 5.70 (0.00) | 5.52 (0.00) | 5.87 (0.00) |
| $(\pi;\pi^*)^3$ | 2.99 (0.00) | 2.99 (0.00) | 2.80 (0.00) | 2.40 (0.00) |
| $(\pi;\sigma^*)^3$ | 5.28 (0.00) | 5.32 (0.00) | 5.25 (0.00) | 5.51 (0.00) |
| $(\sigma;\pi^*)^3$ | 5.21 (0.00) | 5.21 (0.00) | 4.90 (0.00) | 5.10 (0.00) |

**Table S2:** Vertical excitation energies of most important excited states computed with four different electronic structure methods at $S_0$ optimized geometry.

| States | CC(2) | ADC(2) | $\omega$B97X | $\omega$B97X-D |
|---|---|---|---|---|
| $(\pi;\pi^*)^1$ | 3.79 (0.50) | 3.75 (0.46) | 4.09 (0.62) | 3.93 (0.61) |
| $(\pi;\sigma^*)^1$ | 5.44 (0.02) | 5.47 (0.00) | 5.70 (0.00) | 5.48 (0.00) |
| $(\sigma;\pi^*)^1$ | 5.75 (0.00) | 5.70 (0.00) | 5.79 (0.00) | 5.65 (0.12) |
| $(\pi;\pi^*)^3$ | 2.99 (0.00) | 2.99 (0.00) | 2.44 (0.00) | 2.50 (0.00) |
| $(\pi;\sigma^*)^3$ | 5.28 (0.00) | 5.32 (0.00) | 5.40 (0.00) | 5.22 (0.00) |
| $(\sigma;\pi^*)^3$ | 5.21 (0.00) | 5.21 (0.00) | 5.01 (0.00) | 4.92 (0.00) |

**Table S3:** Vertical excitation energies of most important excited states computed with four different electronic structure methods at $S_0$ optimized geometry.

| States | CC(2) | ADC(2) | CAM-B3LYP | BHLYP |
|---|---|---|---|---|
| $(\pi;\pi^*)^1$ | 3.79 (0.50) | 3.75 (0.46) | 3.91 (0.61) | 4.06 (0.68) |
| $(\pi;\sigma^*)^1$ | 5.44 (0.02) | 5.47 (0.00) | 5.45 (0.00) | 5.60 (0.00) |
| $(\sigma;\pi^*)^1$ | 5.75 (0.00) | 5.70 (0.00) | 5.67 (0.00) | 5.90 (0.00) |
| $(\pi;\pi^*)^3$ | 2.99 (0.00) | 2.99 (0.00) | 2.38 (0.00) | 2.06 (0.00) |
| $(\pi;\sigma^*)^3$ | 5.28 (0.00) | 5.32 (0.00) | 5.17 (0.00) | 5.27 (0.00) |
| $(\sigma;\pi^*)^3$ | 5.21 (0.00) | 5.21 (0.00) | 4.87 (0.00) | 4.98 (0.00) |

**Table S4:** Difference of vertical excitation energizes of HM610 between wave function methods (CC2 and ADC(2)) sf-XC-S-TD-DFT method (M062X and $\omega$B97 functionals)

| ADC(2)-CC2 | M062X-CC2 | $\omega$B97-CC2 | M062X-ADC(2) | $\omega$B97-ADC(2) |
|---|---|---|---|---|
| -0.04 | 0.17 | 0.38 | 0.21 | 0.42 |
| -0.05 | 0.19 | 0.41 | 0.24 | 0.46 |
| -0.03 | 0.02 | 0.19 | 0.05 | 0.22 |
| 0.03 | 0.02 | 0.41 | -0.01 | 0.38 |
| -0.05 | -0.06 | 0.29 | -0.01 | 0.34 |
| -0.05 | -0.15 | 0.23 | -0.1 | 0.28 |
| -0.01 | 0.08 | 0.52 | 0.09 | 0.53 |
| 0 | -0.19 | -0.59 | -0.19 | -0.59 |
| 0 | -0.09 | -0.23 | -0.09 | -0.23 |
| 0 | -0.14 | -0.54 | -0.14 | -0.54 |
| 0 | -0.11 | -0.55 | -0.11 | -0.55 |
| 0 | -0.1 | -0.2 | -0.1 | -0.2 |
| 0 | -0.31 | -0.22 | -0.31 | -0.22 |
| 0 | -0.15 | -0.18 | -0.15 | -0.18 |
| 0 | -0.07 | 0.19 | -0.07 | 0.19 |
| 0 | -0.25 | -0.04 | -0.25 | -0.04 |
| 0 | -0.26 | -0.08 | -0.26 | -0.08 |

**Table S5:** Difference of vertical excitation energizes of HM610 between wave function methods (CC2 and ADC(2)) sf-XC-S-TD-DFT method ($\omega$B97X and $\omega$B97X-D functionals)

| ADC(2)-CC2 | $\omega$B97X-CC2 | $\omega$B97X-D-CC2 | $\omega$B97X -ADC(2) | $\omega$B97X-D-ADC(2) |
|---|---|---|---|---|
| -0.04 | 0.3 | 0.14 | 0.34 | 0.18 |
| -0.05 | 0.33 | 0.16 | 0.38 | 0.21 |
| -0.03 | 0.14 | 0.03 | 0.17 | 0.06 |
| 0.03 | 0.26 | 0.04 | 0.23 | 0.01 |
| -0.05 | 0.21 | 0.07 | 0.26 | 0.12 |
| -0.05 | 0.14 | -0.09 | 0.19 | -0.04 |
| -0.01 | 0.38 | 0.11 | 0.39 | 0.12 |
| 0 | -0.55 | -0.49 | -0.55 | -0.49 |
| 0 | -0.23 | -0.28 | -0.23 | -0.28 |
| 0 | -0.54 | -0.49 | -0.54 | -0.49 |
| 0 | -0.5 | -0.43 | -0.5 | -0.43 |
| 0 | -0.24 | -0.29 | -0.24 | -0.29 |
| 0 | -0.26 | -0.3 | -0.26 | -0.3 |
| 0 | -0.27 | -0.36 | -0.27 | -0.36 |
| 0 | 0.08 | -0.1 | 0.08 | -0.1 |
| 0 | -0.12 | -0.26 | -0.12 | -0.26 |
| 0 | -0.17 | -0.31 | -0.17 | -0.31 |

**Table S6:** Difference of vertical excitation energizes of HM610 between wave function methods (CC2 and ADC(2)) sf-XC-S-TD-DFT method (CAM-B3LYP and BHLYP functionals)

| ADC(2)-CC2 | CAM-B3LYP-CC2 | BHLYP-CC2 | CAM-B3LYP -ADC(2) | BHLYP-ADC(2) |
|---|---|---|---|---|
| -0.04 | 0.12 | 0.27 | 0.16 | 0.31 |
| -0.05 | 0.12 | 0.28 | 0.17 | 0.33 |
| -0.03 | 0.01 | 0.09 | 0.04 | 0.12 |
| 0.03 | 0.01 | 0.16 | -0.02 | 0.13 |
| -0.05 | -0.03 | 0.07 | 0.02 | 0.12 |
| -0.05 | -0.08 | 0.15 | -0.03 | 0.2 |
| -0.01 | 0.03 | 0.19 | 0.04 | 0.2 |
| 0 | -0.61 | -0.93 | -0.61 | -0.93 |
| 0 | -0.33 | -0.46 | -0.33 | -0.46 |
| 0 | -0.6 | -0.64 | -0.6 | -0.64 |
| 0 | -0.56 | -0.75 | -0.56 | -0.75 |
| 0 | -0.35 | -0.32 | -0.35 | -0.32 |
| 0 | -0.38 | -0.48 | -0.38 | -0.48 |
| 0 | -0.41 | -0.3 | -0.41 | -0.3 |
| 0 | -0.15 | -0.05 | -0.15 | -0.05 |
| 0 | -0.36 | -0.3 | -0.36 | -0.3 |
| 0 | -0.4 | -.0.27 | -0.4 | -0.27 |

**Table S7:** Difference of vertical excitation energizes of HM610 computed in gas phase and DMSO solvent using PCM model for M062X and CAM-B3LYP functionals

| States | M062X | M062X (DMSO) | Difference | CAM-B3LYP | CAM-B3LYP (DMSO) | Difference |
|---|---|---|---|---|---|---|
| $S_1$ | 3.96 (0.66) | 3.94 (0.62) | -0.02 | 3.91 (0.61) | 3.88 (0.57) | -0.03 |
| $S_2$ | 4.40 (0.09) | 4.40 (0.14) | 0.00 | 4.33 (0.14) | 4.34 (0.18) | 0.01 |
| $S_3$ | 5.43 (0.00) | 5.47 (0.11) | 0.04 | 5.42 (0.07) | 5.39 (0.09) | -0.03 |
| $S_4$ | 5.46 (0.11) | 5.52 (0.01) | 0.06 | 5.45 (0.00) | 5.49 (0.00) | 0.04 |
| $S_5$ | 5.52 (0.00) | 5.55 (0.00) | 0.03 | 5.55 (0.08) | 5.54 (0.06) | -0.01 |
| $S_6$ | 5.60 (0.00) | 5.60 (0.03) | 0.00 | 5.67 (0.00) | 5.67 (0.00) | 0.00 |
| $S_7$ | 5.95 (0.04) | 5.96 (0.20) | 0.01 | 5.90 (0.20) | 5.92 (0.17) | 0.02 |
| $T_1$ | 2.80 (0.00) | 2.80 (0.00) | 0.00 | 2.38 (0.00) | 2.39 (0.00) | 0.01 |
| $T_2$ | 3.53 (0.00) | 3.50 (0.00) | -0.03 | 3.29 (0.00) | 3.26 (0.00) | -0.03 |
| $T_3$ | 3.86 (0.00) | 3.85 (0.00) | -0.01 | 3.40 (0.00) | 3.39 (0.00) | -0.01 |
| $T_4$ | 4.31 (0.00) | 4.30 (0.00) | -0.01 | 3.86 (0.00) | 3.87 (0.00) | 0.01 |
| $T_5$ | 4.71 (0.00) | 4.72 (0.00) | 0.01 | 4.46 (0.00) | 4.46 (0.00) | 0.00 |
| $T_6$ | 4.90 (0.00) | 4.92 (0.00) | 0.02 | 4.83 (0.00) | 4.76 (0.00) | -0.07 |
| $T_7$ | 5.13 (0.00) | 5.05 (0.00) | -0.08 | 4.87 (0.00) | 4.90 (0.00) | 0.03 |
| $T_8$ | 5.25 (0.00) | 5.29 (0.00) | 0.04 | 5.17 (0.00) | 5.21 (0.00) | 0.04 |
| $T_9$ | 5.47 (0.00) | 5.52 (0.00) | 0.05 | 5.36 (0.00) | 5.41 (0.00) | 0.05 |
| $T_{10}$ | 5.63 (0.00) | 5.68 (0.00) | 0.05 | 5.49 (0.00) | 5.54 (0.00) | 0.05 |

**Table S8:** Character of vertical excitation energizes of HM610 computed in gas phase and DMSO solvent using PCM model for M062X and CAM-B3LYP functionals.

| States | M062X | M062X (DMSO) | CAM-B3LYP | CAM-B3LYP (DMSO) |
|---|---|---|---|---|
| $S_1$ | $(\pi;\pi^*)^1$ | $(\pi;\pi^*)^1$ | $(\pi;\pi^*)^1$ | $(\pi;\pi^*)^1$ |
| $S_2$ | $(\pi;\pi^*)^1$ | $(\pi;\pi^*)^1$ | $(\pi;\pi^*)^1$ | $(\pi;\pi^*)^1$ |
| $S_3$ | $(\pi;\sigma^*)^1$ | $(\pi;\pi^*)^1$ | $(\pi;\pi^*)^1$ | $(\pi;\pi^*)^1$ |
| $S_4$ | $(\pi;\pi^*)^1$ | $(\pi;\sigma^*)^1$ | $(\pi;\sigma^*)^1$ | $(\pi;\sigma^*)^1$ |
| $S_5$ | $(\sigma;\pi^*)^1$ | $(\sigma;\pi^*)^1$ | $(\pi;\pi^*)^1$ | $(\pi;\pi^*)^1$ |
| $S_6$ | $(\pi;\pi^*)^1$ | $(\pi;\pi^*)^1$ | $(\sigma;\pi^*)^1$ | $(\sigma;\pi^*)^1$ |
| $S_7$ | $(\pi;\sigma^*)^1$ | $(\pi;\pi^*)^1$ | $(\pi;\pi^*)^1$ | $(\pi;\pi^*)^1$ |
| $T_1$ | $(\pi;\pi^*)^3$ | $(\pi;\pi^*)^3$ | $(\pi;\pi^*)^3$ | $(\pi;\pi^*)^3$ |
| $T_2$ | $(\pi;\pi^*)^3$ | $(\pi;\pi^*)^3$ | $(\pi;\pi^*)^3$ | $(\pi;\pi^*)^3$ |
| $T_3$ | $(\pi;\pi^*)^3$ | $(\pi;\pi^*)^3$ | $(\pi;\pi^*)^3$ | $(\pi;\pi^*)^3$ |
| $T_4$ | $(\pi;\pi^*)^3$ | $(\pi;\pi^*)^3$ | $(\pi;\pi^*)^3$ | $(\pi;\pi^*)^3$ |
| $T_5$ | $(\pi;\pi^*)^3$ | $(\pi;\pi^*)^3$ | $(\pi;\pi^*)^3$ | $(\pi;\pi^*)^3$ |
| $T_6$ | $(\sigma;\pi^*)^3$ | $(\sigma;\pi^*)^3$ | $(\pi;\pi^*)^3$ | $(\pi;\pi^*)^3$ |
| $T_7$ | $(\pi;\pi^*)^3$ | $(\pi;\pi^*)^3$ | $(\sigma;\pi^*)^3$ | $(\sigma;\pi^*)^3$ |
| $T_8$ | $(\pi;\sigma^*)^3$ | $(\pi;\sigma^*)^3$ | $(\pi;\sigma^*)^3$ | $(\pi;\sigma^*)^3$ |
| $T_9$ | $(\pi;\pi^*)^3$ | $(\pi;\pi^*)^3$ | $(\pi;\pi^*)^3$ | $(\pi;\pi^*)^3$ |
| $T_{10}$ | $(\pi;\pi^*)^3$ | $(\pi;\pi^*)^3$ | $(\pi;\pi^*)^3$ | $(\pi;\pi^*)^3$ |

**Table S9:** Spin-orbit couplings ($cm^{-1}$) in **HM610** employing M062X (TDDFT) functional at $S_0$ optimized geometry with PCM model (DMSO)

| $\hat{H}_{DKH}$ | $\|T_1\rangle$ | $\|T_2\rangle$ | $\|T_3\rangle$ | $\|T_4\rangle$ | $\|T_5\rangle$ | $\|T_6\rangle$ | $\|T_7\rangle$ | $\|T_8\rangle$ | $\|T_9\rangle$ | $\|T_{10}\rangle$ |
|---|---|---|---|---|---|---|---|---|---|---|
| $\langle S_0\|$ | 1 | 1 | 0 | 1 | 3 | 31 | 5 | 102 | 2 | 1 |
| $\langle S_1\|$ | 0 | 0 | 0 | 1 | 1 | 7 | 1 | 15 | 1 | 0 |
| $\langle S_2\|$ | 0 | 0 | 9 | 1 | 1 | 3 | 1 | 18 | 1 | 0 |
| $\langle S_3\|$ | 3 | 3 | 7 | 2 | 1 | 4 | 2 | 37 | 2 | 1 |
| $\langle S_4\|$ | 15 | 14 | 37 | 8 | 5 | 1 | 11 | 7 | 3 | 8 |
| $\langle S_5\|$ | 9 | 2 | 7 | 2 | 2 | 0 | 2 | 1 | 12 | 6 |
| $\langle S_6\|$ | 1 | 1 | 2 | 0 | 1 | 7 | 0 | 3 | 1 | 1 |
| $\langle S_7\|$ | 1 | 1 | 1 | 0 | 1 | 7 | 0 | 3 | 0 | 0 |

**Table S10:** Spin-orbit couplings ($cm^{-1}$) in **HM610** employing CAM-B3LYP (TDDFT) functional at $S_0$ optimized geometry with PCM model (DMSO)

| $\hat{H}_{DKH}$ | $\|T_1\rangle$ | $\|T_2\rangle$ | $\|T_3\rangle$ | $\|T_4\rangle$ | $\|T_5\rangle$ | $\|T_6\rangle$ | $\|T_7\rangle$ | $\|T_8\rangle$ | $\|T_9\rangle$ | $\|T_{10}\rangle$ |
|---|---|---|---|---|---|---|---|---|---|---|
| $\langle S_0\|$ | 0 | 1 | 1 | 0 | 2 | 1 | 28 | 95 | 2 | 1 |
| $\langle S_1\|$ | 1 | 0 | 1 | 1 | 0 | 1 | 7 | 16 | 1 | 0 |
| $\langle S_2\|$ | 0 | 0 | 1 | 1 | 0 | 1 | 3 | 21 | 1 | 0 |
| $\langle S_3\|$ | 1 | 1 | 1 | 1 | 0 | 1 | 6 | 41 | 2 | 1 |
| $\langle S_4\|$ | 16 | 19 | 37 | 13 | 0 | 16 | 1 | 0 | 3 | 8 |
| $\langle S_5\|$ | 1 | 2 | 5 | 2 | 0 | 2 | 5 | 8 | 0 | 1 |
| $\langle S_6\|$ | 12 | 4 | 4 | 5 | 1 | 3 | 0 | 1 | 13 | 7 |
| $\langle S_7\|$ | 1 | 1 | 0 | 0 | 0 | 1 | 8 | 6 | 0 | 0 |

**Table S11:** Coupled states in HM610 computed at M062X/x2c-TZVPPall level of theory on top of $S_0$ optimized geometry

| Coupled states | Orbitals | Character | \|SOC\| |
|---|---|---|---|
| $S_0T_6$ | $(85; 79 \rightarrow 86)$ | $(\pi)^1;(\sigma;\pi^*)^3$ | 33 |
| $S_0T_8$ | $(85; 85 \rightarrow 89)$ | $(\pi)^1; (\pi;\sigma^*)^3$ | 98 |
| $S_3T_3$ | $(85 \rightarrow 89; 85 \rightarrow 87)$ | $(\pi;\sigma^*)^1; (\pi;\pi^*)^3$ | 36 |
| $S_4T_8$ | $(85 \rightarrow 87; 85 \rightarrow 89)$ | $(\pi;\pi^*)^1; (\pi;\sigma^*)^3$ | 38 |

**Table S12:** Coupled states in HM610 computed at *ω*B97/x2c-TZVPPall level of theory on top of $S_0$ optimized geometry

| Coupled states | Orbitals | Character | \|SOC\| |
|---|---|---|---|
| $S_0T_7$ | $(85; 79 \rightarrow 86)$ | $(\pi)^1;(\sigma;\pi^*)^3$ | 29 |
| $S_0T_8$ | $(85; 85 \rightarrow 89)$ | $(\pi)^1; (\pi;\sigma^*)^3$ | 110 |
| $S_3T_8$ | $(83 \rightarrow 86; 85 \rightarrow 89)$ | $(\pi;\pi^*)^1; (\pi;\sigma^*)^3$ | 37 |
| $S_4T_2$ | $(85 \rightarrow 89; 85 \rightarrow 87)$ | $(\pi;\sigma^*)^1;(\pi;\pi^*)^3$ | 34 |

**Table S13:** Coupled states in HM610 computed at $\omega$B97X/x2c-TZVPPall level of theory on top of $S_0$ optimized geometry

| Coupled states | Orbitals | Character | \|SOC\| |
|---|---|---|---|
| $S_0T_7$ | $(85; 79 \rightarrow 86)$ | $(\pi)^1;(\sigma;\pi^*)^3$ | 30 |
| $S_0T_8$ | $(85; 85 \rightarrow 89)$ | $(\pi)^1; (\pi;\sigma^*)^3$ | 101 |
| $S_3T_8$ | $(83 \rightarrow 86; 85 \rightarrow 89)$ | $(\pi;\pi^*)^1; (\pi;\sigma^*)^3$ | 38 |
| $S_4T_3$ | $(85 \rightarrow 89; 84 \rightarrow 87)$ | $(\pi;\sigma^*)^1;(\pi;\pi^*)^3$ | 30 |

**Table S14:** Coupled states in HM610 computed at $\omega$B97X-D/x2c-TZVPPall level of theory on top of $S_0$ optimized geometry

| Coupled states | Orbitals | Character | \|SOC\| |
|---|---|---|---|
| $S_0T_7$ | $(85; 79 \rightarrow 86)$ | $(\pi)^1;(\sigma;\pi^*)^3$ | 27 |
| $S_0T_8$ | $(85; 85 \rightarrow 89)$ | $(\pi)^1; (\pi;\sigma^*)^3$ | 88 |
| $S_3T_8$ | $(83 \rightarrow 86; 85 \rightarrow 89)$ | $(\pi;\pi^*)^1; (\pi;\sigma^*)^3$ | 39 |
| $S_4T_3$ | $(85 \rightarrow 89; 84 \rightarrow 87)$ | $(\pi;\sigma^*)^1;(\pi;\pi^*)^3$ | 38 |

**Table S15:** Coupled states in HM610 computed at CAM-B3LYP/x2c-TZVPPall level of theory on top of $S_0$ optimized geometry

| Coupled states | Orbitals | Character | \|SOC\| |
|---|---|---|---|
| $S_0T_7$ | $(85; 79 \rightarrow 86)$ | $(\pi)^1;(\sigma;\pi^*)^3$ | 30 |
| $S_0T_8$ | $(85; 85 \rightarrow 89)$ | $(\pi)^1; (\pi;\sigma^*)^3$ | 91 |
| $S_3T_8$ | $(85 \rightarrow 87; 85 \rightarrow 89)$ | $(\pi;\pi^*)^1; (\pi;\sigma^*)^3$ | 40 |
| $S_4T_3$ | $(85 \rightarrow 89; 84 \rightarrow 87)$ | $(\pi;\sigma^*)^1;(\pi;\pi^*)^3$ | 34 |

**Table S16:** Coupled states in HM610 computed at BHLYP/x2c-TZVPPall level of theory on top of $S_0$ optimized geometry

| Coupled states | Orbitals | Character | \|SOC\| |
|---|---|---|---|
| $S_0T_7$ | $(85; 79 \rightarrow 86)$ | $(\pi)^1;(\sigma;\pi^*)^3$ | 31 |
| $S_0T_8$ | $(85; 85 \rightarrow 89)$ | $(\pi)^1; (\pi;\sigma^*)^3$ | 96 |
| $S_3T_8$ | $(83 \rightarrow 86; 85 \rightarrow 89)$ | $(\pi;\pi^*)^1; (\pi;\sigma^*)^3$ | 36 |
| $S_4T_2$ | $(85 \rightarrow 89; 85 \rightarrow 87)$ | $(\pi;\sigma^*)^1;(\pi;\pi^*)^3$ | 34 |

**Table S17:** Coupled states in HM610 computed at M062X/x2c-TZVPPall level of theory on top of $S_0$ optimized geometry employing PCM model (DMSO)

| Coupled states | Orbitals | Character | \|SOC\| |
|---|---|---|---|
| $S_0T_6$ | $(85; 79 \rightarrow 86)$ | $(\pi)^1;(\sigma;\pi^*)^3$ | 31 |
| $S_0T_8$ | $(85; 85 \rightarrow 89)$ | $(\pi)^1; (\pi;\sigma^*)^3$ | 102 |
| $S_3T_3$ | $(85 \rightarrow 87; 84 \rightarrow 87)$ | $(\pi;\pi^*)^1; (\pi;\pi^*)^3$ | 37 |
| $S_4T_3$ | $(85 \rightarrow 89; 84 \rightarrow 87)$ | $(\pi;\sigma^*)^1; (\pi;\pi^*)^3$ | 38 |

**Table S18:** Coupled states in HM610 computed at CAM-B3LYP/x2c-TZVPPall level of theory on top of $S_0$ optimized geometry employing PCM model (DMO)

| Coupled states | Orbitals | Character | \|SOC\| |
|---|---|---|---|
| $S_0T_7$ | $(85; 79 \rightarrow 86)$ | $(\pi)^1;(\sigma;\pi^*)^3$ | 28 |
| $S_0T_8$ | $(85; 85 \rightarrow 89)$ | $(\pi)^1; (\pi;\sigma^*)^3$ | 95 |
| $S_3T_8$ | $(85 \rightarrow 87; 85 \rightarrow 89)$ | $(\pi;\pi^*)^1; (\pi;\sigma^*)^3$ | 41 |
| $S_4T_3$ | $(85 \rightarrow 89; 84 \rightarrow 87)$ | $(\pi;\sigma^*)^1;(\pi;\pi^*)^3$ | 37 |

**Table S19:** Frontier Molecular Orbitals of HM610 ranging from HOMO-7 to LUMO+3 computed with M062X and *ω*B97 functionals.

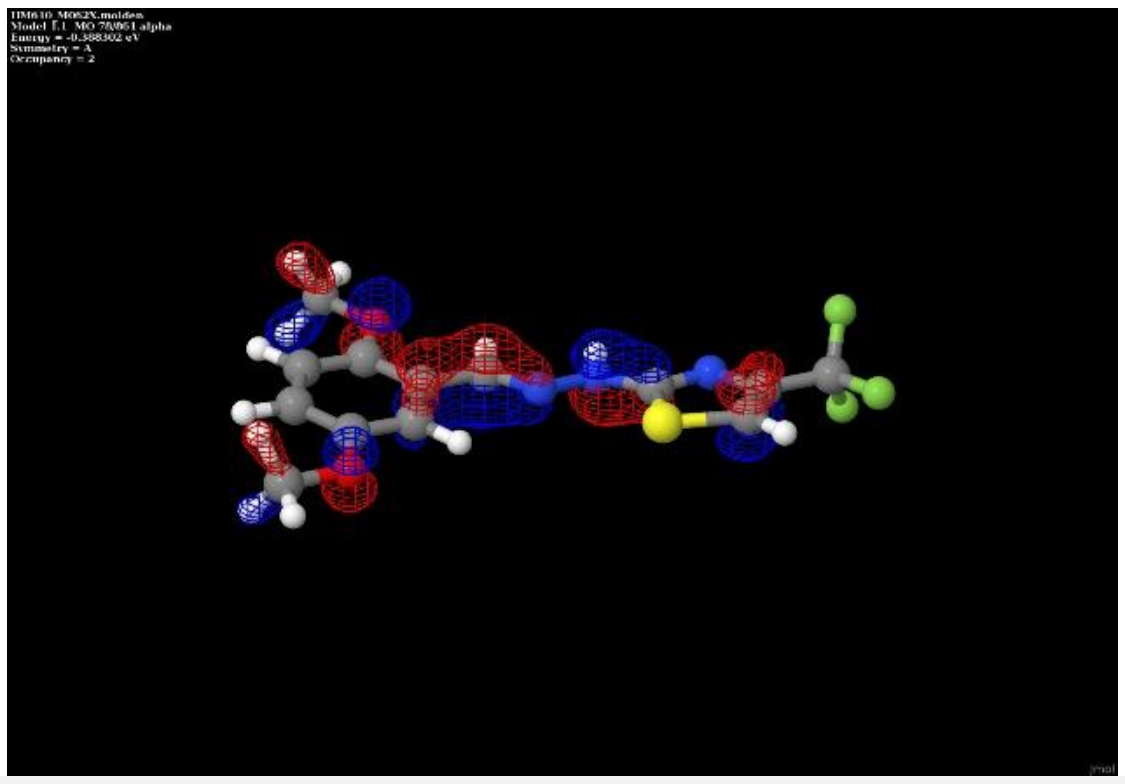

**HOMO-7 (M062X) (orbital:78)**

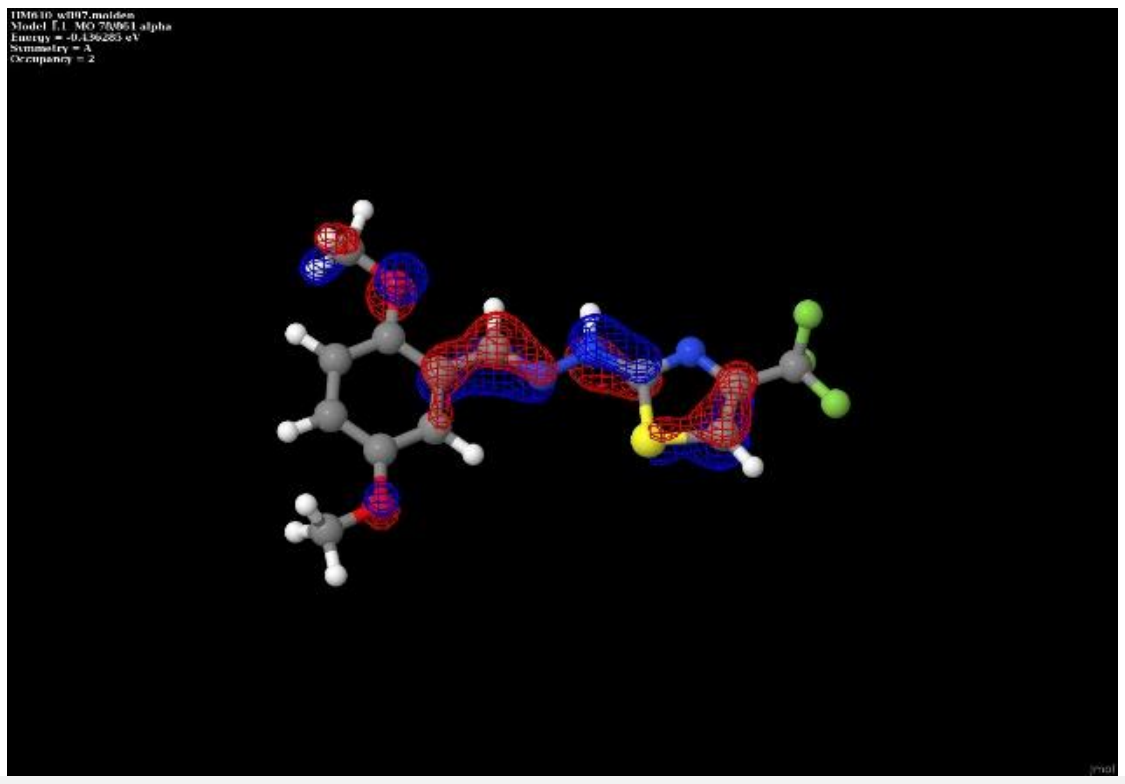

**HOMO-7 (*ω*B97) (orbital:78)**

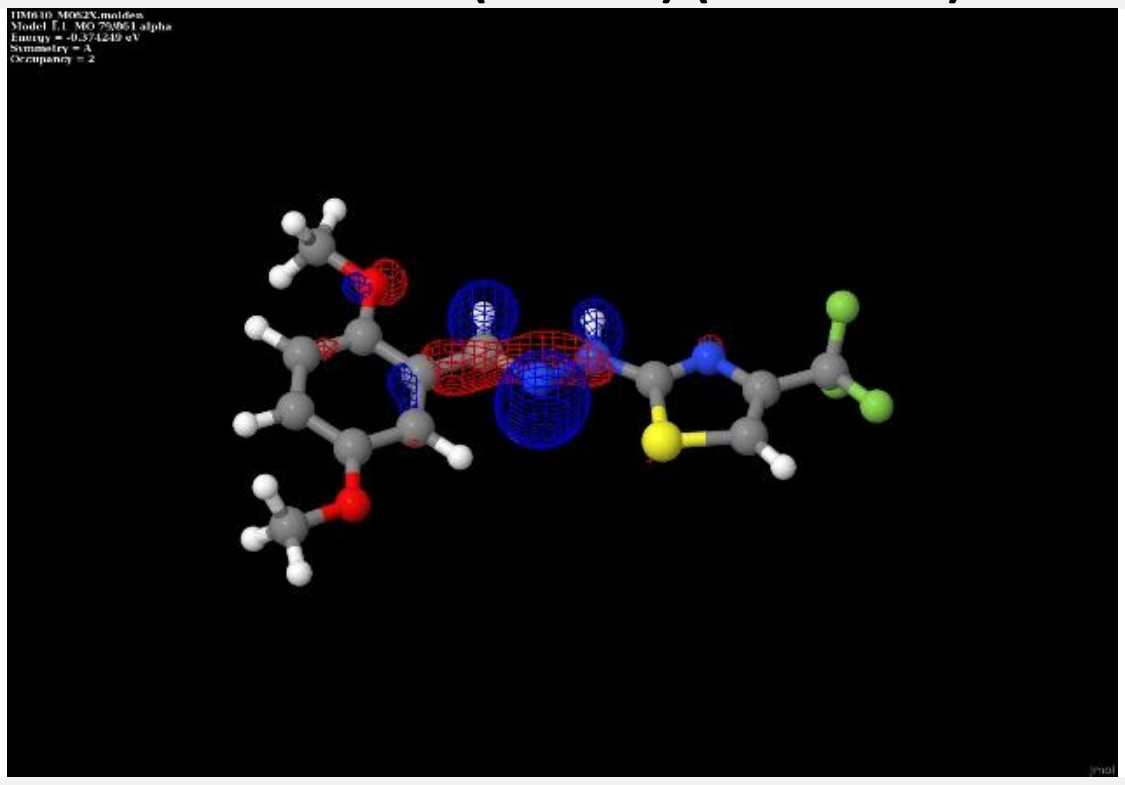

**HOMO-6 (M062X) (orbital:79)**

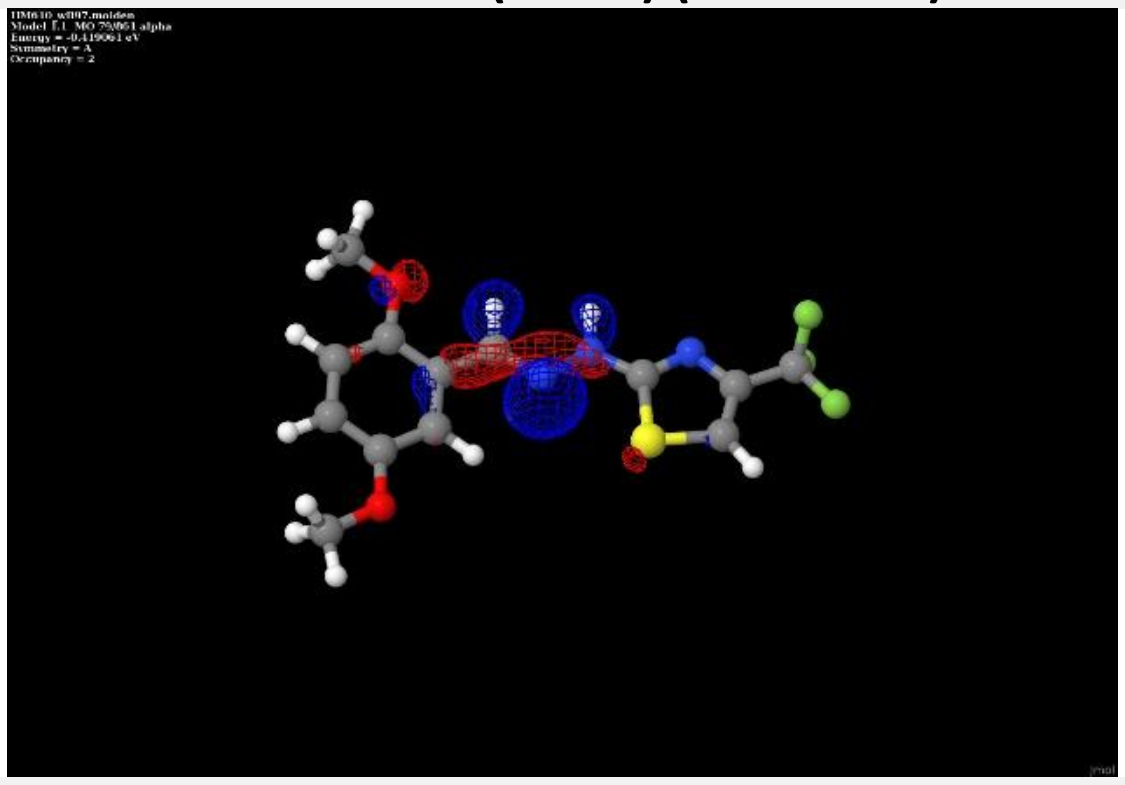

**HOMO-6 (*ω*B97) (orbital:79)**

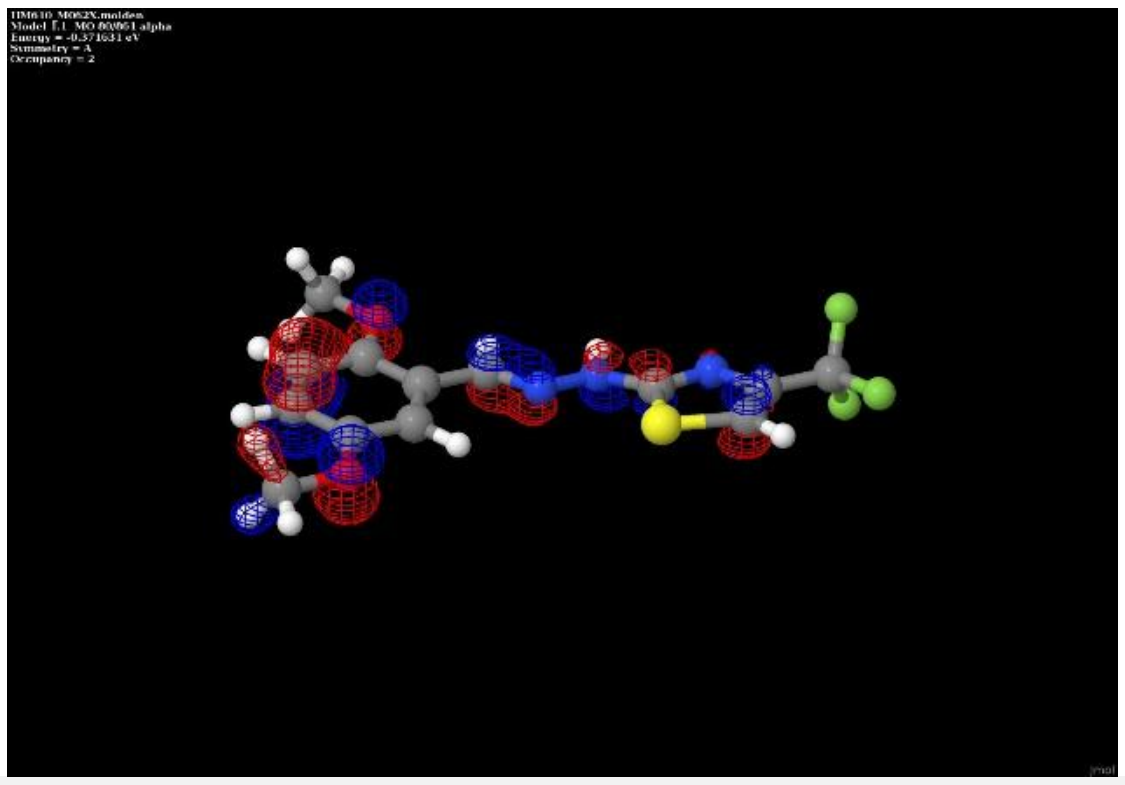

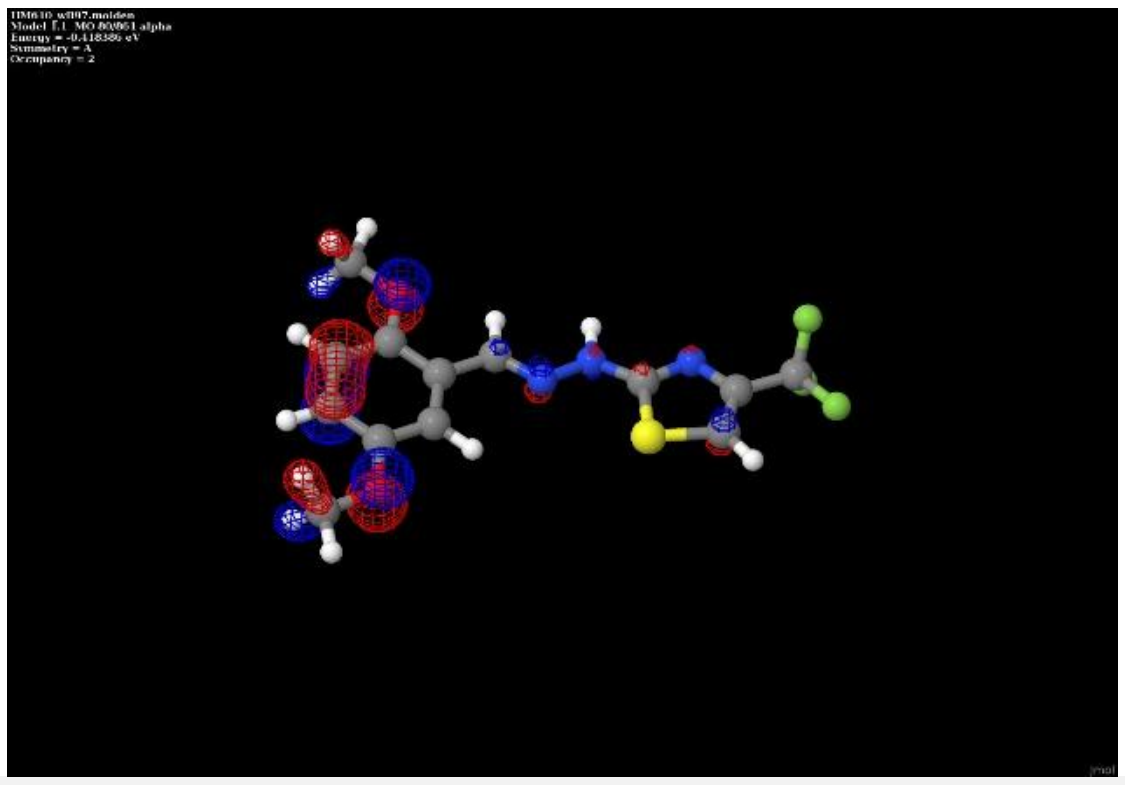

**HOMO-5 (M062X) (orbital:80)** | **HOMO-5 ($\omega$B97) (orbital:80)**

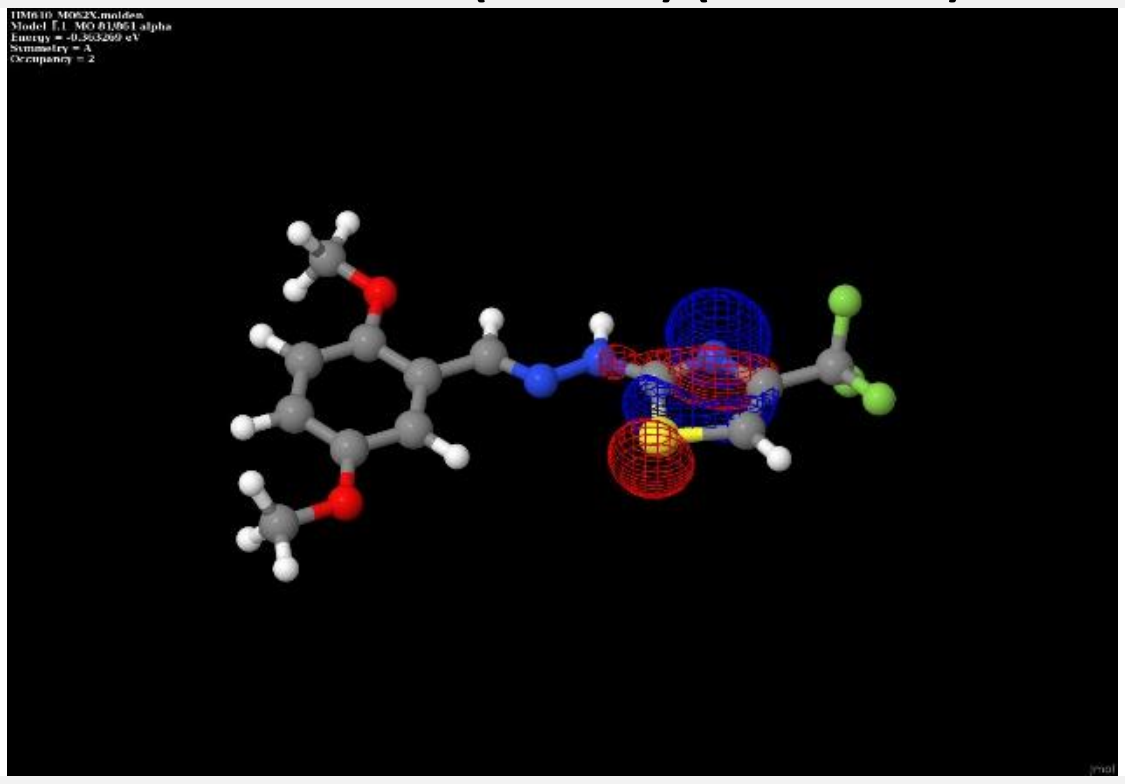

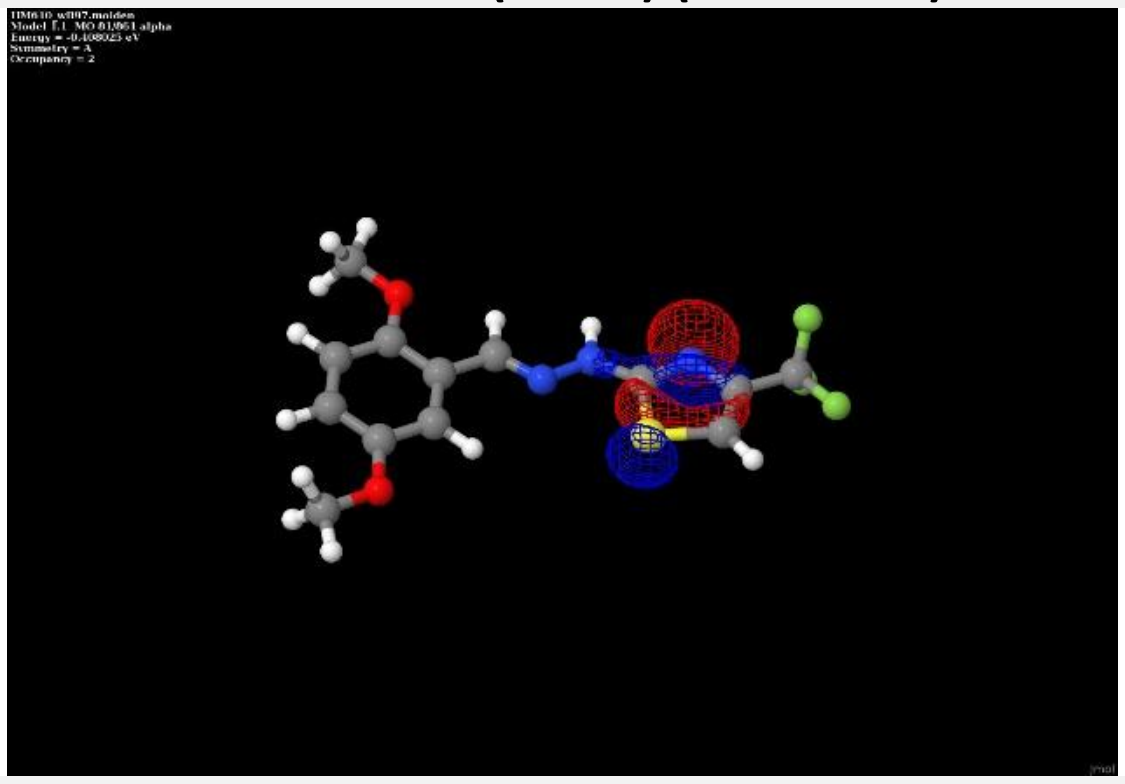

**HOMO-4 (M062X) (orbital:81)** | **HOMO-4 ($\omega$B97) (orbital:81)**

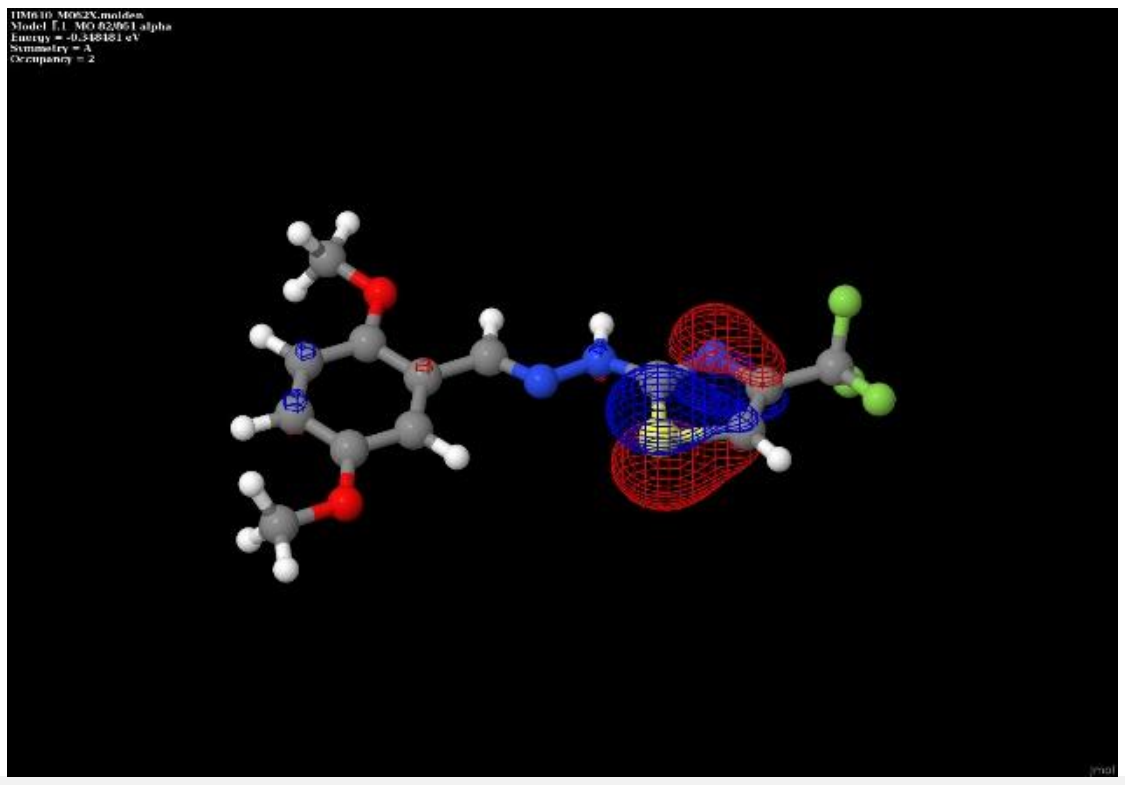

HOMO-3 (M062X) (orbital:82)

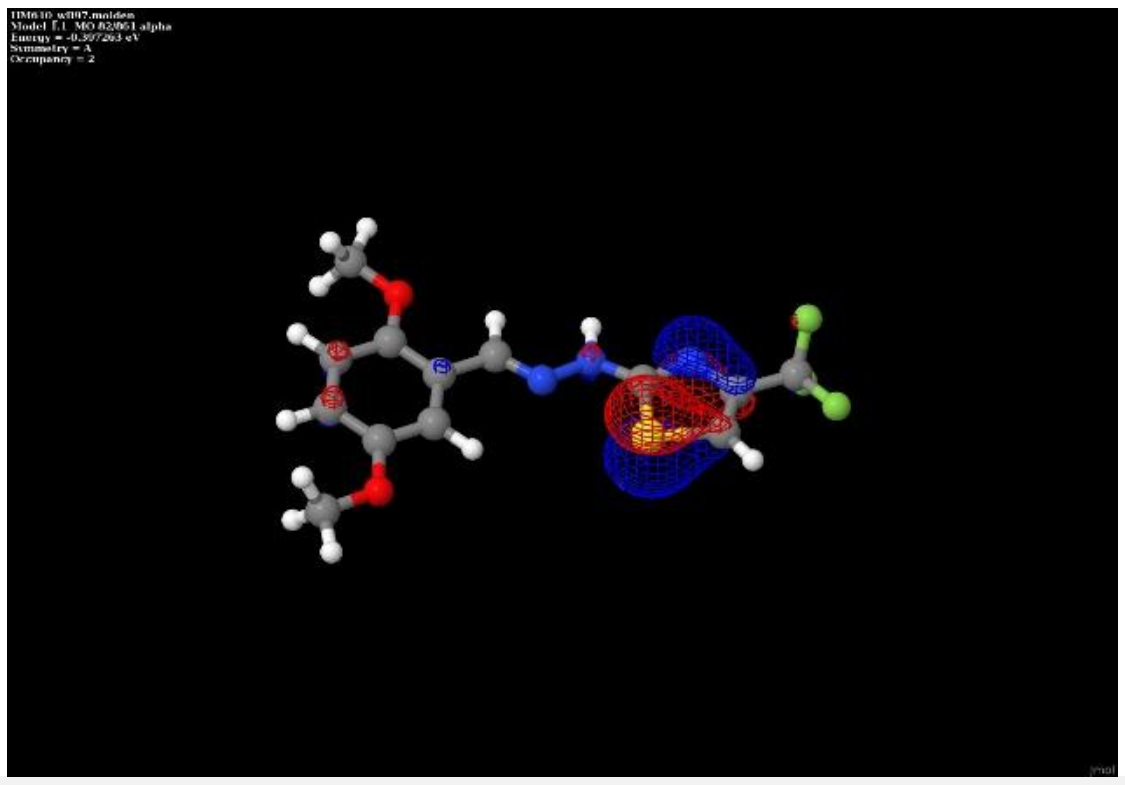

HOMO-3 ($\omega$B97) (orbital:82)

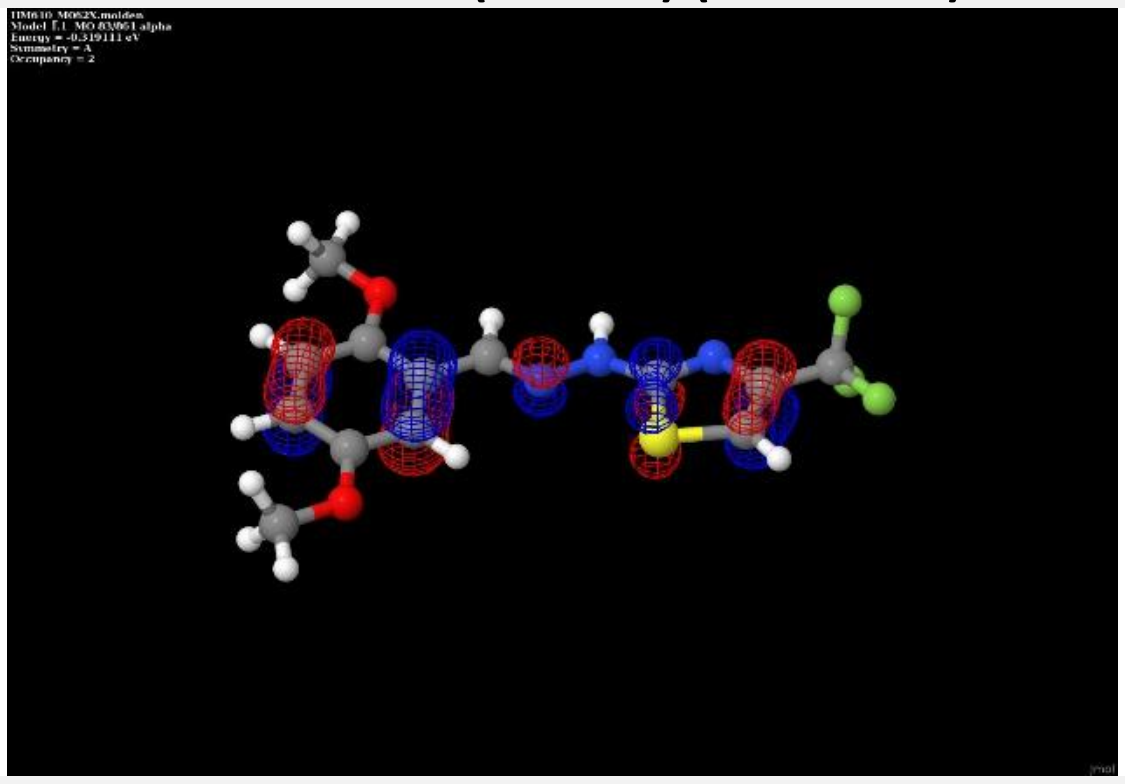

HOMO-2 (M062X) (orbital:83)

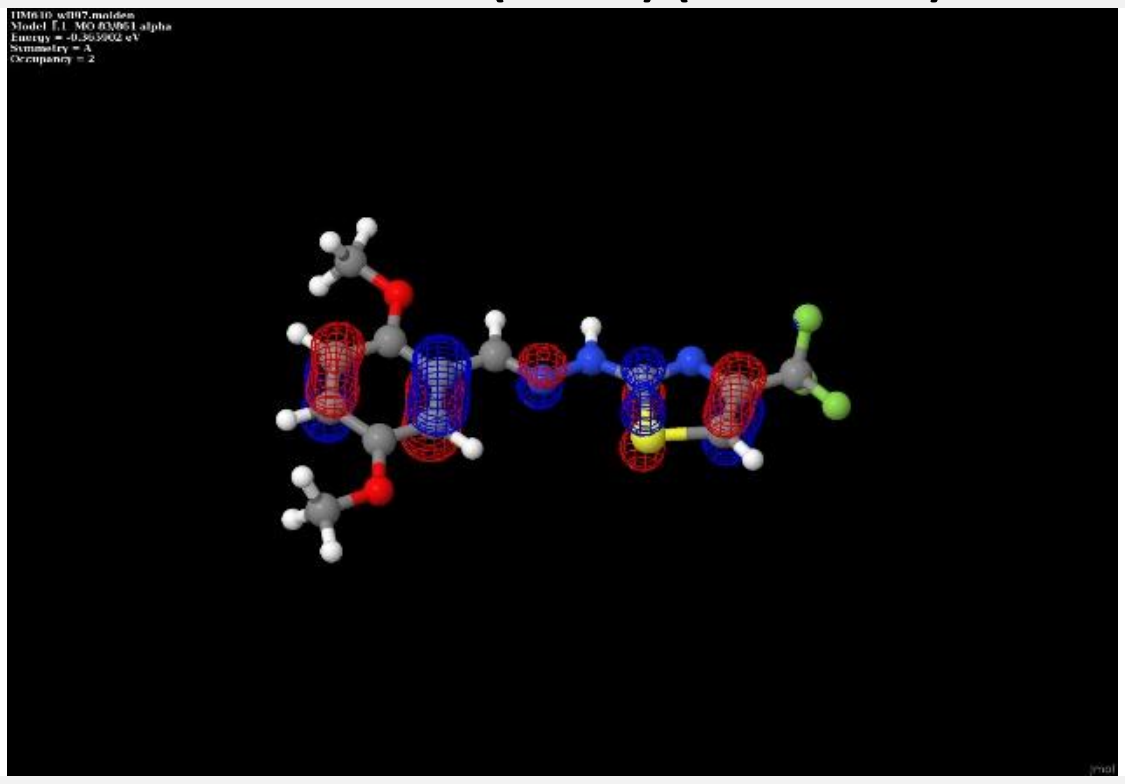

HOMO-2 ($\omega$B97) (orbital:83)

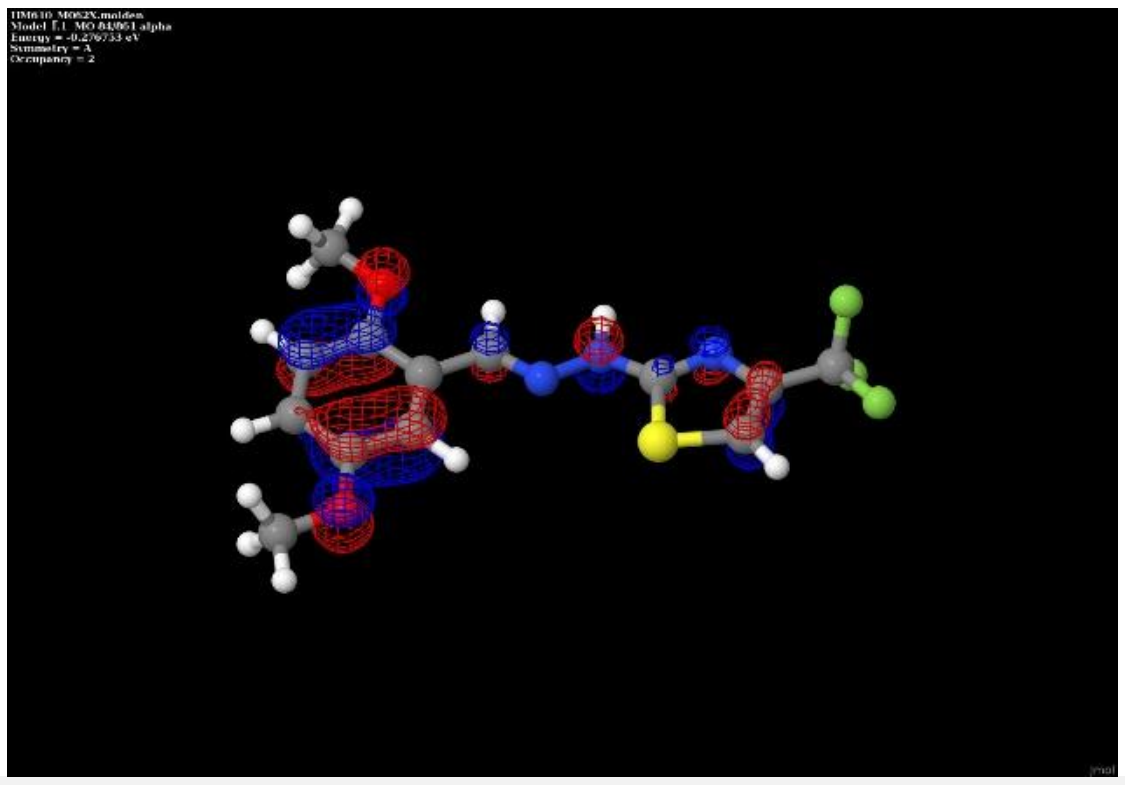

HOMO-1 (M062X) (orbital:84)

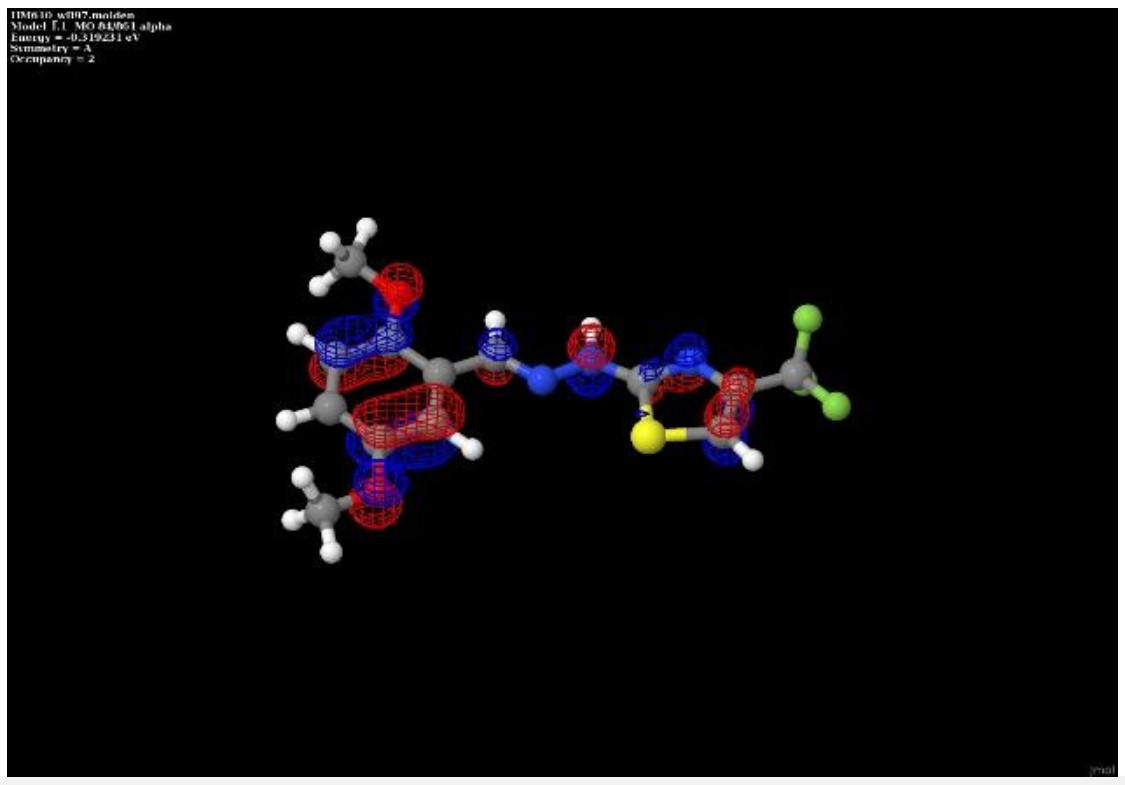

HOMO-1 (ωB97) (orbital:84)

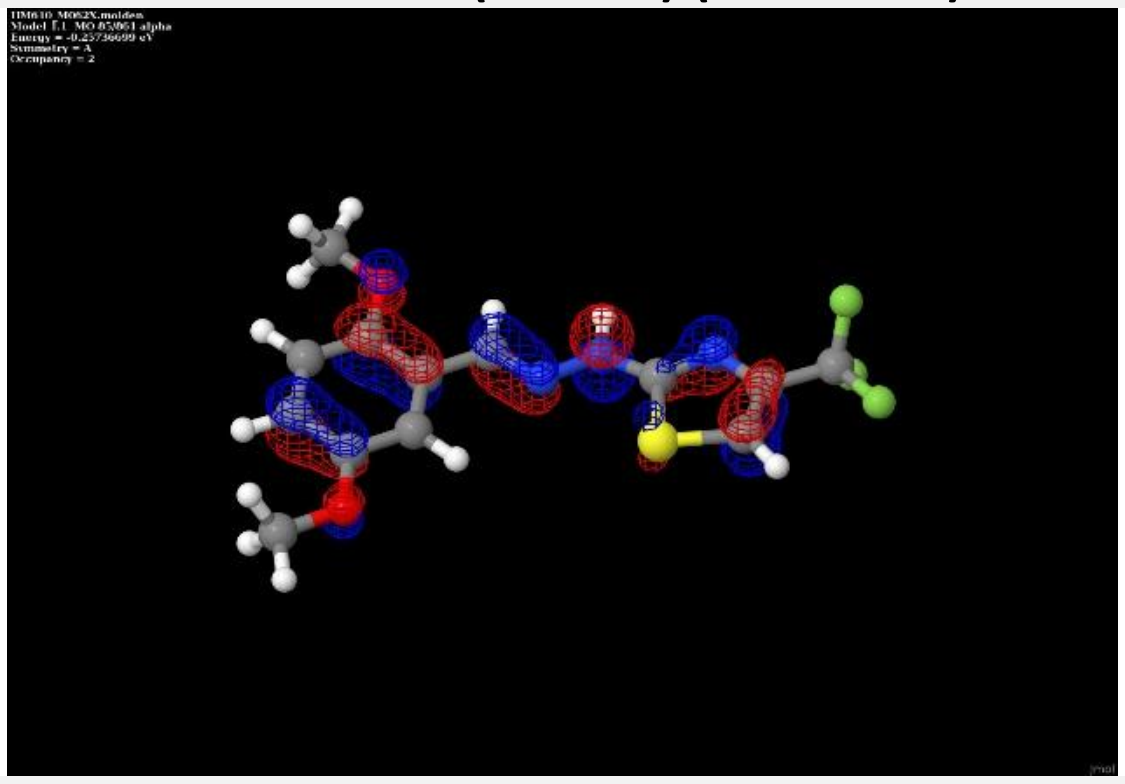

HOMO (M062X) (orbital:85)

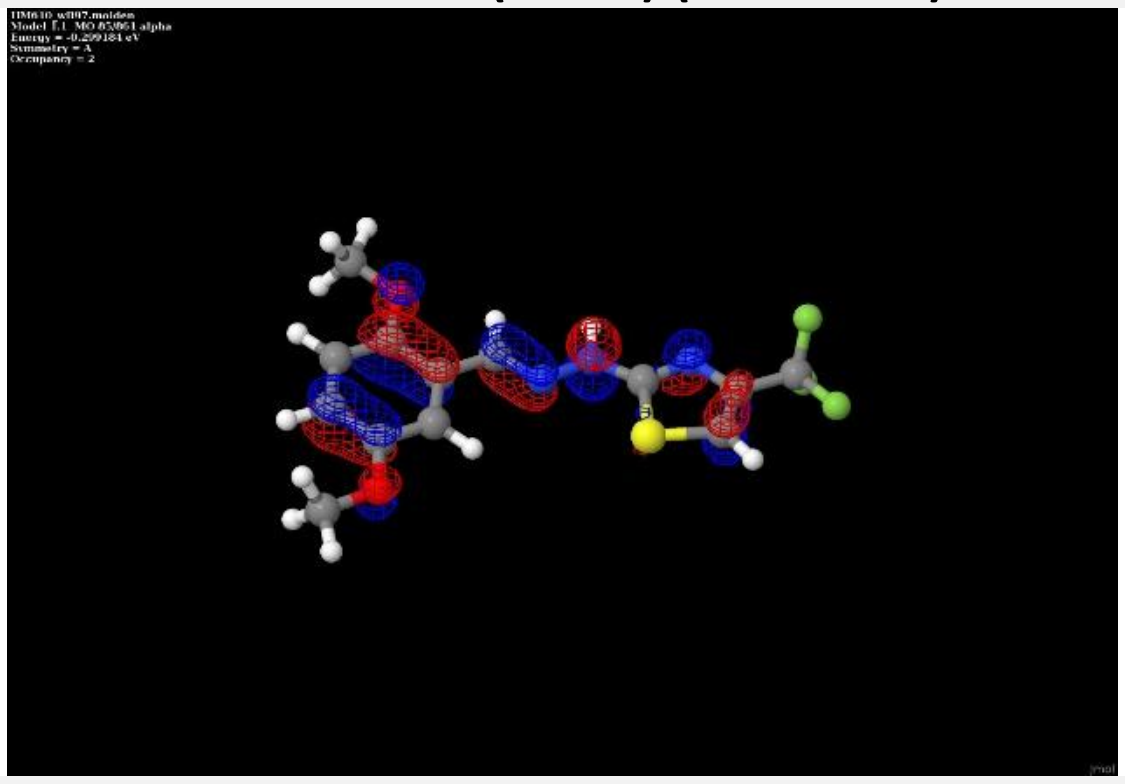

HOMO (ωB97) (orbital:85)

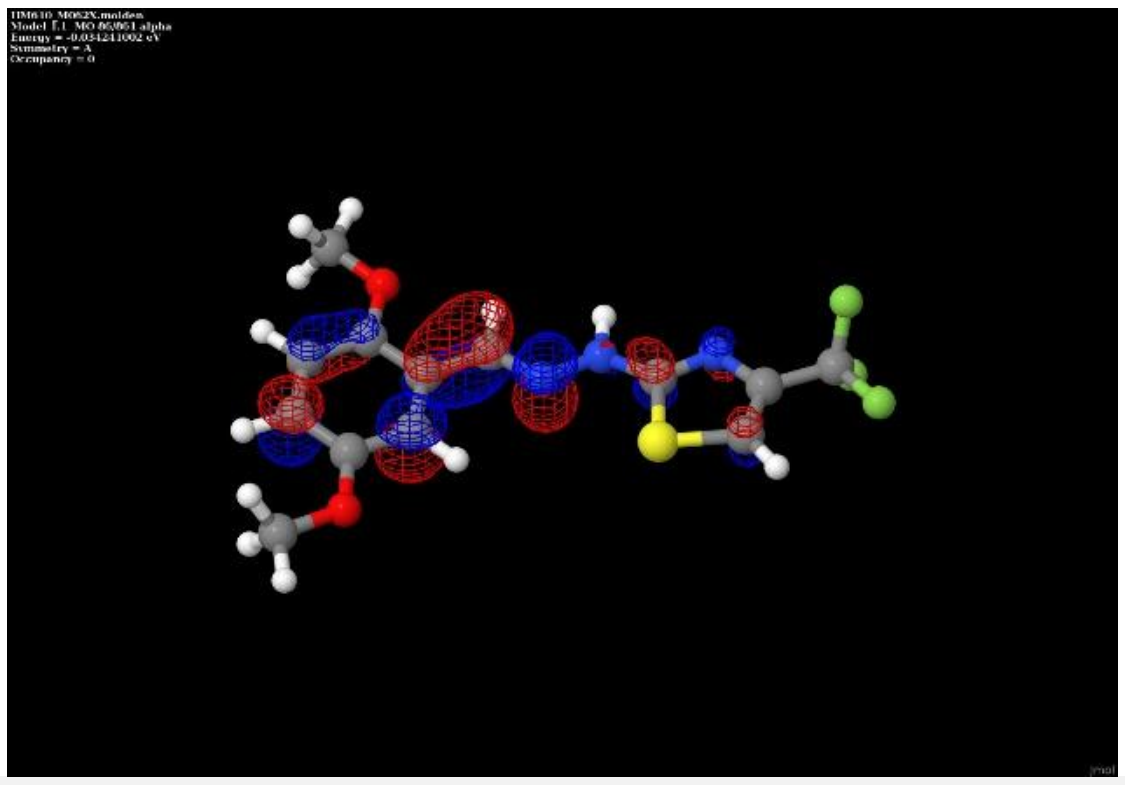

**LUMO (M062X) (orbital:86)**

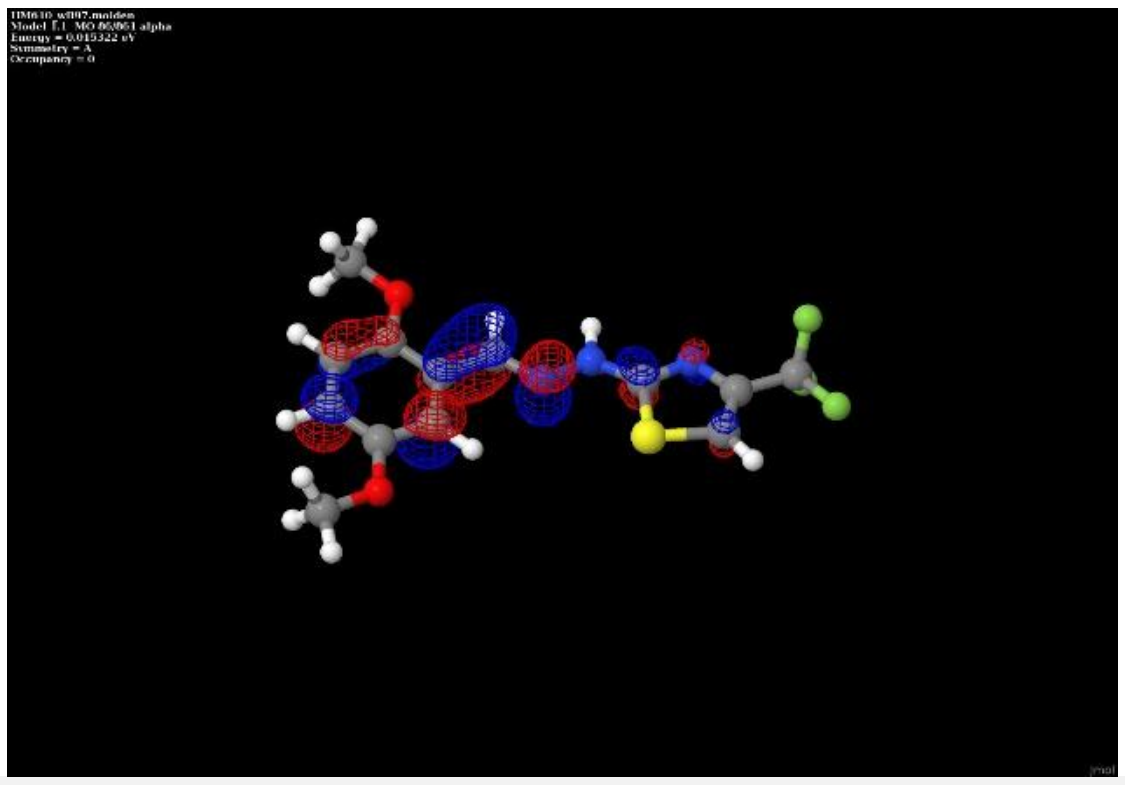

**LUMO ($\omega$B97) (orbital:86)**

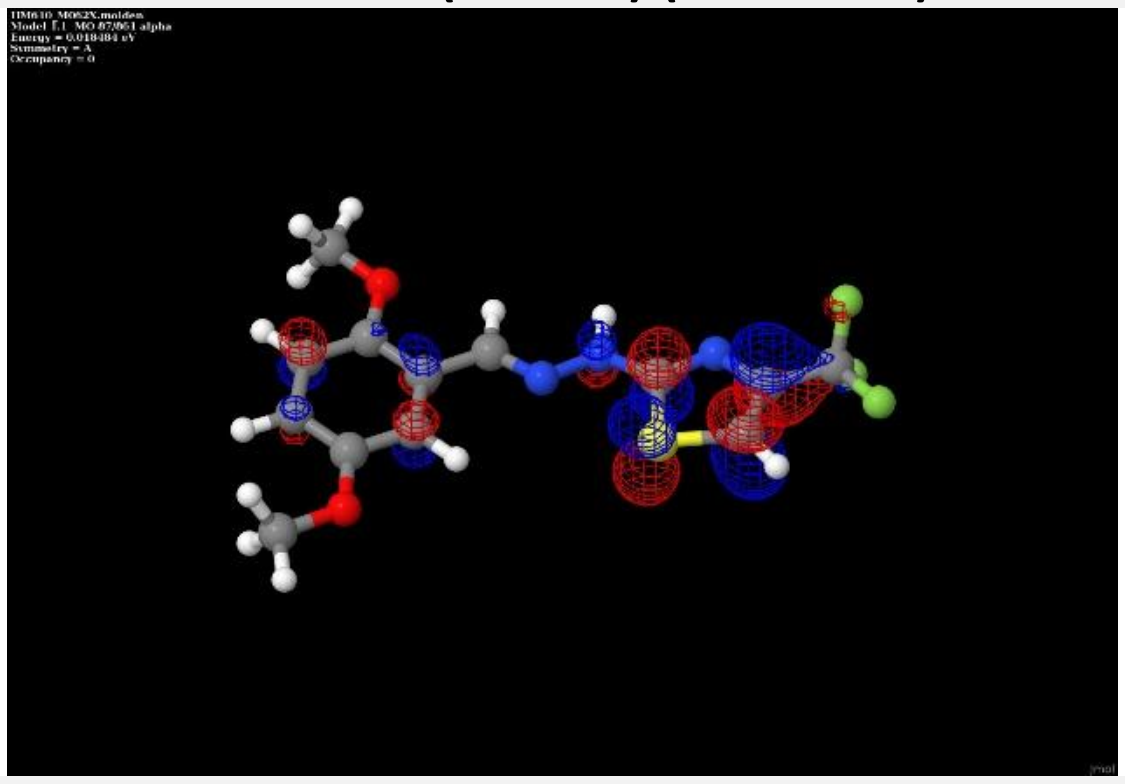

**LUMO+1 (M062X) (orbital:87)**

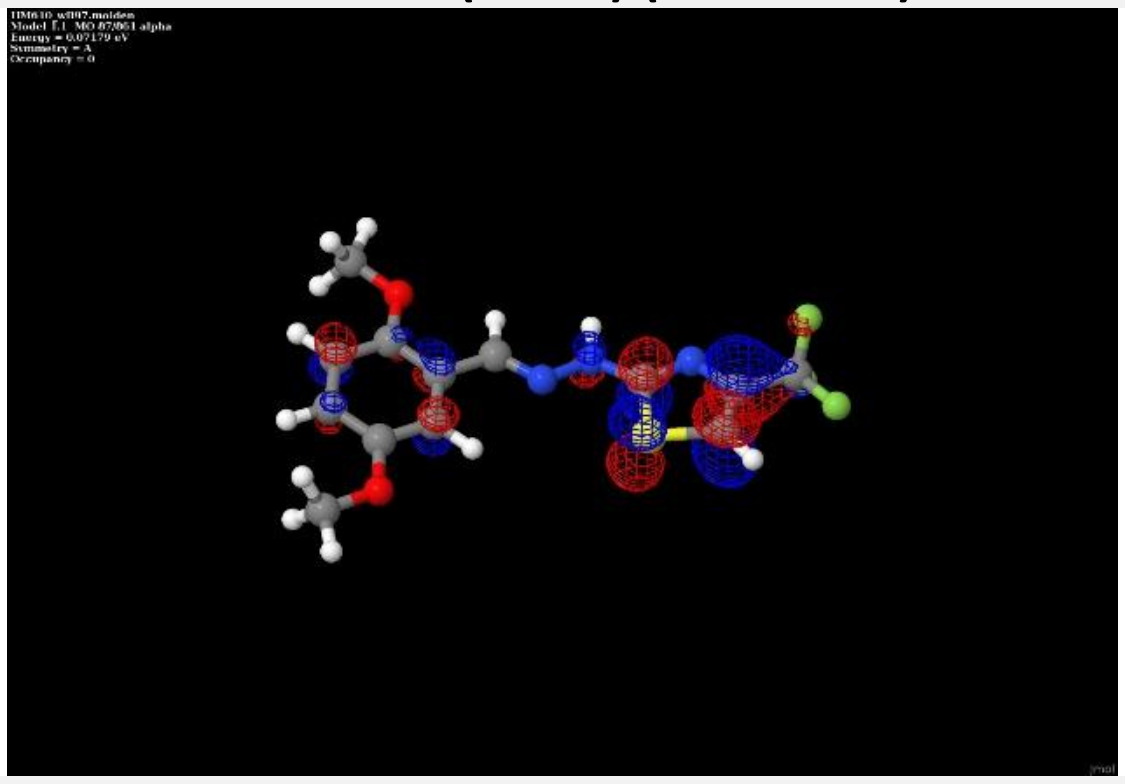

**LUMO+1 ($\omega$B97) (orbital:87)**

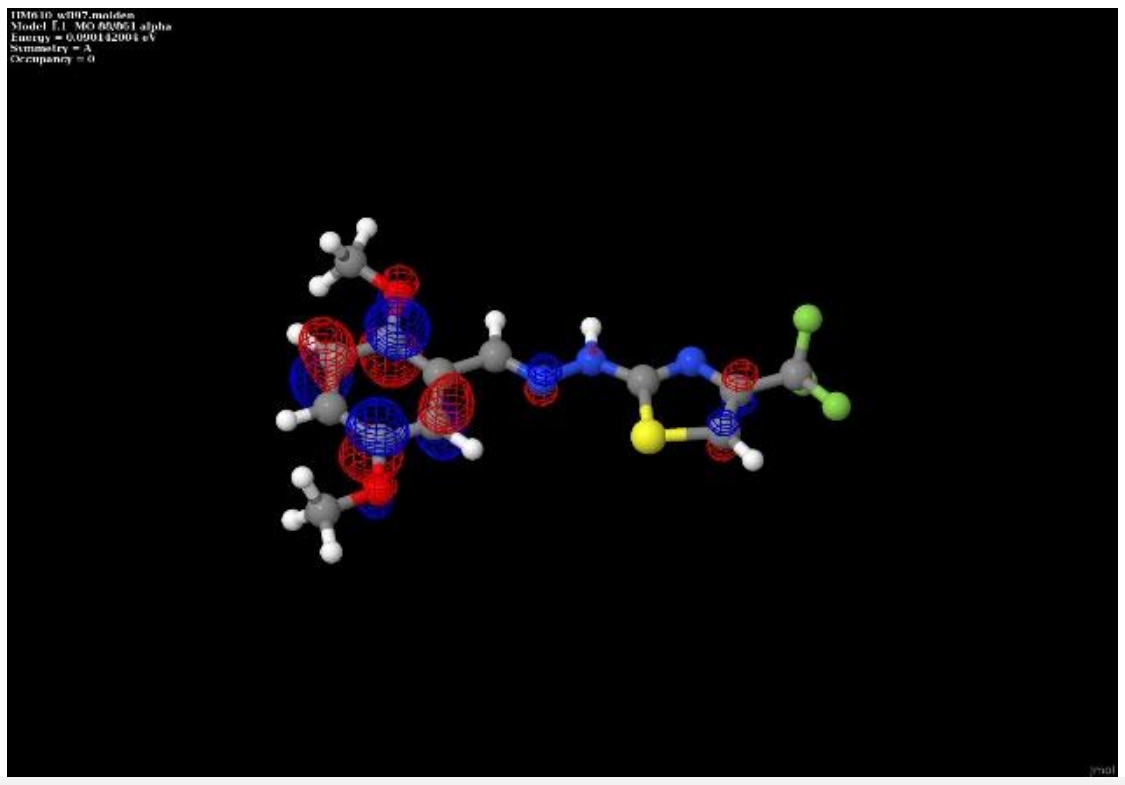

**LUMO+2 (M062X) (orbital:88)**

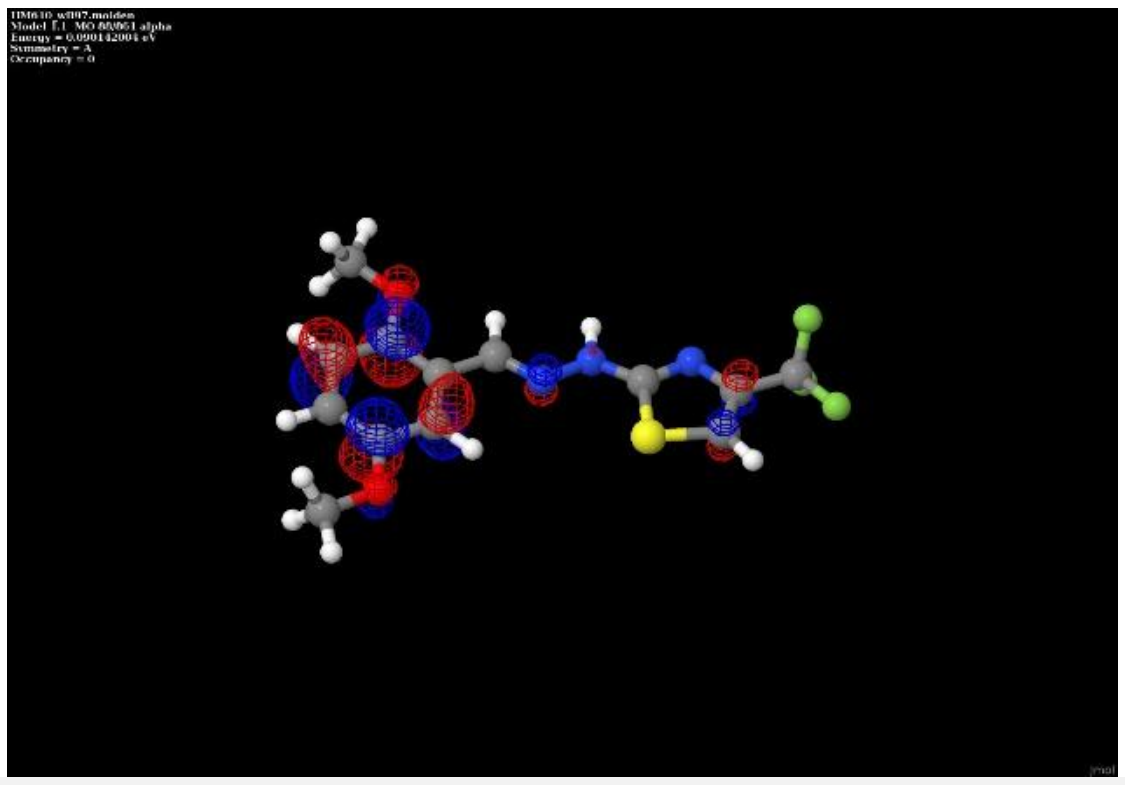

**LUMO+2 ($\omega$B97) (orbital:88)**

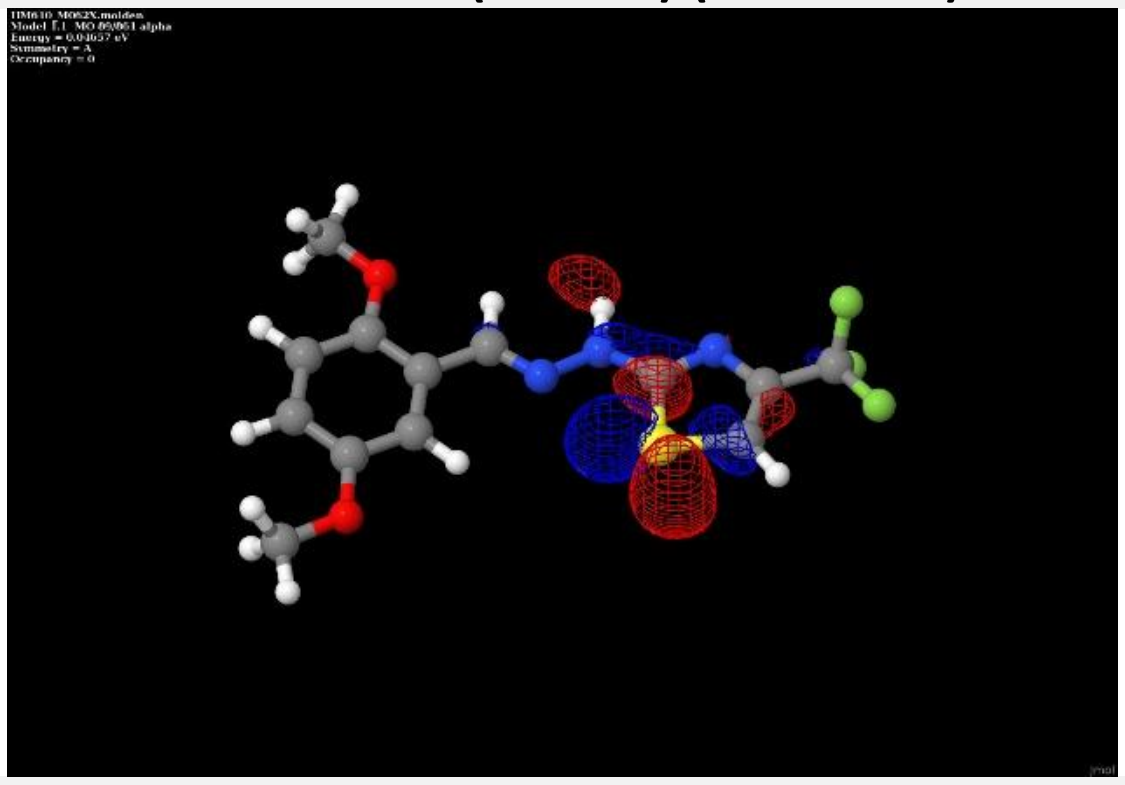

**LUMO+3 (M062X) (orbital:89)**

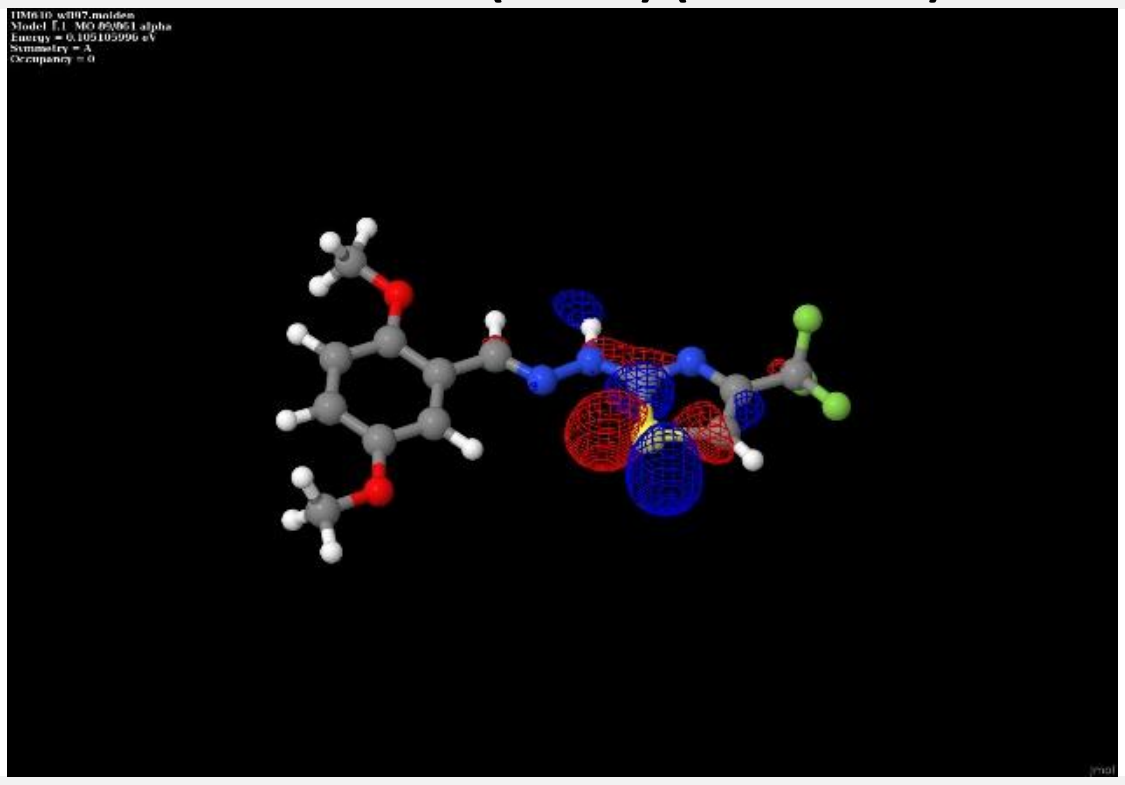

**LUMO+3 ($\omega$B97) (orbital:89)**

**Table S20:** Orbital entropies in HM610 computed by DMRG.

| Orbital Entropies |
|---|
| 0.00443976 |
| 0.00306474 |
| 0.08831756 |
| 0.01637977 |
| 0.08739358 |
| 0.03228039 |
| 0.11237942 |
| 0.21611789 |
| 0.18910291 |
| 0.21714082 |
| 0.23206128 |
| 0.19493116 |
| 0.18837806 |
| 0.02623684 |
| 0.00347528 |
| 0.11460917 |
| 0.03853718 |
| 0.01489764 |
| 0.11628968 |
| 0.00417322 |

**Table S21:** $^{1}$H Chemical shifts computed at different DFT level of theories.

| Number of atoms | Chemical Shift<br>B3LYP/cc-pVQZ | Chemical Shift<br>B3LYP/cc-pVQZ (CDCl3) | Chemical Shift<br>B3LYP/ Def2QZVPP (CDCl3) | Chemical Shift<br>M062X/cc-pVQZ (CDCl3) | Chemical Shift<br>M062X/ Def2QZVPP (CDCl3) | Chemical shift (Exp) |
|---|---|---|---|---|---|---|
| H21 | 6.92 | 7.13 | 7.12 | 7.67 | 7.62 | 6.87 |
| H22 | 8.41 | 8.69 | 8.73 | 9.12 | 9.14 | 10.26 $^{N-H}$ |
| H23 | 6.94 | 7.14 | 7.15 | 7.69 | 7.62 | 6.94 |
| H24 | 8.09 | 8.01 | 8.02 | 8.59 | 8.58 | 7.49 |
| H25 | 8.39 | 8.51 | 8.54 | 8.92 | 8.93 | 8.27$^{(Azomethine)}$ |
| H26 | 7.11 | 7.29 | 7.29 | 7.76 | 7.74 | 7.14 |
| H29 | 4.10 | 4.14 | 4.14 | 4.12 | 4.09 | 3.85 |
| H30 | 3.68 | 3.77 | 3.77 | 3.71 | 3.68 | 3.85 |
| H31 | 3.69 | 3.78 | 3.77 | 3.70 | 3.67 | 3.85 |
| H32 | 4.15 | 4.22 | 4.23 | 4.20 | 4.20 | 3.86 |
| H33 | 3.76 | 3.84 | 3.83 | 3.77 | 3.76 | 3.86 |
| H34 | 3.72 | 3.78 | 3.78 | 3.71 | 3.70 | 3.86 |

***Numbering of atoms mentioned in above table is shown in Fig S2**

**Table S22:** $^{13}$C Chemical shifts computed at different DFT level of theories.

| Number of atoms | Chemical Shift B3LYP/cc-pVQZ | Chemical Shift B3LYP/cc-pVQZ (CDCl3) | Chemical Shift B3LYP/ Def2QZVPP (CDCl3) | Chemical Shift M062X/cc-pVQZ (CDCl3) | Chemical Shift M062X/ Def2QZVPP (CDCl3) | Chemical shift (Exp) |
|---|---|---|---|---|---|---|
| C2 | 179.33 | 179.66 | 180.29 | 180.82 | 178.5 | 170.7 |
| C4 | 154.52 | 152.21 | 152.73 | 163.00 | 163.8 | 140.7 |
| C5 | 118.83 | 120.08 | 120.03 | 133.58 | 134.2 | 112.2 |
| C8 | 142.68 | 143.86 | 144.19 | 157.94 | 159.0 | 139.4 |
| C9 | 133.81 | 132.64 | 132.80 | 145.09 | 145.3 | 119.2 |
| C10 | 122.95 | 120.82 | 120.91 | 135.01 | 136.0 | 117.2 |
| C11 | 165.40 | 164.30 | 164.78 | 175.73 | 177.2 | 152.5 |
| C12 | 117.36 | 118.35 | 118.54 | 131.79 | 133.3 | 111.2 |
| C13 | 115.95 | 116.97 | 117.16 | 129.79 | 130.0 | 110.3 |
| C14 | 162.45 | 162.20 | 162.70 | 173.50 | 174.8 | 153.7 |
| C17 | 133.52 | 133.32 | 133.89 | 130.14 | 131.0 | 122.4 |
| C27 | 57.90 | 58.20 | 58.50 | 57.59 | 57.7 | 56.2 |
| C28 | 57.89 | 58.18 | 58.45 | 57.46 | 57.2 | 55.8 |

***Numbering of atoms mentioned in above table is shown in Fig S2**

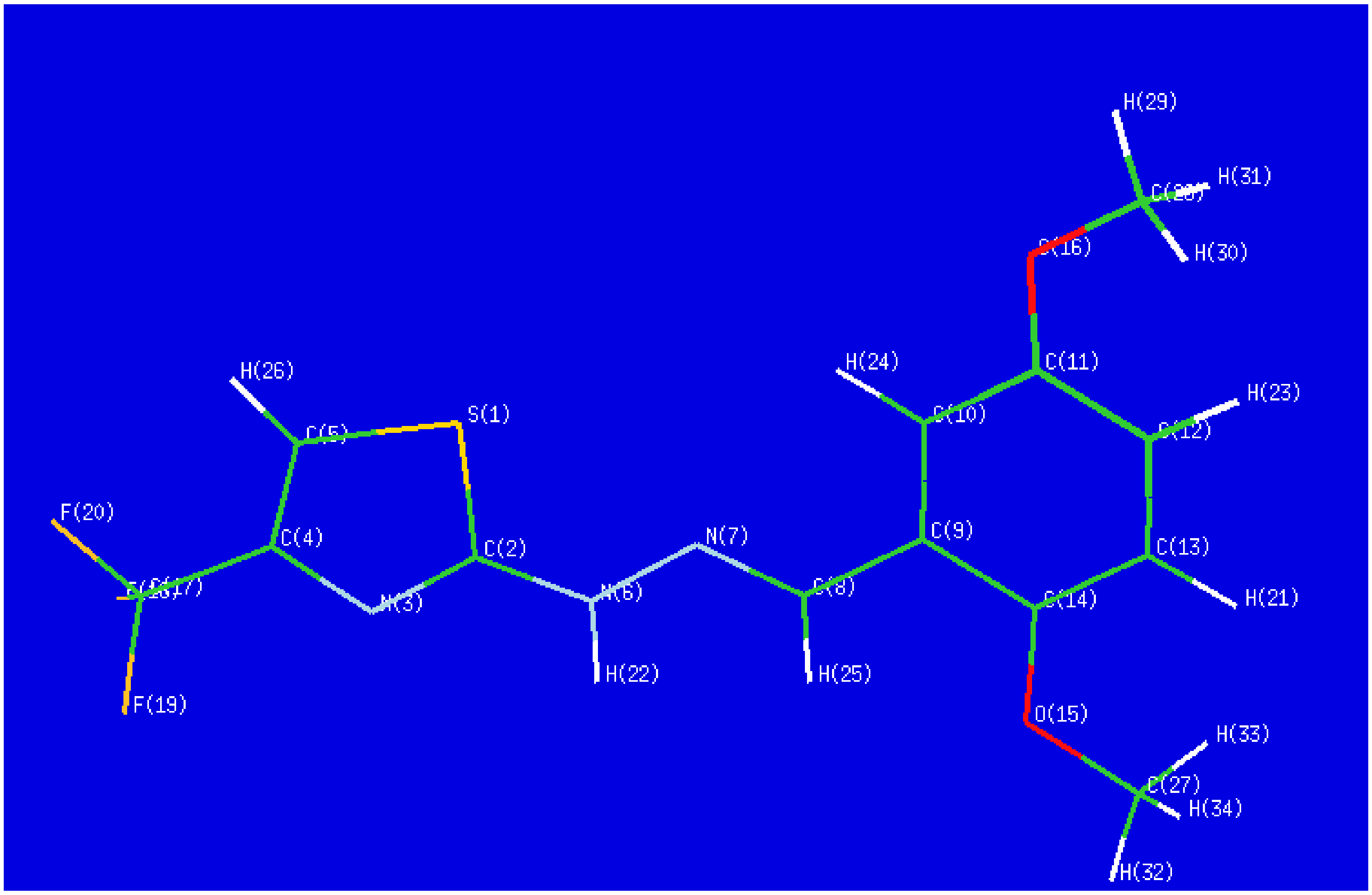

**Figure S2: Molecular Structure of 2-(2-(2,5-dimethoxybenzylidene)hydrazineyl)-4-(trifluoromethyl)thiazole (HM610)**

**Table S23**. Total SCF energies, zero-point vibrational energies, and thermal contributions to Gibbs energy for all compounds presented in Figure 4, in their optimized geometries for singlet and triplet spin states, calculated at the DFT M062X/def2-TZVP level.

| | **R1** | **R2** | **R3** | **Spin multiplicity** | **Energy (UM062X) (Ha)** | **Zero-Point Energy (Ha)** | **Thermal corrections to Gibbs Free Energy (Ha)** |
|---|---|---|---|---|---|---|---|
| 1 | H | H | H | Singlet | -948.8627945 | 0.175892 | 0.135563 |
| | | | | Triplet | -948.7845603 | 0.173495 | 0.132338 |
| 2 | MeO | MeO | H | Singlet | -1177.911932 | 0.242013 | 0.194759 |
| | | | | Triplet | -1177.835133 | 0.239616 | 0.191807 |
| 3 | H | H | $CF_3$ | Singlet | -1285.955572 | 0.180998 | 0.135033 |
| | | | | Triplet | -1285.876031 | 0.178636 | 0.131574 |
| 4 | MeO | MeO | MeO | Singlet | -1292.439815 | 0.275162 | 0.224756 |
| | | | | Triplet | -1292.362617 | 0.272786 | 0.221683 |
| 5 | $CF_3$ | $CF_3$ | $CF_3$ | Singlet | -1960.135994 | 0.191412 | 0.133838 |
| | | | | Triplet | -1960.058465 | 0.189069 | 0.130746 |
| HM610 | MeO | MeO | $CF_3$ | Singlet | -1515.005122 | 0.247077 | 0.194281 |
| | | | | Triplet | -1514.927018 | 0.244779 | 0.191186 |

**Table S24**. TDDFT results for all compounds presented in Figure 4, in their optimized geometries for singlet and triplet spin states, calculated at the TDDFT M062X/def2-TZVP level employing 20 states in the CIS procedure.

| | **R1** | **R2** | **R3** | **Spin multiplicity** | **Excited state number** | **ΔE (eV)** | **λ (nm)** | **f** |
|---|---|---|---|---|---|---|---|---|
| 1 | H | H | H | Singlet | 3 | 4.0935 | 302.88 | 0.6832 |
| | | | | Triplet | 1 | 3.4186 | 362.68 | 0.0476 |
| 2 | MeO | MeO | H | Singlet | 4 | 3.9091 | 317.17 | 0.6881 |
| | | | | Triplet | 1 | 2.7283 | 454.44 | 0.0433 |
| 3 | H | H | $CF_3$ | Singlet | 3 | 4.187 | 296.12 | 0.7283 |
| | | | | Triplet | 1 | 3.4502 | 359.35 | 0.0115 |
| 4 | MeO | MeO | MeO | Singlet | 4 | 3.8552 | 321.60 | 0.6612 |
| | | | | Triplet | 1 | 2.7257 | 454.86 | 0.0468 |
| 5 | $CF_3$ | $CF_3$ | $CF_3$ | Singlet | 3 | 3.9917 | 310.60 | 0.7080 |
| | | | | Triplet | 1 | 3.2499 | 381.50 | 0.0066 |
| HM610 | MeO | MeO | $CF_3$ | Singlet | 4 | 3.9503 | 313.86 | 0.6645 |
| | | | | Triplet | 1 | 2.6728 | 463.88 | 0.0444 |

## Cartesian Coordinates for All Optimized Molecular Structures

### HM610 $S_0$ optimized at TD-DFT/M062X/def2-TZVP

34

-1515.00504436076 a.u.

| | | | |
|---|---|---|---|
| S | 1.9374634 | 1.4365523 | -0.0554943 |
| C | 1.9349104 | -0.2841726 | 0.1666034 |
| N | 3.0975361 | -0.8523622 | 0.2104556 |
| C | 4.0678925 | 0.1003335 | 0.0643642 |
| C | 3.6598396 | 1.3780652 | -0.0881238 |
| N | 0.7697263 | -0.9777184 | 0.2908500 |
| N | -0.4075626 | -0.3448321 | 0.1377513 |
| C | -1.4652586 | -1.0536384 | 0.1316404 |
| C | -2.7858695 | -0.4427872 | -0.0202593 |
| C | -2.9531518 | 0.9320466 | -0.0063417 |
| C | -4.2110665 | 1.5031716 | -0.1625311 |
| C | -5.3158121 | 0.6807794 | -0.3342626 |
| C | -5.1601212 | -0.7038089 | -0.3478148 |
| C | -3.9083099 | -1.2730335 | -0.1909824 |
| O | -3.6723901 | -2.6137404 | -0.1967318 |
| O | -4.2595930 | 2.8607301 | -0.1310650 |
| C | 5.4974538 | -0.3435744 | 0.0729779 |
| F | 5.8241006 | -0.9481021 | 1.2180206 |
| F | 5.7528357 | -1.2071688 | -0.9135105 |
| F | 6.3234122 | 0.6981029 | -0.0819878 |
| H | -6.0348834 | -1.3219269 | -0.4848226 |
| H | 0.8308184 | -1.9884818 | 0.3417999 |
| H | -6.3062306 | 1.0932585 | -0.4583699 |
| H | -2.0937122 | 1.5750693 | 0.1287120 |
| H | -1.4203076 | -2.1380741 | 0.2294925 |
| H | 4.2598705 | 2.2616713 | -0.2184088 |
| C | -4.7710694 | -3.4722908 | -0.3977424 |
| C | -5.5182621 | 3.4709762 | -0.2907173 |
| H | -5.3466962 | 4.5425381 | -0.2363544 |
| H | -5.9591983 | 3.2235732 | -1.2606701 |
| H | -6.2076753 | 3.1738830 | 0.5048737 |
| H | -4.3755904 | -4.4840013 | -0.3717852 |
| H | -5.5163094 | -3.3539179 | 0.3932024 |
| H | -5.2408527 | -3.2898019 | -1.3679959 |

**TMS $S_0$ optimized at TD-DFT/ M062X/def2-TZVP**

17
-449.14541374160 a.u.

| | | | |
|---|---|---|---|
| Si | 0.1206952 | -0.0000004 | -0.0006788 |
| C | 0.2868928 | 0.0000025 | 1.8656390 |
| H | 1.3358349 | 0.0000224 | 2.1673030 |
| H | -0.1858740 | -0.8816975 | 2.3019654 |
| H | -0.1859078 | 0.8816838 | 2.3019665 |
| C | -1.6950432 | 0.0000017 | -0.4638507 |
| H | -1.8274672 | 0.0000233 | -1.5472662 |
| H | -2.2008358 | -0.8818112 | -0.0665127 |
| H | -2.2008434 | 0.8817937 | -0.0664761 |
| C | 0.9459081 | 1.5293841 | -0.7013140 |
| H | 0.8891577 | 1.5413654 | -1.7912267 |
| H | 0.4677422 | 2.4386202 | -0.3325636 |
| H | 2.0001955 | 1.5698412 | -0.4209524 |
| C | 0.9459040 | -1.5293879 | -0.7013106 |
| H | 0.8891528 | -1.5413719 | -1.7912232 |
| H | 2.0001913 | -1.5698482 | -0.4209493 |
| H | 0.4677347 | -2.4386211 | -0.3325575 |

**Compound 1 $S_0$ from Figure 4 optimized at DFT/M062X/def2-TZVP**

23

-948.8627945 a.u.

| | | | |
|---|---|---|---|
| C | -0.99523400 | 0.12566200 | -0.52672000 |
| C | -2.13890000 | 0.85404300 | -0.20736300 |
| C | -2.00697500 | 2.06655000 | 0.47374200 |
| C | -0.75323800 | 2.53236900 | 0.82377700 |
| C | 0.38561500 | 1.79972900 | 0.50173500 |
| C | 0.26177200 | 0.59558300 | -0.17410100 |
| H | -2.89741700 | 2.62968800 | 0.72028600 |
| H | 1.36438300 | 2.16916000 | 0.77833900 |
| H | 1.14286600 | 0.02080300 | -0.42764300 |
| C | -3.44828300 | 0.32830500 | -0.59454500 |
| H | -3.46721600 | -0.62784400 | -1.12570600 |
| N | -4.52578100 | 0.95106300 | -0.32817500 |
| N | -5.68990000 | 0.41244300 | -0.71230700 |
| C | -6.87094500 | 1.04688000 | -0.44100100 |
| S | -6.91038200 | 2.55974300 | 0.39758000 |
| N | -8.01768100 | 0.55979600 | -0.79116700 |
| C | -8.63171800 | 2.52722600 | 0.26037200 |
| C | -9.02110400 | 1.40770900 | -0.38910500 |
| H | -9.23883600 | 3.32124700 | 0.65950400 |
| H | -5.73685800 | -0.47308800 | -1.20310100 |
| H | -10.04560500 | 1.14435200 | -0.60662400 |
| H | -1.09262800 | -0.81567600 | -1.05519500 |
| H | -0.65832300 | 3.47256600 | 1.35144700 |

**Compound 1 $T_1$ from Figure 4 optimized at DFT/M062X/def2-TZVP**

23

-948.7845603 a.u.

| | | | |
|---|---|---|---|
| C | -0.96216900 | 0.26708600 | -0.87223600 |
| C | -2.20644600 | 0.90600600 | -0.67608700 |
| C | -2.31099400 | 1.85664500 | 0.36309700 |
| C | -1.22348700 | 2.13933300 | 1.16497700 |
| C | -0.00639100 | 1.49389600 | 0.96305900 |
| C | 0.11497600 | 0.55782400 | -0.06250600 |
| H | -3.25205500 | 2.36994500 | 0.51399600 |
| H | 0.84124900 | 1.72011100 | 1.59585100 |
| H | 1.06031600 | 0.05669300 | -0.22636700 |
| C | -3.30980800 | 0.58242100 | -1.50259000 |
| H | -3.18767800 | -0.10266200 | -2.33611000 |
| N | -4.53827200 | 1.22623200 | -1.39322400 |
| N | -5.37345200 | 0.66310100 | -0.53404500 |
| C | -6.62126300 | 1.18973200 | -0.31137800 |
| S | -7.15148800 | 2.60752300 | -1.13828900 |
| N | -7.45905000 | 0.66143400 | 0.52497500 |
| C | -8.63851000 | 2.47869800 | -0.27834600 |
| C | -8.61223200 | 1.40044200 | 0.54229200 |
| H | -9.43202200 | 3.19119000 | -0.42558900 |
| H | -5.11858500 | -0.16581500 | -0.00351700 |
| H | -9.42051800 | 1.09587100 | 1.19022200 |
| H | -0.86558100 | -0.45988600 | -1.66982700 |
| H | -1.31892700 | 2.87248700 | 1.95567100 |

**Compound 2 $S_0$ from Figure 4 optimized at DFT/M062X/def2-TZVP**

31
-1177.911932 a.u.

| | | | |
|---|---|---|---|
| C | -0.90154300 | 0.10868300 | -0.45209900 |
| C | -2.06971700 | 0.84187700 | -0.17351300 |
| C | -1.96262600 | 2.07090900 | 0.45658400 |
| C | -0.72367700 | 2.58867800 | 0.81683000 |
| C | 0.42606400 | 1.86242700 | 0.54068700 |
| C | 0.33163300 | 0.62464700 | -0.09323800 |
| H | -2.85632800 | 2.64051700 | 0.67399100 |
| H | 1.40336700 | 2.23669800 | 0.80765600 |
| H | 1.23989100 | 0.07753600 | -0.29810100 |
| C | -3.37098500 | 0.29071100 | -0.55452900 |
| H | -3.38374400 | -0.67941600 | -1.05061100 |
| N | -4.44818400 | 0.92630500 | -0.31194300 |
| N | -5.60933100 | 0.37123500 | -0.68461100 |
| C | -6.79272200 | 1.01020700 | -0.44122400 |
| S | -6.83939500 | 2.55091400 | 0.34550400 |
| N | -7.93749100 | 0.50901500 | -0.78005200 |
| C | -8.56017300 | 2.50986600 | 0.20150900 |
| C | -8.94410700 | 1.36806700 | -0.41150000 |
| H | -9.17100800 | 3.31530400 | 0.57097500 |
| H | -5.64944300 | -0.53057300 | -1.14541600 |
| H | -9.96714100 | 1.09528400 | -0.62456300 |
| O | -0.73900800 | 3.80239600 | 1.42978800 |
| C | 0.49849000 | 4.35440700 | 1.80999600 |
| H | 0.27614800 | 5.31015800 | 2.27689500 |
| H | 1.01600900 | 3.71202500 | 2.52849900 |
| H | 1.14450600 | 4.51692500 | 0.94224200 |
| O | -1.07456200 | -1.09183000 | -1.07318600 |
| C | 0.07611300 | -1.84876600 | -1.36657800 |
| H | 0.61965500 | -2.11137900 | -0.45483500 |
| H | -0.27124200 | -2.75602300 | -1.85355400 |
| H | 0.74399000 | -1.30712100 | -2.04201000 |

**Compound 2 $T_1$ from Figure 4 optimized at DFT/M062X/def2-TZVP**

31

-1177.835133 a.u.

| | | | |
|---|---|---|---|
| C | -0.88141900 | 0.04344000 | -0.72944200 |
| C | -2.11769000 | 0.75201200 | -0.66564600 |
| C | -2.16198700 | 1.94124300 | 0.07966100 |
| C | -1.04202100 | 2.41853400 | 0.74339800 |
| C | 0.15401100 | 1.71384800 | 0.67557800 |
| C | 0.22286600 | 0.52704500 | -0.06501200 |
| H | -3.08053100 | 2.51084600 | 0.13037400 |
| H | 1.04105100 | 2.06269400 | 1.18252200 |
| H | 1.16389700 | -0.00181200 | -0.10403900 |
| C | -3.25276700 | 0.25060300 | -1.33626600 |
| H | -3.18051200 | -0.63111400 | -1.96041900 |
| N | -4.46343600 | 0.93931100 | -1.34296700 |
| N | -5.25404300 | 0.65018600 | -0.32075000 |
| C | -6.48038600 | 1.24954600 | -0.18305700 |
| S | -7.03728200 | 2.41276900 | -1.32890100 |
| N | -7.27823200 | 0.98347900 | 0.80377100 |
| C | -8.47693400 | 2.55515200 | -0.39332200 |
| C | -8.41905800 | 1.73168100 | 0.68142200 |
| H | -9.26813600 | 3.22456600 | -0.68452800 |
| H | -4.97853600 | -0.01373100 | 0.39800200 |
| H | -9.19445900 | 1.63094800 | 1.42616300 |
| O | -1.20684900 | 3.58168100 | 1.43053200 |
| C | -0.08914900 | 4.10150300 | 2.10927600 |
| H | -0.42100900 | 5.02060400 | 2.58490500 |
| H | 0.26974400 | 3.40732400 | 2.87477600 |
| H | 0.72726800 | 4.32658700 | 1.41687700 |
| O | -0.89865500 | -1.09736700 | -1.46507400 |
| C | 0.30015200 | -1.82944200 | -1.56504400 |
| H | 0.63713900 | -2.17019500 | -0.58213300 |
| H | 0.08124400 | -2.69021500 | -2.19107300 |
| H | 1.08915700 | -1.23204200 | -2.02999100 |

**Compound 3 $S_0$ from Figure 4 optimized at DFT/M062X/def2-TZVP**

26

-1285.955572 a.u.

| | | | |
|---|---|---|---|
| C | -1.01613900 | 0.16549100 | -0.63743100 |
| C | -2.14582200 | 0.86564900 | -0.22057800 |
| C | -1.98792100 | 2.05231300 | 0.49895100 |
| C | -0.72076400 | 2.52158700 | 0.79094200 |
| C | 0.40433600 | 1.81751300 | 0.37144800 |
| C | 0.25432400 | 0.63900400 | -0.34309500 |
| H | -2.86788800 | 2.59320800 | 0.82118000 |
| H | 1.39373700 | 2.18968400 | 0.60277700 |
| H | 1.12509800 | 0.08723600 | -0.67184900 |
| C | -3.46922100 | 0.33579300 | -0.54940300 |
| H | -3.50700900 | -0.60158900 | -1.11189200 |
| N | -4.53723400 | 0.93240800 | -0.20045000 |
| N | -5.71500700 | 0.38285800 | -0.53867900 |
| C | -6.87991400 | 0.99425600 | -0.18444500 |
| S | -6.87869800 | 2.48565200 | 0.70137300 |
| N | -8.04380600 | 0.51139100 | -0.48072400 |
| C | -8.60111800 | 2.44958600 | 0.66557400 |
| C | -9.01254100 | 1.34546400 | 0.00662600 |
| H | -9.19945600 | 3.22075300 | 1.11877800 |
| H | -5.77633300 | -0.48554700 | -1.05797400 |
| H | -0.60443100 | 3.44158800 | 1.34885900 |
| H | -1.13469800 | -0.75584500 | -1.19586700 |
| C | -10.44476500 | 0.97411100 | -0.22368400 |
| F | -10.74566000 | -0.20437900 | 0.32741600 |
| F | -10.73061900 | 0.88411100 | -1.52483800 |
| F | -11.26664700 | 1.88751300 | 0.30662700 |

**Compound 3 $T_1$ from Figure 4 optimized at DFT/M062X/def2-TZVP**

26

-1285.876031 a.u.

| | | | |
|---|---|---|---|
| C | -1.00518700 | 0.47952100 | -1.01427200 |
| C | -2.25943800 | 1.06348800 | -0.73001000 |
| C | -2.42160700 | 1.72786200 | 0.50564500 |
| C | -1.37891800 | 1.79118800 | 1.40787000 |
| C | -0.15148100 | 1.20277000 | 1.11494100 |
| C | 0.02665100 | 0.54883300 | -0.10294400 |
| H | -3.37029300 | 2.19788300 | 0.73142700 |
| H | 0.66079900 | 1.25587500 | 1.82725900 |
| H | 0.98045800 | 0.09365600 | -0.33649700 |
| C | -3.31478400 | 0.96581700 | -1.66824600 |
| H | -3.14868300 | 0.50881100 | -2.63880100 |
| N | -4.55018700 | 1.57744000 | -1.46878100 |
| N | -5.43247800 | 0.82408500 | -0.82946800 |
| C | -6.69095500 | 1.29671400 | -0.56758700 |
| S | -7.16151600 | 2.88531400 | -1.06111500 |
| N | -7.58641400 | 0.59792300 | 0.05554600 |
| C | -8.70453900 | 2.58284400 | -0.36572800 |
| C | -8.72797000 | 1.33811400 | 0.16517000 |
| H | -9.49359800 | 3.31485400 | -0.38366600 |
| H | -5.20693000 | -0.11409500 | -0.50839200 |
| H | -1.51766200 | 2.30605400 | 2.34982000 |
| H | -0.86489600 | -0.02807000 | -1.96109200 |
| C | -9.90762400 | 0.72504200 | 0.85641000 |
| F | -9.62782100 | 0.40698000 | 2.12191200 |
| F | -10.31073500 | -0.39198300 | 0.24794800 |
| F | -10.94239000 | 1.57289000 | 0.87229300 |

**Compound 4 $S_0$ from Figure 4 optimized at DFT/M062X/def2-TZVP**

35

-1292.439815 a.u.

| | | | |
|---|---|---|---|
| C | -0.89892100 | 0.10852300 | -0.47679100 |
| C | -2.07092800 | 0.83803300 | -0.20496900 |
| C | -1.97250600 | 2.06390500 | 0.43263000 |
| C | -0.73799100 | 2.58218500 | 0.80704000 |
| C | 0.41568900 | 1.85961500 | 0.53753700 |
| C | 0.32984900 | 0.62509200 | -0.10383600 |
| H | -2.86950600 | 2.63029900 | 0.64490400 |
| H | 1.38970900 | 2.23437300 | 0.81562200 |
| H | 1.24111400 | 0.08096900 | -0.30320100 |
| C | -3.36761100 | 0.28619400 | -0.60053000 |
| H | -3.37391900 | -0.68251300 | -1.09956200 |
| N | -4.44784700 | 0.91934200 | -0.36672600 |
| N | -5.60437000 | 0.36219300 | -0.75258800 |
| C | -6.78786400 | 1.00004100 | -0.51890700 |
| S | -6.84222200 | 2.53619700 | 0.26824000 |
| N | -7.92502500 | 0.49179400 | -0.87244100 |
| C | -8.56728100 | 2.49579500 | 0.10858900 |
| C | -8.93197100 | 1.34323800 | -0.51477200 |
| H | -9.17875500 | 3.30024700 | 0.47420300 |
| H | -5.63867900 | -0.53857600 | -1.21610800 |
| O | -0.76134700 | 3.79241600 | 1.42647200 |
| O | -1.06382500 | -1.08884900 | -1.10580700 |
| C | 0.47136900 | 4.34427300 | 1.82212700 |
| H | 0.24270800 | 5.29678900 | 2.29256500 |
| H | 0.98294200 | 3.69845800 | 2.54181500 |
| H | 1.12560100 | 4.51319100 | 0.96177600 |
| C | 0.09099300 | -1.84200700 | -1.39286800 |
| H | 0.62672500 | -2.10849600 | -0.47764500 |
| H | -0.25020800 | -2.74725000 | -1.88785700 |
| H | 0.76380600 | -1.29537200 | -2.05930200 |
| O | -10.17199400 | 0.93279700 | -0.82602100 |
| C | -11.20946400 | 1.81250200 | -0.45583000 |
| H | -12.13881700 | 1.34670800 | -0.77082300 |
| H | -11.22188600 | 1.96434900 | 0.62730000 |
| H | -11.09469400 | 2.77963800 | -0.95367000 |

**Compound 4 $T_1$ from Figure 4 optimized at DFT/M062X/def2-TZVP**

35

-1292.362617 a.u.

| | | | |
|---|---|---|---|
| C | -0.86477200 | 0.06238000 | -0.76834700 |
| C | -2.11595100 | 0.74645200 | -0.79812700 |
| C | -2.25901200 | 1.90083500 | -0.01166000 |
| C | -1.22042800 | 2.36739100 | 0.77986400 |
| C | -0.00897300 | 1.68660600 | 0.80202400 |
| C | 0.15792100 | 0.53490500 | 0.02271000 |
| H | -3.18987600 | 2.45219600 | -0.02905000 |
| H | 0.81606200 | 2.02777400 | 1.40903800 |
| H | 1.10916500 | 0.02425600 | 0.05573400 |
| C | -3.16852200 | 0.25635100 | -1.59910400 |
| H | -3.01652800 | -0.59408500 | -2.25167000 |
| N | -4.38602500 | 0.92579300 | -1.69611200 |
| N | -5.27014700 | 0.57522200 | -0.77494600 |
| C | -6.51087300 | 1.15591100 | -0.73161000 |
| S | -6.96183300 | 2.37585100 | -1.85842300 |
| N | -7.39556300 | 0.82233300 | 0.15557500 |
| C | -8.49608400 | 2.45596400 | -1.06741700 |
| C | -8.52340500 | 1.56302100 | -0.03692100 |
| H | -9.25938800 | 3.13473100 | -1.40208400 |
| H | -5.05805200 | -0.12256400 | -0.06670800 |
| O | -1.47687600 | 3.49624700 | 1.49535300 |
| O | -0.78519400 | -1.04496800 | -1.54961000 |
| C | -0.44432100 | 4.00256300 | 2.30623000 |
| H | -0.84171300 | 4.89318600 | 2.78578500 |
| H | -0.14979300 | 3.27985200 | 3.07281600 |
| H | 0.43275600 | 4.27282400 | 1.71097900 |
| C | 0.43376800 | -1.75000600 | -1.56294500 |
| H | 0.68135000 | -2.12844200 | -0.56734000 |
| H | 0.29593800 | -2.58560000 | -2.24378500 |
| H | 1.25049800 | -1.11800500 | -1.92233500 |
| O | -9.53044900 | 1.32184200 | 0.81586100 |
| C | -10.69416700 | 2.09269600 | 0.61278400 |
| H | -11.40595800 | 1.78507700 | 1.37321800 |
| H | -10.47713200 | 3.15910300 | 0.72114000 |
| H | -11.11354700 | 1.90839900 | -0.38034700 |

**Compound 5 $S_0$ from Figure 4 optimized at DFT/M062X/def2-TZVP**

32
-1960.135994 a.u.

| | | | |
|---|---|---|---|
| C | -0.95619000 | 0.13259400 | -0.57480700 |
| C | -2.14661500 | 0.77357600 | -0.21628300 |
| C | -2.06324500 | 1.96697200 | 0.50598200 |
| C | -0.83548300 | 2.49029300 | 0.85109800 |
| C | 0.34392700 | 1.84823000 | 0.49157600 |
| C | 0.27684800 | 0.66780600 | -0.22248300 |
| H | -2.97560000 | 2.47194800 | 0.78948800 |
| H | 1.30102400 | 2.26971600 | 0.76905400 |
| H | 1.18320400 | 0.15536900 | -0.51014100 |
| C | -3.46454300 | 0.23247900 | -0.57437000 |
| H | -3.52527600 | -0.69992000 | -1.13211400 |
| N | -4.51995600 | 0.85733200 | -0.22987800 |
| N | -5.70580600 | 0.34730100 | -0.56553100 |
| C | -6.85607900 | 0.99718600 | -0.21161400 |
| S | -6.81621200 | 2.48512100 | 0.67484800 |
| N | -8.02903000 | 0.54590300 | -0.51343600 |
| C | -8.53749100 | 2.49638200 | 0.63231400 |
| C | -8.97784100 | 1.40480800 | -0.02950800 |
| H | -9.11509400 | 3.28401300 | 1.08440600 |
| H | -5.79538500 | -0.51860800 | -1.08623300 |
| C | -0.73726800 | 3.77732200 | 1.62638000 |
| C | -0.98854800 | -1.15786400 | -1.35443000 |
| C | -10.41926900 | 1.07285900 | -0.26658300 |
| F | -0.07354300 | 3.60126300 | 2.77531400 |
| F | -1.93326000 | 4.28279700 | 1.92824000 |
| F | -0.07051600 | 4.70909700 | 0.93413500 |
| F | 0.23187400 | -1.62599900 | -1.60891200 |
| F | -1.60697500 | -1.00923100 | -2.53587000 |
| F | -1.64918300 | -2.12087100 | -0.69357000 |
| F | -10.75238600 | -0.09787000 | 0.28089200 |
| F | -10.70017500 | 0.99404400 | -1.56888300 |
| F | -11.21627300 | 2.00768800 | 0.26286900 |

**Compound 5 $T_1$ from Figure 4 optimized at DFT/M062X/def2-TZVP**

| | | | |
|---|---|---|---|
| | 32 | | |
| | -1960.058465 a.u. | | |
| C | -0.95248900 | 0.14214800 | -0.76425400 |
| C | -2.20980000 | 0.79693400 | -0.70519800 |
| C | -2.35134900 | 1.84594100 | 0.23241100 |
| C | -1.30609800 | 2.20197300 | 1.05138100 |
| C | -0.07967100 | 1.54533200 | 0.98442500 |
| C | 0.08377000 | 0.51527900 | 0.06951700 |
| H | -3.29107100 | 2.37743400 | 0.28744200 |
| H | 0.73248100 | 1.83676800 | 1.63644700 |
| H | 1.02963700 | -0.00324500 | 0.00611800 |
| C | -3.30187100 | 0.44093200 | -1.53334600 |
| H | -3.20435900 | -0.29761900 | -2.31974200 |
| N | -4.50453800 | 1.13201700 | -1.48155300 |
| N | -5.40691300 | 0.62195900 | -0.65993800 |
| C | -6.64119400 | 1.20795900 | -0.53022900 |
| S | -7.05019700 | 2.63224600 | -1.41739300 |
| N | -7.55392300 | 0.73551200 | 0.25664500 |
| C | -8.59228700 | 2.59110900 | -0.66067900 |
| C | -8.66147000 | 1.52898500 | 0.17584100 |
| H | -9.34869000 | 3.32929100 | -0.86538200 |
| H | -5.22733900 | -0.21320200 | -0.10826500 |
| C | -1.45615000 | 3.32535700 | 2.04236800 |
| C | -0.73860600 | -0.98515200 | -1.73760100 |
| C | -9.85595800 | 1.16966800 | 1.00731200 |
| F | -1.19777200 | 2.91000300 | 3.28827000 |
| F | -2.68412300 | 3.84606500 | 2.04139000 |
| F | -0.59926800 | 4.31994700 | 1.78135000 |
| F | 0.48732300 | -1.50143800 | -1.65266900 |
| F | -0.91028500 | -0.58877300 | -3.00739900 |
| F | -1.60278600 | -1.99150200 | -1.53000900 |
| F | -9.56714400 | 1.16995000 | 2.30936900 |
| F | -10.31927900 | -0.04380600 | 0.70377800 |
| F | -10.84894900 | 2.04366300 | 0.81154100 |

**Compound HM610 $S_0$ from Figure 4 optimized at DFT/M062X/def2-TZVP**

34

-1515.005122 a.u.

| | | | |
|---|---|---|---|
| C | -0.92744300 | 0.24878100 | -0.70986300 |
| C | -2.08491500 | 0.93204400 | -0.29427000 |
| C | -1.95684500 | 2.11709500 | 0.41177800 |
| C | -0.70540700 | 2.64063800 | 0.71520900 |
| C | 0.43441200 | 1.96377900 | 0.30349400 |
| C | 0.31858300 | 0.77077200 | -0.40721600 |
| H | -2.84285300 | 2.64738000 | 0.73448400 |
| H | 1.42121600 | 2.34392100 | 0.52319100 |
| H | 1.21998900 | 0.26280100 | -0.71636000 |
| C | -3.39896100 | 0.37483000 | -0.61675700 |
| H | -3.42716400 | -0.56338000 | -1.17032000 |
| N | -4.46883700 | 0.96717200 | -0.26157200 |
| N | -5.64246200 | 0.40001400 | -0.58914600 |
| C | -6.81139300 | 0.99765600 | -0.22947900 |
| S | -6.82007500 | 2.49457000 | 0.64792500 |
| N | -7.97280900 | 0.49943200 | -0.51266200 |
| C | -8.54237300 | 2.43806700 | 0.62745000 |
| C | -8.94642900 | 1.32528100 | -0.02147200 |
| H | -9.14596500 | 3.20470900 | 1.08129400 |
| H | -5.69433800 | -0.47359200 | -1.10061400 |
| C | -10.37601400 | 0.93631800 | -0.23686800 |
| F | -10.65933100 | -0.24263800 | 0.32309900 |
| F | -10.67323800 | 0.83569500 | -1.53496800 |
| F | -11.20433400 | 1.84295400 | 0.29572400 |
| O | -0.69894400 | 3.80771300 | 1.41145700 |
| O | -1.12276100 | -0.90909000 | -1.39937300 |
| C | 0.55247400 | 4.36557900 | 1.73559600 |
| H | 0.34587200 | 5.28114300 | 2.28306000 |
| H | 1.13665900 | 3.68993200 | 2.36703700 |
| H | 1.12618200 | 4.60351600 | 0.83518800 |
| C | 0.01756700 | -1.61683000 | -1.82846800 |
| H | 0.63176700 | -1.92775400 | -0.97912000 |
| H | -0.34826500 | -2.49657000 | -2.35085100 |
| H | 0.62233700 | -1.01408200 | -2.51129400 |

**Compound HM610 $T_1$ from Figure 4 optimized at DFT/M062X/def2-TZVP**

34

-1514.927018 a.u.

| | | | |
|---|---|---|---|
| C | -0.91335400 | 0.26495000 | -1.09128000 |
| C | -2.17345500 | 0.91758800 | -0.94773600 |
| C | -2.27757200 | 1.96180500 | -0.01447600 |
| C | -1.19262200 | 2.35143500 | 0.75551500 |
| C | 0.02765300 | 1.70176200 | 0.60739500 |
| C | 0.15611600 | 0.66051000 | -0.31971600 |
| H | -3.21512700 | 2.48898600 | 0.10314000 |
| H | 0.88872000 | 1.98471700 | 1.19395300 |
| H | 1.11487000 | 0.17267900 | -0.41709000 |
| C | -3.27121200 | 0.50386500 | -1.72949200 |
| H | -3.15499400 | -0.25649300 | -2.49123000 |
| N | -4.50331900 | 1.15325800 | -1.66268400 |
| N | -5.31780000 | 0.67092500 | -0.73555000 |
| C | -6.56194600 | 1.21223100 | -0.55479400 |
| S | -7.09509800 | 2.54317900 | -1.52136900 |
| N | -7.39748000 | 0.77419800 | 0.33358200 |
| C | -8.57762900 | 2.50827400 | -0.65110700 |
| C | -8.54026200 | 1.51801700 | 0.27067900 |
| H | -9.37854000 | 3.19737800 | -0.85690400 |
| H | -5.04661800 | -0.09926800 | -0.12964000 |
| O | -1.41425700 | 3.37547200 | 1.62265300 |
| O | -0.87356200 | -0.73364100 | -2.00879000 |
| C | -0.33422400 | 3.80277600 | 2.41812500 |
| H | -0.71094500 | 4.61648400 | 3.03203600 |
| H | 0.02554500 | 2.99782800 | 3.06547500 |
| H | 0.49315000 | 4.16690600 | 1.80224500 |
| C | 0.35221300 | -1.40259700 | -2.19589100 |
| H | 0.67579400 | -1.89640600 | -1.27561100 |
| H | 0.17801900 | -2.14939100 | -2.96561500 |
| H | 1.13099900 | -0.71105200 | -2.52842100 |
| C | -9.65585000 | 1.18599600 | 1.21389900 |
| F | -10.69960200 | 2.00097300 | 1.02165800 |
| F | -10.08733400 | -0.06613800 | 1.04806000 |
| F | -9.27937300 | 1.30807500 | 2.488478000 |